\documentclass[journal, twoside, final]{IEEEtran}
\pdfoutput=1
\usepackage{amssymb, amsmath, amsthm, amsfonts}
\usepackage{algorithmic}
\usepackage{algorithm}
\usepackage{array}
\usepackage[caption=false, font=footnotesize, labelfont=sf, textfont=sf]{subfig}
\usepackage{textcomp}
\usepackage{stfloats}
\usepackage{url}
\usepackage{verbatim}
\usepackage{graphicx}
\usepackage{cite}
\usepackage{bm, color}
\usepackage{booktabs}
\usepackage{setspace}
\usepackage{longtable, threeparttable}
\usepackage{makecell}
\usepackage{multirow}
\usepackage{ragged2e}
\usepackage{fancyhdr}
\usepackage{bbding}
\usepackage{color}
\usepackage{dsfont}
\usepackage{datetime}
\usepackage{float}
\newtheorem{lemma}{Lemma}

\newtheorem{proposition}{Proposition}

\newtheorem{remark}{Remark}

\allowdisplaybreaks[4]

\makeatletter

\newcommand{\Rmnum}[1]{\expandafter\@slowromancap\romannumeral #1@}
\makeatother

\begin{document}

\title{Edge Learning for Large-Scale Internet of Things With Task-Oriented Efficient Communication}

\author{Haihui Xie, Minghua~Xia,~\IEEEmembership{Senior Member,~IEEE}, Peiran Wu,~\IEEEmembership{Member,~IEEE}, \\ Shuai~Wang,~\IEEEmembership{Member,~IEEE}, and H. Vincent Poor,~\IEEEmembership{Fellow,~IEEE}

	\thanks{Manuscript received August 25, 2022; revised December 6, 2022 and April 11, 2023; accepted April 19, 2023. This work was supported in part by the National Natural Science Foundation of China under Grants 62171486 and U2001213, in part by the Guangdong Basic and Applied Basic Research Project under Grant 2021B1515120067, and in part by the U.S. National Science Foundation under Grant CNS-2128448. The associate editor coordinating the review of this paper and approving it for publication was A. S. Cacciapuoti. {\it (Corresponding author: Minghua Xia.)}}
	\thanks{Haihui Xie is with the School of Electronics and Information Technology, Sun Yat-sen University, Guangzhou 510006, China (e-mail: xiehh6@mail2.sysu.edu.cn).}
	\thanks{Minghua Xia and Peiran Wu are with the School of Electronics and Information Technology, Sun Yat-sen University, Guangzhou 510006, China, and also with the Southern Marine Science and Engineering Guangdong Laboratory, Zhuhai 519082, China (e-mail: \{xiamingh, wupr3\}@mail.sysu.edu.cn).}
   	\thanks{Shuai Wang is with the Guangdong-Hong Kong-Macao Joint Laboratory of Human-Machine Intelligence-Synergy Systems, Shenzhen Institute of Advanced Technology, Chinese Academy of Sciences, Shenzhen, 518055, China (e-mail: s.wang@siat.ac.cn).}
   	\thanks{H. Vincent Poor is with the Department of Electrical and Computer Engineering, Princeton University, Princeton, NJ 08544 USA (e-mail: poor@princeton.edu).}
	\thanks{%
	Color versions of one or more of the figures in this article are available online at https://ieeexplore.ieee.org.
	
	Digital Object Identifier XXX}
}

\markboth{IEEE Transactions on Wireless Communications} %
{Xie \MakeLowercase{\textit{et al.}}: Edge Learning for Large-Scale Internet of Things}
	
	\maketitle

\IEEEpubid{\begin{minipage}{\textwidth} \ \\[12pt] \centering 1536-1276 \copyright\ 2022 IEEE. Personal use is permitted, but republication/redistribution requires IEEE permission. \\
See \url{https://www.ieee.org/publications/rights/index.html} for more information.\end{minipage}}

	\begin{abstract}
		In the Internet of Things (IoT) networks, edge learning for data-driven tasks provides intelligent applications and services. As the network size becomes large, different users may generate distinct datasets. Thus, to suit multiple edge learning tasks for large-scale IoT networks, this paper performs efficient communication under the task-oriented principle by using the collaborative design of wireless resource allocation and edge learning error prediction. In particular, we start with multi-user scheduling to alleviate co-channel interference in dense networks. Then, we perform optimal power allocation in parallel for different learning tasks. Thanks to the high parallelization of the designed algorithm, extensive experimental results corroborate that the multi-user scheduling and task-oriented power allocation improve the performance of distinct edge learning tasks efficiently compared with the state-of-the-art benchmark algorithms. 
	\end{abstract}
	
	 \IEEEpubidadjcol
	
	\begin{IEEEkeywords}
		\noindent Edge learning, multi-user scheduling, Internet of Things, parallel computing.
	\end{IEEEkeywords}
	
	%
	%
	
\section{Introduction} \label{Section1}
\IEEEPARstart{M}{assive} connectivity is a feature of many Internet of Things (IoT) deployments, and in such applications, the widely connected users generate an enormous amount of data. To extract information from these data, IoT users can train learning models to effectively represent different types of data \cite{8664630}. Although these IoT users have a specific capability to train simple learning models, the limited memory, computing, and battery capability deter the application of complicated models, such as deep neural networks \cite{Toward2017Dawy}. To deal with this issue, edge learning techniques have emerged, transferring the burden of complex model updates to an edge server, i.e., leveraging the storage, communication, and computational capabilities at the edge server \cite{8970161}. Moreover, the edge server also allows rapid access to the enormous amount of data distributed over end-user devices for fast model learning, providing intelligent services and applications for IoT users \cite{Edge2020Deng}.
	
	In edge learning, the main design objective is to acquire fast intelligence from the rich but highly distributed data of subscribed IoT users. This critically depends on data processing at edge servers, as well as efficient communication between edge servers and IoT users \cite{2024789118}. However, compared to increasingly high processing speeds at edge servers, communication suffers from the hostility of wireless channels and consequently becomes the bottleneck for ultra-fast edge learning \cite{9606720}. Moreover, the diversity of ubiquitous IoT users and complex transmission environments lead to additional interference. Such interference would significantly deteriorate the reliability and communication latency of the IoT network while uploading a vast amount of data to an edge server \cite{9380667}. To address these issues, the traditional  {\it data-oriented} communication systems are designed to maximize network throughput based on Shannon's theory, which targets transmitting data reliably given the limited radio resources \cite{Spatial2019Cui}. However, such approaches are often ineffective in edge learning, as they rely only on classical source coding and channel coding theory and fail to improve learning performance \cite{Edge2020Li}. Therefore, a paradigm shift for wireless system design is required from data-oriented to {\it task-oriented} communications.
	
	 \IEEEpubidadjcol
	\subsection{Related Works and Motivation}
	
	The initial attempts of task-oriented communications were to design {\it task-aware} transmission phases rather than end-to-end data reconstruction, see, e.g., \cite{4027772, 9344705, 8809196} for designing task-aware reporting phases in the case of distributed inference tasks. Unlike task-aware efficient transmissions, several pioneering works \cite{9606667, 2207-00969, 2206-05949, 9653664, Machine2020Wang} have also studied task-oriented schemes in edge learning systems. The work \cite{9606667} designed a task-oriented communication scheme to realize a trade-off between preserving the relevant information and fitting with bandwidth-limited edge inference nicely. In \cite{2207-00969, 2206-05949}, task-oriented methods were proposed to maximize learning accuracy by jointly designing sensing, communication, and computation. The work \cite{9653664} proposed a task-oriented transmission scheme to accelerate learning processes efficiently by capturing the semantic features from the correlated multimodal data of the IoT users. Nevertheless, these task-oriented communication schemes are mostly heuristics based, and it is necessary to improve their learning performance via additional optimization. 
	
	To overcome the drawback of heuristic methods, the work \cite{Machine2020Wang} proposed a learning-centric power allocation (LCPA) model to guide power allocation efficiently so as to optimize the limited network resources under the task-oriented principle. First, the learning performance was approximated by parameter fitting to capture the shape of learning models. Then, a majorization minimization (MM) algorithm was designed to allocate transmit power efficiently. However, the MM algorithm is inefficient for large-scale IoT networks due to its high computational complexity and lack of co-channel interference (CCI) management. In particular, for massive IoT users, it is necessary to collect multi-modal datasets and process heterogeneous learning tasks concurrently, making it imperative for designing parallel and low-complexity task-oriented communication algorithms. Furthermore, concurrent transmissions of massive IoTs will inevitably yield severe CCI, thus degrading the performance of task-oriented communications \cite{9252948}. But due to the highly-coupled CCIs, the associated power allocation problem is non-separable and non-convex, which is non-trivial to realize the inference management and algorithm parallelization. 
		
	To fill the gap, this paper designs a task-oriented power allocation model for efficient communications in large-scale IoT networks with edge learning. On the one hand, as a task-oriented learning system involves heterogeneous learning tasks, it is necessary to predict the required resources for training different tasks. Therefore, our method is designed as an offline learning procedure that fits historical datasets to a performance prediction model and an online inference procedure that guides the IoT-edge communications with the pre-trained performance model. Note that this performance model can be fine-tuned by exploiting a small amount of real-time data from active IoT users. On the other hand, we formulate a task-oriented power allocation problem to guide communication-efficient data collection for large-scale IoT networks. To alleviate CCI, multi-user scheduling is first performed before power allocation. Then, a highly parallel algorithm is designed for different learning tasks. Lastly, we develop an accelerated algorithm to make the parallel algorithm more efficient. In brevity, Table~\ref{Table0} compares the existing and proposed schemes.

	\begin{table*}[t!]
		\centering
		\small
		\renewcommand\arraystretch{1.2}
		\caption{Comparison of Existing and Proposed Schemes.}
		\label{Table0}
		\begin{threeparttable}[!t]
			\resizebox{\linewidth}{!}{
				\begin{tabular}{!{\vrule width1.2pt} c !{\vrule width1.2pt} c|c|c|c|c|c|c|c!{\vrule width1.2pt}}
				\Xhline{1.2pt} 
					\textbf{Type} & \textbf{Scheme} &  
					\textbf{\begin{tabular}[c]{@{}c@{}}Learning \\ Efficiency\tnote{a} \end{tabular}} 
					& \textbf{\begin{tabular}[c]{@{}c@{}}Multi-task \\ Multi-modal\tnote{b} \end{tabular}}
					& \textbf{\begin{tabular}[c]{@{}c@{}}Algorithm \\ Complexity \end{tabular}}
					&  \textbf{\begin{tabular}[c]{@{}c@{}}Parallelism \\ Acceleration \end{tabular}}
					&
					\textbf{\begin{tabular}[c]{@{}c@{}}Optimal \\ Objective \end{tabular}}
					& \textbf{\begin{tabular}[c]{@{}c@{}}Task \\ Fairness \end{tabular}}
					& \textbf{\begin{tabular}[c]{@{}c@{}}Multi-user \\ Scheduling \end{tabular}}
					\\ \Xhline{1.2pt} 
					\multirow{3}{*}{\textbf{\begin{tabular}[c]{@{}c@{}} Task-aware \end{tabular}}}  & \cite{4027772}          &   +       & \XSolidBrush    & +++                            & +         & \XSolidBrush      & \XSolidBrush                                                                  & \XSolidBrush
					\\ \cline{2-9}
					& \cite{9344705}          &   +       & \XSolidBrush   & +++                            & +         & \XSolidBrush  & \XSolidBrush                                                                  & \XSolidBrush
					\\ \cline{2-9}
					& \cite{8809196}   &  +              & \XSolidBrush       & +++         & +     & \XSolidBrush    & \XSolidBrush     &    \XSolidBrush                                                                                                  \\ \hline
					\multirow{5.5}{*}{\textbf{\begin{tabular}[c]{@{}c@{}} Task-oriented \end{tabular}}}  &
					\begin{tabular}[c]{@{}c@{}} \cite{9606667} \end{tabular} &            ++                                                                     & \Checkmark      & ++                         & +                                                                   &  \XSolidBrush                                                                   & \XSolidBrush & \XSolidBrush
					\\ \cline{2-9}
					& \cite{2207-00969, 2206-05949}      & ++                                                                                                                            & \Checkmark                                                               & ++                                                     & +                                                                     & \XSolidBrush                                                                     & \XSolidBrush & \XSolidBrush
					\\ \cline{2-9}
					& \cite{9653664}       &       ++                                                    &           \Checkmark                                                     & ++                                                                  & +                                                             & \XSolidBrush                                                                                                 & \XSolidBrush  & \XSolidBrush
					\\ \cline{2-9}
					& \cite{Machine2020Wang}       &       +++                                                    &           \XSolidBrush                                                      & +++                                                                 & N/A                                                             & Min-max                                                                                                                                     & \XSolidBrush  & \XSolidBrush
					\\ \cline{2-9}
					& \textbf{Proposed}                                                & +++                                                                                                                          & \Checkmark    & +                                                              & +++                                                                      & \begin{tabular}[c]{@{}c@{}}Weighted \\ sum \end{tabular}                                                               &  \Checkmark  & \Checkmark                                      \\ \hline
					\Xhline{1.2pt} 
				\end{tabular}
			}
			{\scriptsize
			\begin{tablenotes}
					\item[a] The symbols ``+, ++, +++'' indicate low, moderate, and high capability, respectively.
					\item[b] The tick ``\Checkmark'' indicates a functionality supported, whereas the cross ``\XSolidBrush" indicates not supported.
			\end{tablenotes}
			}	
		\end{threeparttable}
	\end{table*}

	\subsection{Summary of Main Results}
	Aiming at efficient communication for task-oriented edge learning, this paper starts with a multi-user scheduling strategy to mitigate CCI. In particular, a relaxation-and-rounding algorithm is exploited to identify scheduled users efficiently, and an approximate closed-form solution is obtained. Secondly, a parallel algorithm with Gauss-Seidel methods is developed. By a set of variable decompositions, we realize a highly parallel iteration. Thirdly, we design an accelerated algorithm to speed up this parallel algorithm. Finally, extensive experimental results demonstrate the efficiency of our design. In summary, the main contributions are as follows:
		\begin{itemize}
			\item[1)] A task-oriented power allocation model is proposed to process multiple distinct datasets at the edge. Moreover, a multi-user scheduling strategy is performed before power allocation to mitigate CCI for large-scale IoT networks efficiently.
			\item[2)] A highly parallel algorithm is designed for the task-oriented power allocation problem. By variable decompositions and eliminating auxiliary variables, the power allocation in the presence of CCI is realized efficiently in parallel. 
			\item[3)] An accelerated algorithm is developed to make the parallel algorithm more efficient for large-scale IoT networks. Specifically, this algorithm utilizes the Lipschitz continuous property of the learning error and the identity mapping of the gain matrix to improve the convergence.
			\item[4)] Extensive experimental results show that the multi-user scheduling strategy can mitigate CCI in large-scale IoT networks. Moreover, our parallel and accelerated algorithms efficiently solve task-oriented power allocation problems with a significantly shorter computation time than the existing algorithms.
		\end{itemize}

\subsection{Organization}
	The rest of this paper is organized as follows. Section~\Rmnum{2} describes the system model and formulates a task-oriented power allocation problem. Section~\Rmnum{3} performs multi-user scheduling to mitigate CCI and designs a parallel algorithm for solving the task-oriented power allocation problem. Section~\Rmnum{4} develops an accelerated algorithm for large-scale IoT networks. Section~\Rmnum{5} discusses the experimental results, and finally, Section~\Rmnum{6} concludes the paper.
	
	{\it Notation}:
	Scalars, column vectors, and matrices are denoted by regular italic letters, lower- and upper-case letters in bold typeface, respectively. The symbol $\bm{\mathsf{1}}$ indicates a column vector with all entries being unity. The superscripts $(\cdot)^T$ and $(\cdot)^H$ denote the transpose and Hermitian transpose of a vector or matrix, respectively, and the subscript $\|\bm{x}\|_2$ denotes the two-norm of~$\bm{x}$. The abbreviation $\mathcal{CN} ( \bm{0}, \, \varrho\bm{I} )$ stands for a multi-variable complex Gaussian distribution with mean vector $\bm{0}$ and variance matrix $\varrho \bm{I}$. The notation $\left| \mathcal{X} \right|$ denotes the cardinality of the set $\mathcal{X}$, and $\mathcal{Y} \setminus \mathcal{X}$ denotes the complement of set $\mathcal{Y}$ except for $\mathcal{X}$. The matrix operation $\bm{A}(\mathcal{K}_{i}, \, \mathcal{K}_{j})$ denotes a sub-matrix of size $|\mathcal{K}_{i}| \times |\mathcal{K}_{j}|$ that includes the rows and columns in $\bm{A}$ specified by the sets of indices $\mathcal{K}_{i}$ and $\mathcal{K}_{j}$, respectively. The arithmetic operations $\bm{x} \succeq \bm{y}$, $\bm{x} \circ \bm{y}$, and $\left\langle \bm{x}, \, \bm{y} \right\rangle$ denote that each element of $\bm{x}$ is greater than or equal to the counterpart of $\bm{y}$, the Hadamard product, and the inner product of two vectors, respectively. The Landau notation $\mathcal{O} ( \cdot )$ denotes the order of arithmetic operations. Further, we define a binary function $\lfloor w \rceil = 1$ if $w \geq 0.5$, and $\lfloor w \rceil = 0$ otherwise, and $\lfloor \bm{w} \rceil \triangleq \left[ \lfloor w_{1} \rceil, \, \lfloor w_{2} \rceil, \, \cdots, \, \lfloor w_{K} \rceil \right]^T \in \mathbb{R}^{K \times 1}$ is computed for each element of $\bm{w}$. Finally, the floor function $\left\lfloor x \right\rfloor \triangleq \max\{ n \in \mathbb{Z} : n \leq x \}$, where $\mathbb{Z}$ is the set of integers.

\section{System Model and Problem Formulation}
	In this section, we first describe a task-oriented edge learning system. Then, we formulate the task-oriented power allocation problem with multi-user scheduling.	

	\subsection{System Model}
\begin{figure}[t!]
			\centering
			\includegraphics[width=0.45\textwidth]{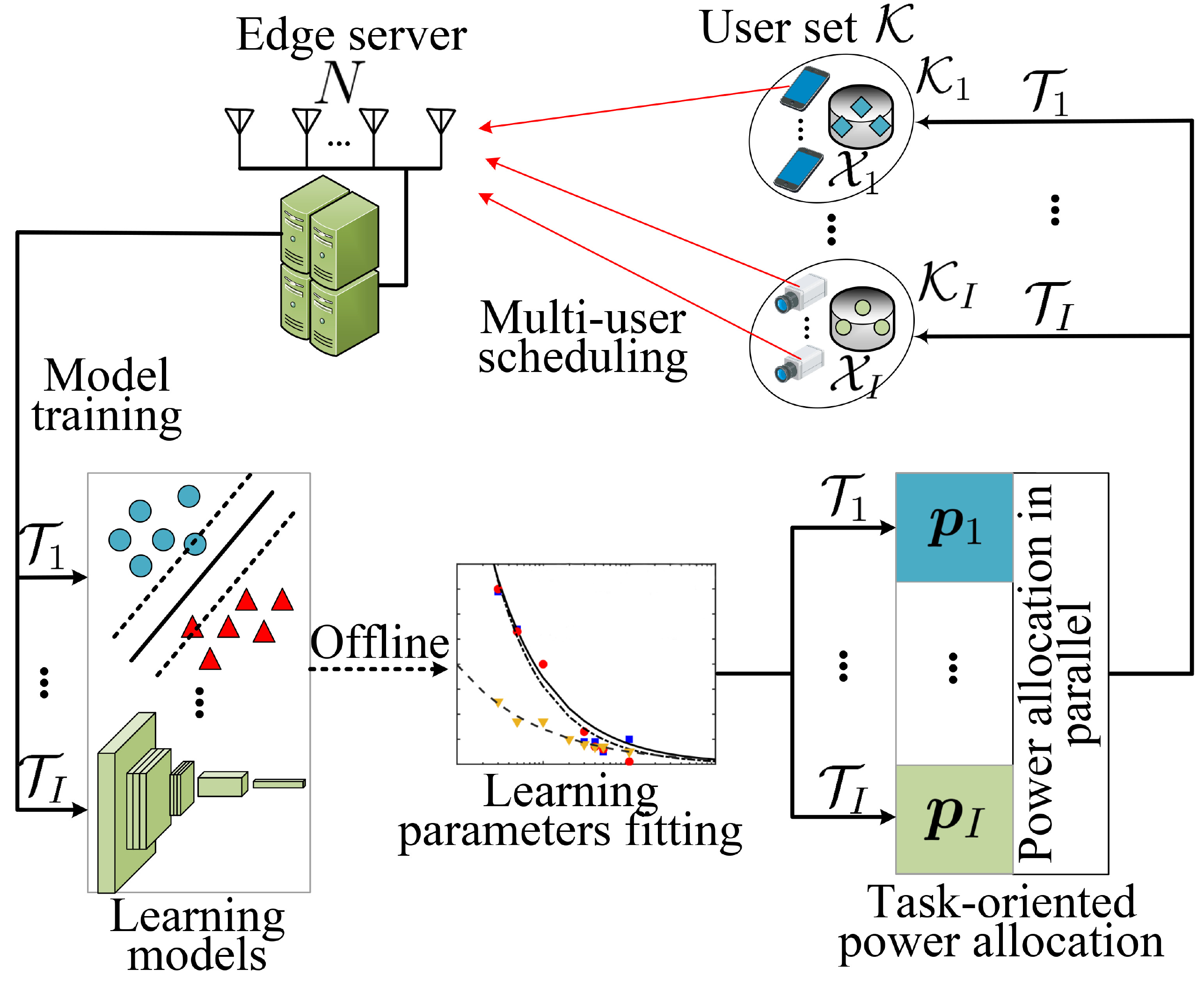}
			\caption{The system model of task-oriented edge learning.}
			\label{Fig1}
		\end{figure}

	Figure~\ref{Fig1} illustrates a task-oriented edge learning system consisting of an edge server equipped with $N$ antennas, $I$ different learning tasks $\{ \mathcal{T}_{1}, \, \mathcal{T}_{2}, \, \cdots, \, \mathcal{T}_{I} \}$ with corresponding user sets $\mathcal{K} \triangleq \{ \mathcal{K}_{1}, \, \mathcal{K}_{2}, \, \cdots, \, \mathcal{K}_{I} \}$ and power allocation parameters $\bm{p} \triangleq \left[ \bm{p}_{1}^{T}, \, \bm{p}_{2}^{T}, \, \cdots, \, \bm{p}_{I}^{T} \right]^{T} \in \mathbb{R}^{K}$, where $\bm{p}_{i} \triangleq \left[ p_{i_{1}}, \, p_{i_{2}}, \cdots, \, p_{|\mathcal{K}_{i}|} \right]^{T}$, $i \in \mathcal{I} \triangleq \{ 1, \, 2, \, \cdots, \, I \}$, $i_{j} \in \mathcal{K}_{i}$, $j \in \{1, \, 2, \cdots, |\mathcal{K}_{i}| \}$, $K \triangleq |\mathcal{K}|$, and $p_{k}$, $k \in \mathcal{K}$, denotes the transmit power of the $k^{\rm th}$ user. As for the $I$ different learning tasks in Fig.~\ref{Fig1}, each of them concerns a set of transmitted data, a multi-user scheduling algorithm, a learning model, a process of parameter fitting for a learning model, and a task-oriented power allocation problem. 
	
	To improve the performance of edge learning, multi-user scheduling is adopted to alleviate CCI in large-scale IoT networks, and task-oriented power allocation is performed to implement efficient communications. To maximize the network utility function of long-term average data rates, by recalling the seminal Shannon formula, the achievable data rate of user $k$ in the presence of multi-user scheduling can be expressed as \cite{Spatial2019Cui}
		\begin{equation} \label{S2-EQ-1}
			R_{k} = \log_{2} \Bigg( 1 + \dfrac{ G_{k, k} p_{k} }{ \sum_{ \ell \in \Pi_{\mathcal{S}} ( \mathcal{K} ) \setminus k } G_{k, \ell} p_{\ell} + \sigma^{2} } \Bigg), \, k \in \Pi_{\mathcal{S}} (\mathcal{K}_{i}),
		\end{equation}
	where $\Pi_{\mathcal{S}} ( \cdot )$ denotes a projection function of multi-user scheduling; $\sigma^{2}$ is the variance of additive white Gaussian noise; $G_{k, \ell}$ represents the composite channel gain from the $\ell^{\rm th}$ user to the edge server when decoding data of the $k^{\rm th}$ user, computed as $G_{k, k} = \rho_{k} \| \bm{h}_{k} \|_{2}^{2}$ if $\ell = k$, and $G_{k, \ell} = { \rho_{\ell} | \bm{h}_{k}^{H} \bm{h}_{ \ell } |^{2} }/{ \| \bm{h}_{k} \|^{2}_{2} }$ if $\ell \neq k$, with $\bm{h}_{k} \in \mathbb{C}^{N \times 1}$ being the complex-valued channel fast-fading vector from the $k^{\rm th}$ user to the edge server and $\rho_{k}$ being the path loss of the $k^{\rm th}$ user. By \eqref{S2-EQ-1}, the number of samples transmitted by user $k$ for the learning task $\mathcal{T}_{i}$ at the edge server can be computed as
		\begin{equation} \label{S2-EQ-2}
			D_{i} = \sum_{ k \in \Pi_{\mathcal{S}} ( \mathcal{K}_{i} ) } \left\lfloor \dfrac{ B T R_{k} }{ V_{i} } \right\rfloor + A_{i} \approx \sum_{k \in \Pi_{\mathcal{S}} (\mathcal{K}_{i})}  \dfrac{ B T R_{k} }{ V_{i} } + A_{i},
		\end{equation}
	where $B$ is the total bandwidth in Hz; $T$ is the transmission time in seconds; $V_{i}$ is the number of bits for each data, and $A_{i}$ is the initial number of historical data for the $i^{\rm th}$ pre-trained task.
	
	This paper considers the average channel over a long transmission period instead of assuming a static channel. The reason is twofold. On the one hand, to fine-tune diverse learning models, it is essential to require a relatively long transmission time with tens or hundreds of seconds to obtain a large number of datasets. On the other hand, the effect of multi-user scheduling can only be disclosed in the context of a long-term channel average rather than an instantaneous channel realization. Assume that the transmission period consists of different time slots. The channels are quasi-static during each time slot and vary in consecutive time slots. Therefore, $G_{k, k}$ and $G_{k, \ell}$ in \eqref{S2-EQ-1} could denote the average channels gain during these slots. 

	\subsection{Problem Formulation}
	To establish a connection between wireless resource allocation and the performance of machine learning, the work \cite{Machine2020Wang} conceived a non-linear exponential function $\Theta_{i} ( D_{i} | a_{i}, \, b_{i} ) \triangleq a_{i}  D_{i}^{ - b_{ i } }$ to capture the shape of the learning error function, where $a_{i}$ and $b_{i}$ are tuning parameters that denote the model complexity and account for the non-independent and identically distributed (n.i.i.d.) parallel datasets, respectively. In practice, the values of $a_{i}$ and $b_{i}$ are obtained by fitting the learning error function from the historical dataset. This fitted function matches the experimental data of the machine learning model very well. In line with this idea and multi-user scheduling, we formulate a task-oriented power allocation problem:
		\begin{subequations}
			\begin{align}
				\mathcal{P}1: \min_{\bm{p}, \, \Pi_{\mathcal{S}}} & \sum_{ i \in \mathcal{I} } \lambda_{i} \times a_{i} D_{i}^{ - b_{ i } } \label{S1-EQ-3} \\
				{\rm s.t.} & \sum_{ k \in \mathcal{K} } p_{k} = P, \, p_{k} \geq 0, \, \forall k \in \mathcal{K}, \label{S1-EQ-3a} \\
				& \Pi_{\mathcal{S}} ( \mathcal{K} ) \subseteq \mathcal{K}, \label{S1-EQ-3b} \\
				& p_{k} = 0, \, \forall k \in \mathcal{K} \setminus \Pi_{\mathcal{S}} ( \mathcal{K} ) , \label{S1-EQ-3c}
			\end{align}
		\end{subequations}	
	where $\lambda_{i} \triangleq A_{i} V_{i} / (\sum_{j \in \mathcal{I}} A_{j} V_{j})$ is a weight of diverse datasets. In general, the power allocation of all scheduled users shall satisfy $\sum_{ k \in \mathcal{K} } p_{k} \leq P$, i.e., not exceeding the total power budget~$P$. As a larger value of $\sum_{ k \in \mathcal{K} } p_{k}$ always improves the learning performance, \eqref{S1-EQ-3a} is obtained \cite{Machine2020Wang}. \eqref{S1-EQ-3b} means to schedule a subset of users, and \eqref{S1-EQ-3c} implies no power is allocated to inactive~users. 

Compared to the conventional min-max objective function used in \cite{Machine2020Wang}, it only focuses on the worst learning task, even if the task is not critical for real-world application. Thus, it is not suitable for the multi-task multi-modal scenario considered in this paper. Instead, the weighted sum model in \eqref{S1-EQ-3} can optimize multiple tasks simultaneously. In particular, the objective function can adapt to different learning tasks by adjusting the weight factors $\lambda_i, i \in \mathcal{I}$ in \eqref{S1-EQ-3}.
	\begin{remark}[On the learning loss model]
		In theory, the training procedure of any smooth learning network can be modeled as a Gibbs distribution of networks characterized by a temperature parameter $T_g$. The asymptotic generalization loss $\epsilon_i$ as the number of samples $D_i$ for the $i^{\rm th}$ learning task goes to infinity can be expressed as \cite[Eq.~$3.12$]{456056}
			\begin{align}\label{A2}
				&\epsilon_i=\epsilon_{i,\mathrm{min}}+\left(\frac{T_g}{2}+\frac{\mathrm{Tr}(\bm{U}_i\bm{V}_i^{-1})}{2W_i}\right)W_iD_i^{-1}, \quad \text{as } D_i\rightarrow +\infty,
			\end{align}
		where $\epsilon_{i,\mathrm{min}}\geq 0$ is the minimum error for the considered learning system, $W_i$ is the number of parameters, and $D_i$ is the number of samples. The matrices $\bm{U}_i$ and $\bm{V}_i$ denote the second-order and first-order derivatives of the generalization loss with respect to the parameters of model $i$. By setting $a_i = \left(\frac{T_g}{2}+\frac{\mathrm{Tr}(\bm{U}_i\bm{V}_i^{-1})}{2W_i}\right)W_i$, $b_i = -1$ and $\epsilon_{i,\mathrm{min}} = 0$, \eqref{A2} reduces to the proposed learning loss model $\Theta_{i} ( D_{i} | a_{i}, \, b_{i} ) \triangleq a_{i}  D_{i}^{ - b_{ i } }$, implying the proposed model holds in the asymptotic sense. In practice, $\epsilon_{i,\mathrm{min}}$ in \eqref{A2} cannot always approach zero as the number of samples reaches infinite, even for some simple learning models. For ease of mathematical tractability, we set $\epsilon_{i,\mathrm{min}}=0$ in this paper by assuming that the learning model is powerful enough such that given an infinite amount of data, the learning loss becomes zero.
	\end{remark}
\section{Algorithm Development in Parallel}	
	In this section, we first describe a multi-user scheduling algorithm given a power allocation. Then, we design a parallel algorithm to solve the power allocation problem.

	\subsection{Multi-user Scheduling Algorithm}
	In the task-oriented learning system, multi-user scheduling is an effective strategy for solving the massive-connectivity problem. Traditional approaches to multi-user scheduling are almost non-convex algorithms, e.g., greedy heuristic search. In particular, as the number of scheduling cases increases exponentially with the number of users, it is hard to enumerate all possible subsets of users explicitly \cite{Joint2020Shi}. To deal with this problem, we introduce binary variables $w_{k}$, $k \in \mathcal{K}$, to replace $\Pi_{\mathcal{S}}$ defined immediately after \eqref{S2-EQ-1}. More specifically, $w_{k} = 1$ if $k \in \Pi_{ \mathcal{S} } \left( \mathcal{K} \right)$, and $w_{k} = 0$ otherwise. As a result, given $p_{k}$, inserting \eqref{S2-EQ-1}-\eqref{S2-EQ-2} into $\mathcal{P}1$, the multi-user scheduling problem in the $i^{\rm th}$ group, $i \in \mathcal{I}$, is formulated as
		\begin{subequations}
			\begin{align}
				\mathcal{P}2: \min_{ \bm{w} } \, & a_{i} \Bigg( \dfrac{B T}{ V_{i} } \sum_{ k \in \mathcal{K}_{i} } w_{k} R_{k} + A_{i} \Bigg)^{ - b_{ i } } \\
				{\rm s.t.} \ & w_{k} \in \{ 0, \, 1 \}, \label{S3-EQ-5a}  \\
				& \sum_{ k \in \mathcal{K}_{i} } w_{k} \leq N_{i}, \label{S3-EQ-5b}
			\end{align}
		\end{subequations}
	where $R_{k} = \log_{2} ( 1 + { G_{k, k} p_{k} } / (\sum_{ \ell \in \mathcal{K} \setminus k } w_{\ell} G_{k, \ell} p_{\ell} + \sigma^{2}))$ as per \eqref{S2-EQ-1}, $\bm{w} \triangleq \left[ w_{1}, \, w_{2}, \, \cdots, \, w_{K} \right]^{T}$, and \eqref{S3-EQ-5b} is derived from \eqref{S1-EQ-3b} with $N_{i}$ being the maximal allowed number of active users for the $i^{\rm th}$ learning task.
	
	To solve $\mathcal{P}2$, we adopt a relaxation-and-rounding algorithm \cite{Energy2020Zeng}. First, we relax the binary constraint \eqref{S3-EQ-5a} as the real-valued constraint $0 < w_{k} \leq 1$. Then, we provide the following Proposition~\ref{S3-P1} to obtain an approximate closed-form solution to the relaxed version of $\mathcal{P}2$.
		\begin{proposition}[Multi-user scheduling algorithm] \label{S3-P1}
			Given $p_{k}, \, k \in \mathcal{K}_{i}$, the multi-user scheduling variable $w_{k}$ is analytically determined by
				\begin{equation} \label{S3-P1-EQ-6}
					\tilde{w}_{k} = \min \left( \max\left( \dfrac{ \tilde{G}_{k, k} p_{k} }{ \delta_{k} \left( \exp \left( \frac{ G_{k, k} p_{k} }{ \delta_{k} + G_{k, k} p_{k} } + \nu_{i} \right) - 1 \right) }, \, \epsilon \right), \, 1  \right),
				\end{equation}
			where $\delta_{k} \triangleq \sum_{ \ell \in \mathcal{K} \setminus k } \tilde{G}_{k, \ell} p_{\ell} + \sigma^{2}$ with $\tilde{G}_{k, \ell} \triangleq w_{k} {G}_{k, \ell} $; $\nu_{i} > 0$ is a tuning parameter for controlling the sparsity of the solution, and $\epsilon > 0$ is a small positive number close to zero. When the multi-user scheduling algorithm converges, $\tilde{\bm{w}} \triangleq \left[ \tilde{\bm{w}}_{1}^{T}, \, \tilde{\bm{w}}_{2}^{T}, \, \cdots, \, \tilde{\bm{w}}_{I}^{T} \right]^{T}$ with $\tilde{ \bm{w} }_{i} \triangleq \left[ \tilde{ w }_{i_1}, \, \tilde{ w }_{i_2}, \, \cdots, \, \tilde{ w }_{i_{|\mathcal{K}_{i}|}} \right]^{T}$ is the optimal solution, in which each element is rounded off to the nearest integer $1$ or $0$, i.e., $\lfloor \tilde{\bm{w}} \rceil$.
		\end{proposition}
		\begin{proof}
			See Appendix~\ref{SA-A}.
		\end{proof}
	Proposition~\ref{S3-P1} shows that the multi-user scheduling decision is analytically determined, with extremely low computational complexity proportional to the number of users, i.e., $\mathcal{O} (K)$. Also, it is noted that our multi-user scheduling strategy is fair with respect to different learning tasks, but the fairness among users is not accounted for since it is beyond the scope of this paper.

	\subsection{Parallel Power Allocation}
	After the multi-user scheduling is performed by Proposition~\ref{S3-P1}, the task-oriented power allocation problem can be rewritten as
	\begin{subequations}
		\begin{align}
			\mathcal{P}3: \min_{ \bm{p} } \, & \sum_{ i \in \mathcal{I} } \lambda_{i} \times a_{i} \Bigg(  \dfrac{B T}{ V_{i} } \sum_{ k \in \mathcal{K}_{i} } \tilde{w}_{k} \tilde{R}_{k} + A_{i} \Bigg)^{ - b_{ i } } \\
			{\rm s.t.} \ & \sum_{ k \in \mathcal{K} } ( 1 - \tilde{w}_{k} ) p_{k} \leq \epsilon, \label{S3-EQ-8a} \\
			& \eqref{S1-EQ-3a} , \nonumber
		\end{align}
	\end{subequations}
	where $\tilde{R}_{k} \triangleq \log_{2} ( 1 + { G_{k, k} p_{k} } / ( \sum_{ \ell \in \mathcal{K} \setminus k } \tilde{w}_{\ell} G_{k, \ell} p_{\ell} + \sigma^{2}) )$; \eqref{S3-EQ-8a} is the relaxation of \eqref{S1-EQ-3c}, which means little power is reserved for inactive users.
	
	The optimization problem $\mathcal{P}3$ is non-convex; even worse, its computational complexity rises with the number of users and tasks. To address these issues, we propose a parallel first-order algorithm. As there is a dependency on the power and CCI terms amongst different learning tasks, it is hard to realize parallelization for various tasks in a straightforward manner. An efficient strategy is to introduce auxiliary variables to separate these terms independently. The resulting problem involves a set of sub-problems by variable decompositions, and these sub-problems are easier to be solved in parallel \cite{ADMM2018Lu, ADMM2012He}.
	
	Now, we begin to extract the relevant sub-problems. First, to divide the interference term, we introduce additional variables~$\bm{\delta}$, defined as
		\begin{equation} \label{SA-EQ-B2}
			\bm{\delta} \triangleq \bm{\Delta} \bm{p} + \sigma^{2} \bm{\mathsf{1}},
		\end{equation}
	where $\bm{\delta} = \left[ \bm{\delta}_{1}^{T}, \, \bm{\delta}_{2}^{T}, \, \cdots, \, \bm{\delta}_{I}^{T} \right]^{T}$ with $\bm{\delta}_{i} \triangleq [ \delta_{i_1}, \, \delta_{i_2}, \, \cdots, \, \delta_{ i_{|\mathcal{K}_{i}|} } ]^{T}$, and $\bm{\Delta} \triangleq \tilde{\bm{G}} - \tilde{\bm{D}}$ with $\tilde{\bm{G}} \triangleq \left[ \begin{array}{ccc} \bm{G} ( 1, \, : )^{T} \circ \tilde{ \bm{w} } & \cdots & \bm{G} ( K, \, : )^{T} \circ \tilde{ \bm{w} } \end{array} \right]^{T}$ and $\tilde{\bm{D}} \triangleq \left[ \begin{array}{ccc} \bm{D} ( :, \, 1 ) \circ \tilde{ \bm{w} } & \cdots & \bm{D} ( :, \, K ) \circ \tilde{ \bm{w} } \end{array} \right]$. Here, the $(k, \, \ell)^{\rm th}$ element of $\bm{G}$ is made up of $G_{k, \, \ell}$, and the $k^{\rm th}$ diagonal element of the diagonal matrix $\bm{D}$ is made up of $G_{k, \, k}$. By partitioning users $\mathcal{K}$ into $I$ groups of users $\{ \mathcal{K}_{i} \}_{i = 1}^{I}$ and introducing a set of variables $\{ \bm{z}_{i} \}_{i = 1}^{I}$ and $\{ P_{i} \}_{i = 1}^{I}$, we have
		\begin{subequations} 
			\begin{align}
				& \bm{\Delta} \left( :, \, \mathcal{K}_{i} \right) \bm{p}_{i} = \bm{z}_{i}, \label{SA-EQ-B3a} \\
				& \sum_{i \in \mathcal{I}} \bm{z}_{i} = \bm{\delta} - \sigma^{2} \bm{\mathsf{1}}, \label{SA-EQ-B3b} \\
				& \bm{\mathsf{1}}^{T} \bm{p}_{i} = P_{i}, \label{SA-EQ-B3c} \\
				& \sum_{i \in \mathcal{I}} P_{i} = P, \label{SA-EQ-B3d}
			\end{align}
		\end{subequations}
	where $\bm{z}_{i} \triangleq \left[ z_{1, i}, \, z_{2, i}, \, \cdots, \, z_{K, i} \right]^{T}$, and $\bm{p}_{i} \triangleq [ p_{i_1}, \, p_{i_2}, \, \cdots, \, p_{ i_{ \left| \mathcal{K}_{i} \right| } } ]^{T}$. It is noteworthy that $\bm{z}_i$ in \eqref{SA-EQ-B3a}-\eqref{SA-EQ-B3b} and $P_{i}$ in \eqref{SA-EQ-B3c}-\eqref{SA-EQ-B3d} are auxiliary variables. As a result, $\mathcal{P}3$ can be transformed into:
		\begin{subequations}
			\begin{align}
				\mathcal{P}4: \min_{\{ \bm{p}_{i}, \, \bm{\delta}_{i}, \, \bm{z}_{i}, \, P_{i} \}_{i \in \mathcal{I}} } \, & \sum_{ i \in \mathcal{I} } \lambda_{i} \times \Phi_{i} ( \bm{p}_{ i } | \bm{\delta}_{i} ) \label{S3-P2-EQ-9a} \\
				{\rm s.t.} \ & \bm{\delta} \succeq \sigma^{2} \bm{\mathsf{1}}, \, \bm{p}_{i} \succeq \bm{0}, \label{S3-P2-EQ-9e} \\
				& \eqref{S3-EQ-8a}, \, \eqref{SA-EQ-B3a}, \, \eqref{SA-EQ-B3b}, \, \eqref{SA-EQ-B3c}, \, \eqref{SA-EQ-B3d}, \nonumber
			\end{align}
		\end{subequations}
	where \eqref{S3-P2-EQ-9e}  is naturally satisfied as $ \sum_{ \ell \in \mathcal{K} \setminus k } \tilde{G}_{k, \ell} p_{\ell} \geq 0$ and $p_{\ell} \geq 0$. Specially, $\Phi_{i} ( \bm{p}_{ i } | \bm{\delta}_{i} )$ in \eqref{S3-P2-EQ-9a} is explicitly given by
		\begin{equation} \nonumber
			\Phi_{i} ( \bm{p}_{ i } | \bm{\delta}_{i} ) \triangleq a_{i} \left( \dfrac{B T}{ V_{i} } \sum_{ k \in \mathcal{K}_{i} } \tilde{w}_{k} \log_{2} \left( 1 + \dfrac{ G_{k, k} p_{k} }{ \delta_{k} } \right) + A_{i} \right)^{ - b_{ i } }.
		\end{equation}
	It is clear that $\mathcal{P}4$ separates the interference term and power constraint by introducing auxiliary variables; thus, it is beneficial to the parallelization of algorithm design. However, as there are multiple auxiliary variables and constraints in $\mathcal{P}4$, they will linearize the augmented terms and slow down the convergence. Even worse, convergence may not be guaranteed if there are more than two variables. 
	
	To address this issue, we propose a method to eliminate auxiliary variables \cite[pp. 249-251]{Parallel1997Dimitri}. First, the augmented Lagrangian function (ALF) of $\mathcal{P}4$ can be written as
	\begin{small}	
		\begin{align}
			& L \left( \{ \bm{p}_{ i } \}_{i = 1}^{I}, \, \{ P_{i} \}_{i = 1}^{I}, \, \{ \bm{z}_{i} \}_{i = 1}^{I}, \, \{ \bm{\delta}_{i} \}_{i = 1}^{I}; \, \{ \bm{\alpha}_{i} \}_{i = 1}^{I}, \, \{ \beta_{i} \}_{i = 1}^{I} \right) \nonumber \\
			& = \sum_{i \in \mathcal{I}} \lambda_{i} \Phi_{i} ( \bm{p}_{ i } | \bm{\delta}_{i} ) + \sum_{i \in \mathcal{I}} \beta_{i} \left( \bm{\mathsf{1}}^{T} \bm{p}_{i} - P_{i} \right) + \, \dfrac{\mu}{2} \sum_{i \in \mathcal{I}} \left( \bm{\mathsf{1}}^{T} \bm{p}_{i} - P_{i} \right)^{2} \nonumber \\
			& \quad{} + \, \sum_{i = 1}^{I} \left\langle \bm{\alpha}_{i}, \bm{\Delta} \left( :, \, \mathcal{K}_{i} \right) \bm{p}_{ i } - \bm{z}_{i} \right\rangle + \dfrac{\mu}{2} \sum_{i = 1}^{I} \left\| \bm{\Delta} \left( :, \, \mathcal{K}_{i} \right) \bm{p}_{ i } - \bm{z}_{i} \right\|^{2}_{2}, \label{S1-EQ-10}
		\end{align}
	\end{small}
\hspace{-8pt}	where $\mu$ is an increasing positive sequence $\{ \mu (t) \}$ about the iteration. From \eqref{S1-EQ-10}, it is observed that \eqref{SA-EQ-B3b} and \eqref{SA-EQ-B3d} are not directly considered in the ALF since different tasks are correlated. As will be shown shortly, this new ALF term allows the sub-problems to be solved in parallel. It is noteworthy that this algorithm differs from the conventional alternating direction method of multipliers (ADMM) algorithms, and its convergence is guaranteed \cite[p. 255]{Parallel1997Dimitri}. Given \eqref{S1-EQ-10}, we have the following proposition.
		\begin{proposition} \label{S3-T1}
			For the ALF given by \eqref{S1-EQ-10}, we can make the following iteration concerning variables $P_{i}$ and $\bm{z}_{i}$:
				\begin{subequations}
					\begin{align}
						P_{i} ( t ) &= \bm{\mathsf{1}}^{T} \bm{p}_{i} ( t ) - \dfrac{1}{I} \left( \bm{\mathsf{1}}^{T} \bm{p} ( t ) - P \right), \label{S3-T1-EQ-11a} \\
						\bm{z}_{i} (\bm{\delta} (t)) &= \bm{\Delta} \left( :, \, \mathcal{K}_{i} \right) \bm{p}_{i} (t) - \dfrac{1}{I} \left( \bm{\Delta} \bm{p} (t) - \bm{\delta} (t) + \sigma^{2} \bm{\mathsf{1}} \right). \label{S3-T1-EQ-11b}
					\end{align}
				\end{subequations}
			
			The relative dual variables are updated by
				\begin{subequations}
					\begin{align}
						& \beta ( t + 1 ) = \beta ( t ) + \dfrac{ \mu(t) }{I} \left( \bm{\mathsf{1}}^{T} \bm{p} ( t + 1 ) - P \right), \label{S3-T1-EQ-12a} \\
						& \bm{\alpha} ( t + 1 ) = \bm{\alpha} ( t ) + \dfrac{\mu(t)}{I} \left( \bm{\Delta} \bm{p} (t + 1) - \bm{\delta} ( t + 1 ) + \sigma^{2} \bm{\mathsf{1}} \right), \label{S3-T1-EQ-12b}
					\end{align}
				\end{subequations}	
			and $\beta_{i} ( t + 1 ) = \beta ( t + 1 )$ and $\bm{\alpha}_{i} ( t + 1 ) = \bm{\alpha} ( t + 1 )$, for all $i = 1, \cdots, I$.			
		\end{proposition}
		
		\begin{proof}
			See Appendix~\ref{SA-C}.
		\end{proof}
			
	By Proposition~\ref{S3-T1}, it is evident that we have eliminated auxiliary variables and decreased the dimension of dual variables. Next, we split the ALF given by \eqref{S1-EQ-10} with respect to $\bm{p}$ and $\bm{\delta}$.

	\subsubsection{\underline{Parallelizable splitting with respect to $\bm{p}$}}
	By Proposition~\ref{S3-T1}, we divide the ALF given by \eqref{S1-EQ-10} into a set of sub-functions, i.e., $L_{i} \left( \bm{p}_{ i }, \, \bm{\delta}; \, \bm{\alpha}, \, \beta \right)$, which denote the ALF of the $i^{\rm th}$ task. To realize the parallel algorithm for different tasks, we obtain $L_{i} \left( \bm{p}_{ i }, \, \bm{\delta}; \, \bm{\alpha}, \, \beta \right)$ given by
		\begin{align}
			&L_{i} \left( \bm{p}_{ i }, \, \bm{\delta}; \, \bm{\alpha}, \, \beta \right) \nonumber \\
			& = \lambda_{i} \Phi_{i} ( \bm{p}_{ i } | \bm{\delta}_{i} ) + \beta \left( \bm{\mathsf{1}}^{T} \bm{p}_{i} - P_{i} \right) + \, \dfrac{\mu}{2} \left( \bm{\mathsf{1}}^{T} \bm{p}_{i} - P_{i} \right)^{2} \nonumber \\
			& \quad{} + \, \left\langle \bm{\alpha}, \bm{\Delta} \left( :, \, \mathcal{K}_{i} \right) \bm{p}_{i} - \bm{z}_{i} (\bm{\delta}) \right\rangle + \dfrac{\mu}{2} \left\| \bm{\Delta} \left( :, \, \mathcal{K}_{i} \right) \bm{p}_{i} - \bm{z}_{i} (\bm{\delta}) \right\|^{2}_{2}.  \label{S3-EQ-14}
		\end{align}

	\subsubsection{\underline{Parallelizable splitting with respect to $\bm{\delta}$}}
	By Proposition~\ref{S3-T1}, it is evident that there are interference terms of \eqref{S3-T1-EQ-11b} and \eqref{S3-T1-EQ-12b}. Thus it is still hard to update $\bm{\delta}$ in parallel. Therefore, we adopt the Gauss-Seidel method to obtain a highly parallelizable iteration for $\bm{\delta}_{i}$ \cite[p. 199]{Parallel1997Dimitri}, as formalized in the following proposition.
		\begin{proposition} \label{S3-P3}
			By \eqref{S1-EQ-10} and Proposition~\ref{S3-T1}, we obtain the following function about $\bm{\delta}$:
				\begin{align}
					&L ( \bm{p}, \, \{ \bm{\delta}_{i} \}_{i = 1}^{I}; \, \{ \bm{\alpha}^{\prime}_{i} \}_{i = 1}^{I} ) \nonumber \\
					& = \sum_{ i \in \mathcal{I} } \lambda_{i} \Phi_{i} ( \bm{p}_{i} | \bm{\delta}_{i} ) + \left\langle \bm{\alpha}, \, \bm{\Delta} \bm{p} + \sigma^{2} \bm{\mathsf{1}} - \bm{\delta} \right\rangle  \nonumber \\
					& \quad {}+ \,  \dfrac{\mu}{2 I} \left\| \bm{\Delta} \bm{p} + \sigma^{2} \bm{\mathsf{1}} - \bm{\delta} \right\|_{2}^{2}, \label{S3-P3-EQ-15}
				\end{align}
			where $\bm{\alpha} \triangleq \left[ {\bm{\alpha}^{\prime}_{1}}^{T}, \, {\bm{\alpha}^{\prime}_{2}}^{T}, \, \cdots, \, {\bm{\alpha}^{\prime}_{I}}^{T} \right]^{T}$.  
		\end{proposition}
		\begin{proof}
			From \eqref{S1-EQ-10}, we obtain the following ALF about $\bm{\delta}$:
				\begin{align}
					&L ( \bm{p}, \, \{ \bm{\delta}_{i} \}_{i = 1}^{I}; \, \{ \bm{\alpha}_{i} \}_{i = 1}^{I} ) \nonumber \\
					& = \sum_{i \in \mathcal{I}} \left( \lambda_{i} \Phi_{i} ( \bm{p}_{ i } | \bm{\delta}_{i} ) + \left\langle \bm{\alpha}_{i} , \, \bm{\Delta} \left( :, \, \mathcal{K}_{i} \right) {\bm{p}}_{i} - \bm{z}_{i} (\bm{\delta}) \right\rangle \right. \nonumber \\
					& \quad{} \left. {}+ \, \dfrac{\mu}{2} \left\| \bm{\Delta} \left( :, \, \mathcal{K}_{i} \right) {\bm{p}}_{i} - \bm{z}_{i} (\bm{\delta}) \right\|_{2}^{2} \right). \label{S3-T3-EQ17}
				\end{align}
			From Proposition~\ref{S3-T1}, we also have $\bm{\alpha}_{i} = \bm{\alpha}$ and $\bm{z}_{i} (\bm{\delta}) = \bm{\Delta} \left( :, \, \mathcal{K}_{i} \right) \bm{p}_{i} - \dfrac{1}{I} \left( \bm{\Delta} \bm{p} - \bm{\delta} + \sigma^{2} \bm{\mathsf{1}} \right).$ Inserting them into \eqref{S3-T3-EQ17} and performing algebraic manipulations, we obtain \eqref{S3-P3-EQ-15}.
		\end{proof}

	With Proposition~\ref{S3-P3}, to realize a parallel algorithm while updating $\bm{\delta}$, we divide $L ( \bm{p}, \, \{ \bm{\delta}_{i} \}_{i = 1}^{I}; \, \{ \bm{\alpha}^{\prime}_{i} \}_{i = 1}^{I} )$ into a set of sub-functions $ L_{i} ( \bm{p}, \, \bm{\delta}_{i}; \, \bm{\alpha}^{\prime}_{i} ) $ as follows:
		\begin{align}
			&L_{i} ( \bm{p}, \, \bm{\delta}_{i}; \, \bm{\alpha}^{\prime}_{i} ) \nonumber \\
			& = \lambda_{i} \Phi_{i} ( \bm{p}_{i} | \bm{ \delta_{i} } ) + \left\langle \bm{\alpha}^{\prime}_{i}, \, \bm{\Delta} \left( \mathcal{K}_{i}, \, : \right) \bm{p} + \sigma^{2} \bm{\mathsf{1}} - \bm{\delta}_{i} \right\rangle \nonumber \\
			& \quad{}+ \, \dfrac{\mu}{2 I} \left\| \bm{\Delta} \left( \mathcal{K}_{i}, \, : \right) \bm{p} + \sigma^{2} \bm{\mathsf{1}} - \bm{\delta}_{i} \right\|_{2}^{2}. \label{S3-P3-13}
		\end{align}	
	By \eqref{S3-P3-13}, it is evident that $\bm{\delta}$ is divided into $I$ blocks corresponding to $I$ different learning tasks, which implies that we can efficiently update $\bm{\delta}_{i}$ in parallel. 

	\subsection{Algorithm Development}
	We have derived the ALF of $\mathcal{P}4$ and obtained a set of sub-functions to realize a parallel algorithm for different learning tasks. Now, we compute partial derivatives of relative variables and then apply the gradient descent algorithms in parallel.
	
	\subsubsection{\underline{Update $\bm{p}_{i}$ with other variables fixed}}
	It is observed that $ L_{i} \left( \bm{p}_{ i }, \, \bm{\delta}; \, \bm{\alpha}, \, \beta \right) $ given by \eqref{S3-EQ-14} is differentiable with respect to $ \bm{p}_{ i }$, and the gradient is computed as
		\begin{equation} \nonumber
			\begin{aligned}
				&\nabla_{\bm{p}_{i}} L_{i} \left( \bm{p}_{ i }, \, \bm{\delta}; \, \bm{\alpha}, \, \beta \right) \nonumber \\
				&= \lambda_{i} \nabla_{ \bm{p}_{ i } } \Phi_{i} ( \bm{p}_{ i } | \bm{\delta}_{i} ) + \bm{\Delta} \left( :, \, \mathcal{K}_{i} \right)^{T} \bm{\alpha} + \beta \bm{\mathsf{1}} + \mu \left( \bm{\mathsf{1}}^{T} \bm{p}_{i} - P_{i} \right) \bm{\mathsf{1}} \\
				&\quad{}+ \, \mu \bm{\Delta} \left( :, \, \mathcal{K}_{i} \right)^{T} \left( \bm{\Delta} \left( :, \, \mathcal{K}_{i} \right) \bm{p}_{ i } - \bm{z}_{i} \right). 
			\end{aligned}
		\end{equation}	
	Then we apply the gradient descent method to obtain the $\bm{p}_{ i } ( t + 1 )$, as explicitly given by
		\begin{align} \label{S3-EQ-13}
			\bm{p}_{ i } ( t + 1 ) & = \max \left( \bm{p}_{ i } (t) - \eta \nabla_{  \bm{p}_{ i } } L_{i} \left( \bm{p}_{ i } ( t ), \, \bm{\delta} ( t ); \, \bm{\alpha} ( t ), \, \beta ( t ) \right) \right. \nonumber \\
			& \quad  \left. {}- \nu ( \bm{\mathsf{1}} - \tilde{ \bm{w} }_{i} ), \, \bm{0} \right),
		\end{align}
	where $\eta$ is the step-size and $\nu ( \bm{\mathsf{1}} - \tilde{ \bm{w} }_{i} )$ denotes a sparsity-regularized term \cite{Activity2019Li}. Moreover, it is seen from Proposition~\ref{S3-T1} that $\bm{\mathsf{1}}^{T} \bm{p}_{i} (t) - P_{i} (t)$ and $\bm{\Delta} \left( :, \, \mathcal{K}_{i} \right) \bm{p}_{i} (t) - \bm{z}_{i} (t)$ can be updated by \eqref{S3-T1-EQ-11a} and \eqref{S3-T1-EQ-11b}, respectively.
	
	\subsubsection{\underline{Update $\bm{\delta}_{i}$ with other variables fixed}} 
	It is seen that $L_{i} ( \bm{p}, \, \bm{\delta}_{i}; \, \bm{\alpha}^{\prime}_{i} )$ given by \eqref{S3-P3-13} is differentiable with respect to $\bm{\delta}_{i}$, and the gradient is computed as
		\begin{equation} \nonumber
			\begin{aligned}
				&\nabla_{ \bm{\delta}_{i} } L_{i} ( \bm{p}, \, \bm{\delta}_{i}; \, \bm{\alpha}^{\prime}_{i} ) \\
				&= \lambda_{i} \nabla_{\bm{ \delta }_{i}} \Phi_{i} ( \bm{p}_{i} | \bm{ \delta }_{i} ) - \bm{\alpha}^{\prime}_{i} - \dfrac{\mu}{I} \left( \bm{\Delta} \left( \mathcal{K}_{i}, \, : \right) \bm{p} + \sigma^{2} \bm{\mathsf{1}} - \bm{\delta}_{i} \right).
			\end{aligned}
		\end{equation}
	Then, we apply the gradient descent method to obtain
		\begin{equation} \label{S3-EQ-15}
			\bm{\delta}_{i} ( t + 1 ) = \max\left( \bm{\delta}_{i} ( t ) - \eta \nabla_{ \bm{\delta}_{i} } L_{i} \left( \bm{p} ( t ), \, \bm{\delta}_{i} ( t ); \, \bm{\alpha}^{\prime}_{i} ( t ) \right), \, \sigma^{2} \bm{\mathsf{1}} \right).
		\end{equation}

	To realize a highly parallelizable iteration of $\bm{p}_{i}$ and $\bm{\delta}_{i}$, as explicitly given by \eqref{S3-EQ-13} and \eqref{S3-EQ-15}, we denote variable blocks $\bm{x}_{p} = \left[ \bm{x}_{p_{1}}^{T}, \, \bm{x}_{p_{2}}^{T}, \, \cdots, \, \bm{x}_{p_{I}}^{T} \right]^{T}$ with $\bm{x}_{p_{i}} \in \mathcal{R}^{\left| \mathcal{K}_{i} \right|}$, and $\bm{x}_{\delta} = \left[ \bm{x}_{\delta_{1}}^{T}, \, \bm{x}_{\delta_{2}}^{T}, \, \cdots, \, \bm{x}_{\delta_{I}}^{T} \right]^{T}$ with $\bm{x}_{\delta_{i}} \in \mathcal{R}^{\left| \mathcal{K}_{i} \right|}$. By using variable blocks $\bm{x}_{\ell_{i}}, \ell \in \{ p, \, \delta \}$, we obtain
		\begin{equation} \label{S3-EQ-22a}
			\bm{x}_{\ell_{i}} (t + 1) = \begin{cases} \bm{p}_{i} (t + 1), & \text { if } \ell = p; \\ \bm{\delta}_{i} (t + 1), & \text { otherwise. }  \end{cases}
		\end{equation}
	
	\subsubsection{\underline{Update relative dual variables with others fixed}} 
	It is obvious that the ALF given by \eqref{S1-EQ-10} is a linear function concerning all dual variables; thus, we have
		\begin{align}
			\bm{\alpha}_{i}^{\prime} ( t + 1 ) & = \bm{\alpha}_{i}^{\prime} ( t ) + \dfrac{\mu ( t )}{I} \Bigg( \bm{\Delta} \left( \mathcal{K}_{i}, \, \mathcal{K}_{i} \right) \bm{p}_{i} ( t + 1 ) - \bm{\delta}_{i}  ( t + 1 ) \nonumber \\
			& \quad {} + \, \sigma^{2} \bm{\mathsf{1}} + \sum_{i \neq j} \left( \bm{\Delta} \left( \mathcal{K}_{i}, \mathcal{K}_{j} \right) \right) \bm{p}_{j} ( t )
			\Bigg). \label{S3-EQ-16a}
		\end{align}
	Using \eqref{S3-EQ-16a} in place of \eqref{S3-T1-EQ-12b} gives a Gauss-Seidel sequence to realize a highly efficient iteration and obtain a real-time message. Apart from the aforementioned dual variables, $\beta (t + 1)$ can be directly updated by \eqref{S3-T1-EQ-12a}.

	In terms of computational complexity, this algorithm involves $K$ scheduling variables, $K$ primal variables, $K$ auxiliary variables, $K ( K - 1 )$ interference terms, and $K + I$ dual variables. Specifically, $K + I$ dual variables come from $K$ interference constraints and $I$ learning tasks. In addition, $K$ primal variables, $K$ auxiliary variables, $K ( K - 1 )$ interference terms, and $K + I$ dual variables can be updated in parallel. Consequently, when the dimension $K$ of users is large, the per-iteration complexity is approximately given by $\mathcal{O} \left( ( K^{2} + K ) / I \right)$.

\begin{figure}
			\centering
			\includegraphics[width=0.5\textwidth]{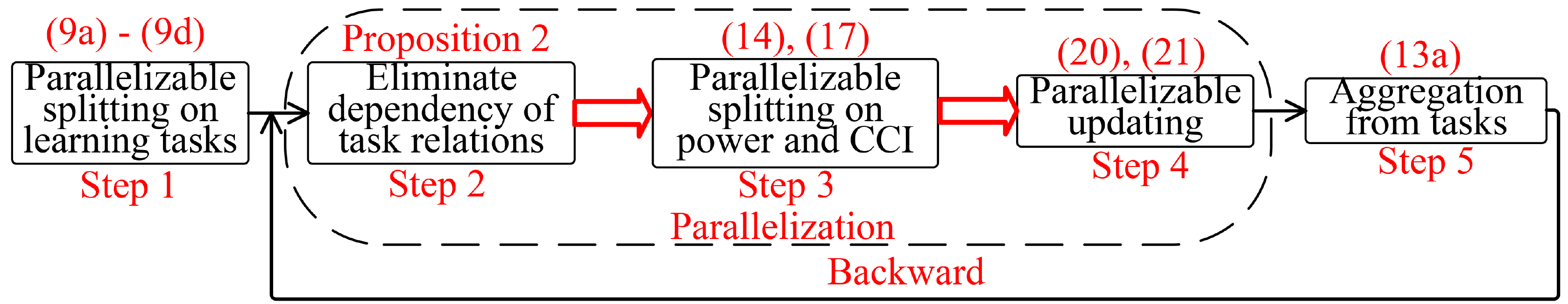}
			\caption{Block diagram of the proposed parallel algorithm.}
			\label{Fig2}
		\end{figure}

	To sum up, Fig.~\ref{Fig2} sketches the block diagram of the proposed parallel algorithm. Also, the detailed steps are formalized in Algorithm~\ref{Algorithm1}, where lines~$3$-$8$ are the main steps of the parallel algorithm, as shown in the parallelization module of Fig.~\ref{Fig2}. Specifically, lines~$4$-$6$ of Algorithm~\ref{Algorithm1} realize the power and CCI optimization, and line~$7$ performs the dual and scheduling variables update in parallel. Then, line~$9$ aggregates messages from different tasks and also constructs an increasing sequence $\mu ( t + 1 ) = \max(  \mu_{\rm s} \mu (t), \, \mu_{\max} )$, which means that the equalities \eqref{SA-EQ-B3a}-\eqref{SA-EQ-B3d} must hold when Algorithm~\ref{Algorithm1} converges.

		\begin{algorithm*}[t!]
		\small
			\caption{The task-oriented power allocation in parallel.}
			\label{Algorithm1}
			\renewcommand{\algorithmicrequire}{\textbf{Input:}}
			\renewcommand{\algorithmicensure}{\textbf{Output:}}
				\begin{algorithmic}[1]
					\REQUIRE Setting $\left( I, \, N, \, K, \, P, \, T, \, B, \, \sigma^{2}, \, \{ \lambda_{i}, a_{i}, \, b_{i}, \, V_{i}, \, A_{i} \}_{ i \in \mathcal{I} } \right)$, channels $\{ \bm{h}_{k} \}_{k \in \mathcal{K}}$, user set $\mathcal{K}$, gain matrix $\bm{G}$, gain diagonal matrix $\bm{D}$, learning rate $\eta$, error tolerance $\varepsilon$, $\mu_{\max}$, and $\mu_{\rm s} > 1$.
					\ENSURE The optimization solution $\hat{ \bm{p} }$;
					\STATE Initialize $t = 0, \, \tilde{ \bm{w} } = \bm{\mathsf{1}}, \, \bm{p} ( 0 ) = P / K \times \bm{\mathsf{1}}, \, \bm{\delta} (0) = \left( \bm{G} - \bm{D} \right) \bm{p} (0) + \sigma^{2} \bm{\mathsf{1}}, \, \bm{\alpha} (0) = 1 / K \times \bm{\mathsf{1}}, \, \beta (0) = 1, $ and $ \mu (0) = 1$;
					\REPEAT
					\FOR {$i \in \mathcal{I}$ in parallel}
					\FOR {$\ell \in \{ p, \, \delta \}$ in parallel}
					\STATE Update $\bm{x}_{\ell_{i}} (t + 1)$ by \eqref{S3-EQ-22a};
					\ENDFOR
					\STATE Update $ \bm{\alpha}_{i}^{\prime} (t + 1)$ by \eqref{S3-EQ-16a}, and $\tilde{ \bm{w} }_{i}$ as per \eqref{S3-P1-EQ-6};
					\ENDFOR
					\STATE Compute $\beta (t + 1)$ as per \eqref{S3-T1-EQ-12a}, and $\mu ( t + 1 ) = \max(  \mu_{\rm s} \mu (t), \, \mu_{\max} )$;
					\STATE Compute ${\rm MSE}$ as per \eqref{S3-EQ-21M};
					\STATE $t = t + 1$;
					\UNTIL{${\rm MSE} \leq \varepsilon$};
					\STATE $\hat{ \bm{p} } = \lfloor \tilde{\bm{w}} \rceil \circ \bm{p} ( t )$.
				\end{algorithmic}
		\end{algorithm*}

	So far, we have developed a parallel algorithm to solve the task-oriented power allocation problem. As multiple variables need to be relaxed for task parallelism, it slows down the convergence. Although Proposition~\ref{S3-T1} can eliminate auxiliary variables, additional relaxed constraints exist to separate the CCI term, such as variables $\bm{\delta}$ and relative dual variables. Also, the per-iteration complexity is usually high, specifically for solving the non-convexity problem of $\Phi_{i} ( \bm{p}_{i} | \bm{\delta}_{i} )$ and the non-unitary matrix $\tilde{\bm{G}} - \tilde{\bm{D}}$, i.e., $(\tilde{\bm{G}} - \tilde{\bm{D}})^{T} (\tilde{\bm{G}} - \tilde{\bm{D}})$ is not an identity mapping. To address these issues, we design an accelerated algorithm in the next section.	

		\begin{figure*}[t!]
			\centering
			\includegraphics[width=0.9\textwidth]{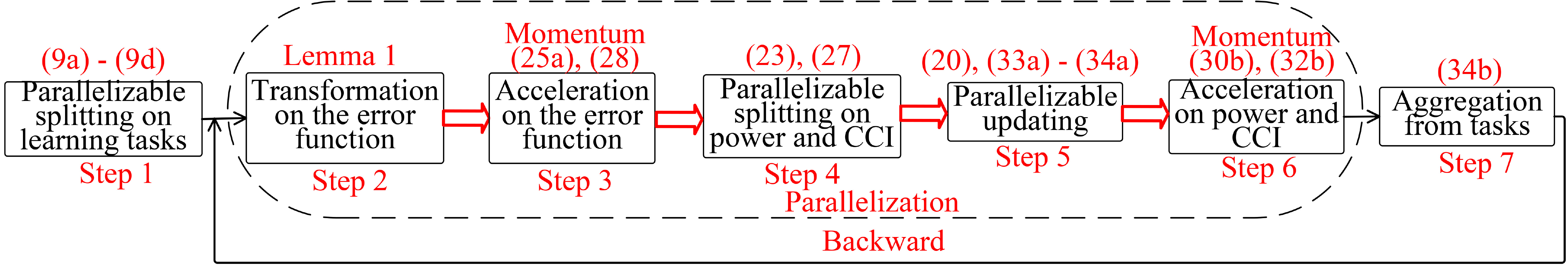}
			\caption{Block diagram of the accelerated algorithm.}
			\label{Fig3}
		\end{figure*}

\section{An accelerated Algorithm: Fast Proximal Algorithms}	
	Now, we design a fast proximal ADMM algorithm with parallelizable splitting \cite{lu2015fast}, and Fig.~\ref{Fig3} sketches its block diagram. Specifically, to improve the convergence rate, we first exploit the smoothness property to linearize $\Phi_{i} ( \bm{p}_{i} | \bm{\delta}_{i} )$ (i.e., Step~$2$ in Fig.~\ref{Fig3}). Accordingly, the smoothness result of $\Phi_{i} ( \bm{p}_{i} | \bm{\delta}_{i} )$ is shown in Lemma~\ref{S3-L1}

		\begin{lemma} \label{S3-L1}
			The function $\Phi_{i} ( \bm{p}_{i} | \bm{\delta}_{i} )$ satisfies the following conditions: 
			\begin{itemize}
				\item[i)] $\Phi_{i} ( \bm{p}_{i} | \bm{\delta}_{i}^{*} )$ is $L_{p_{i}}$-smooth, i.e., $\| \nabla_{\bm{x}} \Phi_{i} ( \bm{x} | \bm{\delta}_{i}^{*} ) - \nabla_{\bm{y}} \Phi_{i} ( \bm{y} | \bm{\delta}_{i}^{*} ) \|_{2} \leq L_{p_{i}} \| \bm{x} - \bm{y} \|_{2}$ for any $\bm{x}, \bm{y}$;
				\item[ii)] $\Phi_{i} ( \bm{p}_{i}^{*} | \bm{\delta}_{i} )$ is $L_{\delta_{i}}$-smooth, i.e., $\| \nabla_{\bm{x}} \Phi_{i} ( \bm{p}_{i}^{*} | \bm{x} ) - \nabla_{\bm{y}} \Phi_{i} ( \bm{p}_{i}^{*} | \bm{y} ) \|_{2} \leq L_{\delta_{i}} \| \bm{x} - \bm{y} \|_{2}$ for any $\bm{x}, \bm{y}$,
			\end{itemize}
			where $\bm{\delta}_{i}^{*}, \, \bm{p}_{i}^{*}, \, i \in \mathcal{I}$ denote their current values stored.
		\end{lemma}
		\begin{proof}
			See Appendix~\ref{SA-E}.
		\end{proof}
	
	Given Lemma~\ref{S3-L1}, the smoothness result enables us to linearize the learning error function $\Phi_{i} ( \bm{p}_{i} | \bm{\delta}_{i} )$. So, we next design an identity mapping of the unitary matrix to improve the convergence rate and solve the related sub-problems more efficiently. 			
			
	\subsection{Parallelization}
	In principle, the essence of our accelerated algorithm is to use an identical transform of matrices to split variable blocks. Using a fast proximal linearized ADMM algorithm with parallelizable splitting \cite{lu2015fast}, we derive ALFs associated with variable blocks $\bm{p}_{i}$ and $\bm{\delta}_{i}$, respectively.

	\subsubsection{\underline{The ALF with respect to $\bm{p}_{i}$ and $\bm{\delta}$}} 
	We first define two block matrices
		\begin{equation} \nonumber
			\bm{A} \triangleq \left[ \begin{array}{cc} \bm{\Delta} \left( :, \, \mathcal{K}_{i} \right) & - {\bm{I}} / {I} \\ \bm{\mathsf{1}}^{T} & \bm{0}^{T} \end{array} \right], \, \bm{A}_{1} \triangleq 
			\left[ \begin{array}{c} \bm{\Delta} \left( :, \, \mathcal{K}_{i} \right) \\ \bm{\mathsf{1}}^{T} \end{array} \right].
		\end{equation}	
	Then, we can rewrite \eqref{SA-EQ-B3a}-\eqref{SA-EQ-B3c} as a linear equation $\bm{A} \bm{x} = \bm{r}$, where $\bm{r} \triangleq \left[ - {\sigma^{2}} / {I} \bm{\mathsf{1}}^{T}, \, P_{i} \right]^{T}$, $\bm{x} \triangleq \left[ \bm{p}_{i}^{T}, \, \bm{\delta}^{T} \right]^{T}$, and $\bm{z}_{i} (\bm{\delta}) = \bm{\delta} / {I} - {\sigma^{2}} / {I} \bm{1}$ given by \eqref{SA-EQ-B2}, \eqref{SA-EQ-B3a}, and \eqref{SA-EQ-B3b}, respectively. Moreover, the ALF given by \eqref{S3-EQ-14} can be rewritten as
	\begin{small}
		\begin{equation} \label{S3-EQ-24}
			L_{i} ( \bm{x}; \, \bm{\lambda}^{\prime} ) = \Phi_{i} ( \bm{x} ) + \langle \bm{\lambda}^{\prime}, \, \bm{A} \bm{x} - \bm{r} \rangle + \dfrac{\mu}{2} \| \bm{A} \bm{x} - \bm{r} \|_{2}^{2} ,
		\end{equation}
	where $\bm{\lambda}^{\prime} \triangleq \left[ \bm{\alpha}^{T}, \, \beta \right]^{T}$ and $\Phi_{i} ( \bm{x} ) \triangleq \lambda_{i} \Phi_{i} ( \bm{p}_{i} | \bm{\delta}_{i} )$. Then, by means of the parallelizable splitting \cite{lu2015fast} and relaxing $L_{i} ( \bm{x}; \, \bm{\lambda}^{\prime} )$, we write an accelerated ALF of \eqref{S3-EQ-24} with respect to $\bm{p}_{i}$ as
		\begin{align}
			& L_{i} \left( \bm{p}_{i} | \nabla_{ \bm{y}_{p_{i}} } \Phi_{i} ( \bm{y}_{p_{i}} ( t + 1 ) | \bm{\delta}_{i} ( t ) ), \, \bm{p}_{i} ( t ), \, \bm{z}_{p_{i}} (t), \, \bm{z}_{i} ( t ); \, \bm{\alpha} (t), \, \beta (t) \right) \nonumber \\
			& = \lambda_{i} \left\langle \nabla_{ \bm{y}_{p_{i}} } \Phi_{i} ( \bm{y}_{p_{i}} ( t + 1 ) | \bm{\delta_{i}} ( t ) ), \, \bm{p}_{i} \right\rangle + \mu ( t ) \left\langle \bm{A}_{1}^{T} ( \bm{A} \bm{z}_{1} ( t ) - \bm{r} ) , \, \bm{p}_{i} \right\rangle \nonumber \\
			& \quad{} + \, \langle \bm{\lambda}^{\prime} ( t ), \, \bm{A}_{1} \bm{p}_{i} \rangle + \frac{1}{2} \left( L_{p_{i}} \theta ( t ) + \mu ( t ) \lambda_{p_{i}}  \right) \left\| \bm{p}_{i} - \bm{z}_{p_{i}} ( t ) \right\|_{2}^{2} \nonumber \\
			& = \left\langle \bm{\alpha} (t), \, \bm{\Delta} \left( :, \, \mathcal{K}_{i} \right) \bm{p}_{i} \right\rangle + \mu ( t ) \left\langle \left( \bm{\mathsf{1}}^{T} \bm{z}_{p_{i}} (t) - P_{i} (t) \right) \bm{\mathsf{1}}, \, \bm{p}_{i} \right\rangle \nonumber \\
			& \quad{} + \, \mu ( t ) \left\langle \bm{\Delta} \left( :, \, \mathcal{K}_{i} \right)^{T} \left( \bm{\Delta} \left( :, \, \mathcal{K}_{i} \right) \bm{z}_{p_{i}} (t) - \dfrac{\bm{z}_{\delta} (t)}{I} + \dfrac{\sigma^{2}}{I} \bm{1} \right), \, \bm{p}_{i} \right\rangle \nonumber \\
			& \quad{} + \, \lambda_{i} \left\langle \nabla_{ \bm{y}_{p_{i}} } \Phi ( \bm{y}_{p_{i}} ( t + 1 ) | \bm{\delta}_{i} ( t ) ), \, \bm{p}_{i} \right\rangle + \beta ( t ) \bm{\mathsf{1}}^{T} \bm{p}_{i} \nonumber \\
			& \quad{} + \, \frac{1}{2} \left( L_{p_{i}} \theta ( t ) + \mu ( t ) \lambda_{p_{i}} \right) \left\| \bm{p}_{i} - \bm{z}_{p_{i}} ( t ) \right\|_{2}^{2}, \, \label{S3-EQ-21}
		\end{align}
	\end{small}
\hspace{-8pt}	where $\bm{z}_{1} \triangleq \left[ \bm{z}_{p_{i}}^{T}, \, \bm{z}_{\delta}^{T} \right]^{T}$, $\bm{z}_{\delta} \triangleq \left[ \bm{z}_{\delta_{1}}^{T}, \, \cdots, \, \bm{z}_{\delta_{I}}^{T} \right]^{T}$; $\bm{z}_{p_{i}}$ and $\bm{z}_{\delta_{i}}$, $i \in \mathcal{I}$, denote the gradient update results of $\bm{p}_{i}$ and $\bm{\delta}_{i}$, respectively. Moreover, $\lambda_{p_{i}} \geq 2 \| \bm{A}_{1} \|_{2}^{2}$ guarantees that \eqref{S3-EQ-21} is a tight majorant surrogate function of \eqref{S3-EQ-24} with respective to $\bm{p}_{i}$ \cite{lu2015fast, ADMM2018Lu}, therefore we have 
		\begin{align}
			\lambda_{p_{i}} & \geq{} 2 K / I \left( \| \bm{\Delta} \left( :, \, \mathcal{K}_{i} \right) \|_{2} + 1 \right)^{2} \nonumber \\
			&\geq{} 2 \left( \| \tilde{ \bm{w}_{i} } \|_{2} \| \bm{\Delta} \left( :, \, \mathcal{K}_{i} \right) \|_{2} + \sqrt{K / I} \right)^{2} \nonumber \\
			& \geq{} 2 \left( \| \bm{\Delta} \left( :, \, \mathcal{K}_{i} \right) \|_{2} + \| \bm{\mathsf{1}} \|_{2} \right)^{2} \geq 2 \| \bm{A}_{1} \|_{2}^{2}. \label{S3-EQ-23}
		\end{align}
	Lastly, the parameters $\bm{y}_{p_{i}} ( t + 1 )$, $\theta ( t + 1 )$, and $\mu ( t + 1 )$ can be updated by
		\begin{subequations}
			\begin{align}
				\bm{y}_{p_{i}} ( t + 1 ) & = ( 1 - \theta ( t ) ) \bm{p}_{i} ( t ) + \theta ( t ) \bm{z}_{p_{i}} ( t ) , \label{S3-EQ-23a} \\
				\theta ( t + 1 ) & = \frac{1}{2} ( - \theta^{2}(t) + \sqrt{ \theta^{4}(t) + 4 \theta^{2}(t) } ) , \label{S3-EQ-23c} \\
				\mu ( t + 1 ) & = 1/{\theta ( t + 1 )}, \label{S3-EQ-27c}
			\end{align}
		\end{subequations}
	where \eqref{S3-EQ-23a} is to accelerate convergence by using the smoothness result given by Lemma~\ref{S3-L1}; \eqref{S3-EQ-23c} is a stepsize of the fast algorithm, and \eqref{S3-EQ-27c} means an increasing sequence explained in Algorithm~\ref{Algorithm1}. With careful choices of $\theta (t)$ and $\mu (t)$, the convergence rate can be accelerated from $\mathcal{O} ( 1 / \tau )$ to $\mathcal{O} ( 1 / \tau^{2} )$ \cite{lu2015fast}, where $\tau$ is the number of iterations needed to converge.

	\subsubsection{\underline{The ALF with respect to $\bm{p}$ and $\bm{\delta}_{i}$}}	
	We first define $\bm{A}^{\prime} \triangleq \left[ \begin{array}{cc} \bm{\Delta} \left( \mathcal{K}_{i}, \, : \right) & - \bm{I} \end{array} \right] $, then we can also rewrite $\bm{\Delta} \left( \mathcal{K}_{i}, \, : \right) \bm{p} + \sigma^{2} \bm{\mathsf{1}} = \bm{\delta}_{i}$ given by \eqref{SA-EQ-B2} as $\bm{A}^{\prime} \bm{x}^{\prime} = \bm{r}^{\prime}$, where $\bm{r}^{\prime} \triangleq - \sigma^{2} \bm{\mathsf{1}}$ and $\bm{x}^{\prime} \triangleq \left[ \bm{p}^{T}, \, \bm{\delta}_{i}^{T} \right]^{T}$. Moreover, the ALF given by \eqref{S3-P3-13} can be re-expressed as
		\begin{equation} \label{S3-EQ-27}
			L_{i} ( \bm{x}^{\prime}; \, \bm{\alpha}^{\prime}_{i} ) \triangleq \Phi_{i} ( \bm{x}^{\prime} ) + \langle \bm{\alpha}^{\prime}_{i}, \, \bm{A}^{\prime} \bm{x}^{\prime} - \bm{r}^{\prime} \rangle + \dfrac{\mu}{2} \| \bm{A}^{\prime} \bm{x}^{\prime} - \bm{r}^{\prime} \|_{2}^{2} ,
		\end{equation}
	where $\Phi_{i} ( \bm{x}^{\prime} ) \triangleq \lambda_{i} \Phi_{i} ( \bm{p}_{i} | \bm{\delta}_{i} )$. By the parallelizable splitting and relaxing $L_{i} ( \bm{x}^{\prime}; \, \bm{\alpha}^{\prime}_{i} )$, we also write compactly another accelerated ALF of \eqref{S3-EQ-27} with respect to $\bm{\delta}_{i}$  as
	\begin{small}	
		\begin{align}
			& L_{i} \left( \bm{\delta}_{i} | \nabla_{\bm{y}_{\delta_{i}} (t + 1)}, \, \bm{p}_{i} ( t + 1 ), \, \bm{\delta}_{i} ( t ), \, \bm{z}_{\delta_{i}} (t); \, \bm{\alpha}^{\prime}_{i} (t), \, \beta (t) \right) \nonumber \\
			& = \lambda_{i} \left\langle \nabla_{\bm{y}_{\delta_{i}} (t + 1)}, \, \bm{\delta}_{i} \right\rangle - \langle \bm{\alpha}^{\prime}_{i} ( t ), \, \bm{\delta}_{i} \rangle - \mu ( t ) \langle \bm{A}^{\prime} \bm{z}_{2} ( t ) - \bm{r}^{\prime}, \, \bm{\delta}_{i} \rangle \nonumber \\
			& \quad{} + \,  \frac{1}{2} \left( L_{\delta_{i}} \theta ( t ) + \mu ( t ) \lambda_{\delta_{i}} \right) \| \bm{\delta}_{i} - \bm{z}_{\delta_{i}} ( t ) \|_{2}^{2} \nonumber \\
			& = \lambda_{i} \left\langle \nabla_{\bm{y}_{\delta_{i}} (t + 1)}, \, \bm{\delta}_{i} \right\rangle + \frac{1}{2} \left( L_{\delta_{i}} \theta ( t ) + \mu ( t ) \lambda_{\delta_{i}} \right) \| \bm{\delta}_{i} - \bm{z}_{\delta_{i}} ( t ) \|_{2}^{2} \nonumber \\
			& \quad{} - \, \left\langle \bm{\alpha}^{\prime}_{i} (t), \, \bm{\delta}_{i} \right\rangle + \mu ( t ) \left\langle \left( \bm{\Delta} \left( \mathcal{K}_{i}, \, : \right) \right) \bm{z}_{p} ( t ) - \bm{z}_{ \delta_{i} } (t) + \sigma^{2} \bm{\mathsf{1}}, \, \bm{\delta}_{i} \right\rangle, \label{S3-EQ-22}
		\end{align}
	\end{small}
\hspace{-8pt}	where $\nabla_{\bm{y}_{\delta_{i}}(t + 1)} \triangleq \nabla_{ \bm{y}_{\delta_{i}} } \Phi ( \bm{p}_{i} ( t + 1 ) | \bm{y}_{\delta_{i}} ( t + 1 ) )$, $\bm{z}_{2} \triangleq \left[ \bm{z}_{p}^{T}, \, \bm{z}_{\delta_{i}}^{T} \right]^{T}$, and $\bm{z}_{p} \triangleq \left[ \bm{z}_{p_{1}}^{T}, \, \cdots, \, \bm{z}_{p_{I}}^{T} \right]^{T}$. Like \eqref{S3-EQ-23}, the choice of $\lambda_{\delta_{i}} \geq 2 \| \bm{I} \|_{2}^{2} = 2$ also guarantees that \eqref{S3-EQ-22} is a tight majorant surrogate function of \eqref{S3-EQ-27}  with respective to $\bm{\delta}_{i}$ \cite{lu2015fast, ADMM2018Lu}. Moreover, $\bm{y}_{\delta_{i}} ( t + 1 )$ is given by
		\begin{equation}
			\bm{y}_{\delta_{i}} ( t + 1 ) = ( 1 - \theta ( t ) ) \bm{\delta}_{i} ( t ) + \theta ( t ) \bm{z}_{\delta_{i}} ( t ), \label{S3-EQ-20a}
		\end{equation}
	whose effect is the same as \eqref{S3-EQ-23a}. As stated above, we can relax $\Phi_{i} ( \bm{p}_{i} | \bm{ \delta_{i} } )$ by Lemma~\ref{S3-L1}, and then Lemma~\ref{S3-L1} allows very large Lipschitz constants $L_{p_{i}}$ and $L_{\delta_{i}}$ for non-convex functions, which are as large as $\mathcal{O} (\tau)$ without affecting the convergence rate. Moreover, we also linearize the augmented terms $1 / {2} \| \bm{A} \bm{x} - \bm{r} \|_{2}^{2}$ and $1 / {2} \| \bm{A}^{\prime} \bm{x}^{\prime} - \bm{r}^{\prime} \|_{2}^{2}$ by $ \lambda_{p_{i}} / 2 \left\| \bm{p}_{i} - \bm{z}_{p_{i}} ( t ) \right\|_{2}^{2}$ and $ \lambda_{\delta_{i}} / 2 \| \bm{\delta}_{i} - \bm{z}_{\delta_{i}} ( t ) \|_{2}^{2}$, respectively. After \eqref{S3-EQ-21} and \eqref{S3-EQ-22} are obtained, we can improve the efficiency for optimizing these sub-functions given by \eqref{S3-EQ-14} and \eqref{S3-P3-13}. 

	\subsection{Algorithm Development}
	We have obtained the parallelizable splitting and derived ALFs of the accelerated algorithm. Now, we compute the partial derivatives of relative variables and apply the gradient descent algorithm to update these variables in parallel.
	
	\subsubsection{\underline{Update $\bm{p}_{i}$ in parallel with other variables fixed}}	
	Here, the ALF given by \eqref{S3-EQ-21} is a quadratic function of $\bm{p}_{i}$, thus it has a closed-form solution with respective to $\bm{p}_{i}$, given by 
		\begin{align}
			&\tilde{ \bm{z} }_{p_{i}} (t + 1) \nonumber \\
			&= - \dfrac{1}{L_{p_{i}} \theta ( t ) + \mu ( t ) \lambda_{p_{i}} } \left( \lambda_{i} \nabla_{ \bm{y}_{p_{i}} } \Phi ( \bm{y}_{p_{i}} ( t + 1 ) | \bm{\delta}_{i} ( t ) ) + \beta (t) \bm{\mathsf{1}} \right. \nonumber \\
			&\quad {}+ \, \mu (t) \bm{\Delta} \left( :, \, \mathcal{K}_{i} \right)^{T} \left( \bm{\Delta} \left( :, \, \mathcal{K}_{i} \right) \bm{z}_{p_{i}} (t) - \dfrac{\bm{z}_{\delta} (t)}{I} + \dfrac{\sigma^{2}}{I} \bm{1} \right) \nonumber \\
			&\quad {}+ \, \left. \bm{\Delta} \left( :, \, \mathcal{K}_{i} \right)^{T} \bm{\alpha} (t) + \mu (t) \left( \bm{\mathsf{1}}^{T} \bm{z}_{p_{i}} ( t ) - P_{i} (t) \right) \bm{\mathsf{1}} \right) + \bm{z}_{p_{i}} (t) , \label{S4-EQ-21}
		\end{align}	
	where $P_{i} (t)$ can be computed by \eqref{S3-T1-EQ-11a}. Then, we obtain
		\begin{subequations} 
			\begin{align}
				\bm{z}_{p_{i}} ( t + 1 ) &= \max( \tilde{ \bm{z} }_{p_{i}} ( t + 1 ) - \nu ( \bm{\mathsf{1}} - \tilde{ \bm{w} }_{i} ), \, \bm{0} ) ,  \label{S3-EQ-24a} \\
				\bm{p}_{i} ( t + 1 ) &= ( 1 - \theta ( t ) ) \bm{p}_{i} ( t ) + \theta ( t ) \bm{z}_{p_{i}} ( t + 1 ), \label{S3-EQ-24b}
			\end{align}
		\end{subequations}
	where \eqref{S3-EQ-24a} and \eqref{S3-EQ-24b} are an orthogonal projection onto sparsity-regularized and accelerated terms, respectively.	

	\subsubsection{\underline{Update $\bm{\delta}_{i}$ in parallel with other variables fixed}}
	Here, the ALF given by \eqref{S3-EQ-22} is also a quadratic function of $\bm{\delta}_{i}$ hence we obtain a closed-form solution as
		\begin{align}
			& \tilde{ \bm{z} }_{\delta_{i}} ( t + 1 ) \nonumber \\
			& = \bm{z}_{\delta_{i}} ( t ) - \dfrac{1}{L_{\delta_{i}} \theta ( t ) + \mu ( t ) \lambda_{\delta_{i}}} \left( \lambda_{i} \nabla_{ \bm{y}_{\delta_{i}} } \Phi ( \bm{p}_{i} ( t + 1 ) | \bm{y}_{\delta_{i}} ( t + 1 ) ) \right. \nonumber \\
			& \quad {} - \, \left. \bm{\alpha}_{i}^{\prime} (t) + \mu (t) \left( \left( \bm{\Delta} \left( \mathcal{K}_{i}, \, : \right) \right) \bm{z}_{p} ( t ) - \bm{z}_{ \delta_{i} } (t) + \sigma^{2} \bm{\mathsf{1}} \right) \right). \label{S4-EQ-23}
		\end{align}
	Next, we have
		\begin{subequations}
			\begin{align}
				\bm{z}_{ \delta_{i} } ( t + 1 ) &= \max( \tilde{ \bm{z} }_{\delta_{i}} ( t + 1 ), \, \sigma^{2} \bm{\mathsf{1}} ), \label{S3-EQ-26a} \\
				\bm{\delta}_{i} ( t + 1 ) &= ( 1 - \theta ( t ) ) \bm{\delta}_{i} ( t ) + \theta ( t ) \bm{z}_{\delta_{i}} ( t + 1 ), \label{S3-EQ-26b}
			\end{align}
		\end{subequations}
	where \eqref{S3-EQ-26a} and \eqref{S3-EQ-26b} denote an orthogonal projection and an accelerated term, respectively. In light of \eqref{S4-EQ-21} and \eqref{S4-EQ-23}, it is obvious that $\bm{p}_{i}$ and $\bm{\delta}_{i}$ can be updated in parallel, thus we have
		\begin{subequations}
			\begin{align}
				\bm{y}_{\ell_{i}} (t + 1) &= \begin{cases} \bm{y}_{p_{i}} (t + 1), & \text { if } \ell = p; \\ \bm{y}_{\delta_{i}} (t + 1), & \text{ otherwise},  \end{cases} \label{S3-EQ-32b} \\
				\bm{z}_{\ell_{i}} (t + 1) &= \begin{cases} \bm{z}_{p_{i}} (t + 1), & \text { if } \ell = p; \\ \bm{z}_{\delta_{i}} (t + 1), & \text{ otherwise},  \end{cases} \label{S3-EQ-32c} 
			\end{align}
		\end{subequations}
	and $\bm{x}_{\ell_{i}} (t + 1), \, \ell \in \{p, \, \delta\}$ is updated by \eqref{S3-EQ-22a}.
	
	\subsubsection{\underline{Update relative dual variables}}
	It is evident that the ALFs given by \eqref{S3-EQ-14} and \eqref{S3-P3-13} are linear functions of all dual variables; hence we have
		\begin{subequations}
			\begin{align}
				\bm{\alpha}_{i}^{\prime} ( t + 1 ) & = \bm{\alpha}_{i}^{\prime} ( t ) + \dfrac{\mu ( t )}{I} \Bigg( \bm{\Delta} \left( \mathcal{K}_{i}, \, \mathcal{K}_{i} \right) \bm{z}_{p_{i}} ( t + 1 ) + \sigma^{2} \bm{\mathsf{1}} \nonumber \\
				& \quad {} - \, \bm{z}_{ \delta_{i} } ( t + 1 ) + \sum_{i \in \mathcal{I} \setminus j} \bm{\Delta} \left( \mathcal{K}_{i}, \, \mathcal{K}_{j} \right) \bm{z}_{p_{j}} ( t )
				\Bigg) , \label{S3-EQ-25a} \\
				\beta ( t + 1 ) & = \beta ( t ) + \dfrac{ \mu(t) }{I} \left( \bm{\mathsf{1}}^{T} \bm{z}_{p} ( t + 1 ) - P \right). \label{S3-EQ-25b}
			\end{align}
		\end{subequations}
	Using \eqref{S3-EQ-25a} in place of \eqref{S3-T1-EQ-12b} leads to a highly parallelizable iteration. 
	
	In terms of computational complexity, the accelerated algorithm is proportional to the parallel algorithm. Thus the per-iteration complexity is also given by $\mathcal{O} \left( ( K^{2} + K ) / I \right)$. Beyond the computational complexity, another important metric to measure the convergence speed is the convergence rate. From \cite[Theorem~$2$]{lu2015fast}, this algorithm improves the convergence rate from $\mathcal{O} ( 1 / \tau )$ to $\mathcal{O} ( 1 / \tau^{2} )$, which makes it more attractive, specifically for large-scale IoT networks. Moreover, this algorithm also allows large Lipschitz constants $L_{p_{i}}$ and $L_{\delta_{i}}$ for relaxing non-convex objective functions without affecting the convergence rate.
	
	To sum up, the procedure is formalized in Algorithm~\ref{Algorithm2}, which is faster than Algorithm~\ref{Algorithm1} due to the acceleration to the error functions (i.e., \eqref{S3-EQ-23a} and \eqref{S3-EQ-20a}) and equality constraints (i.e., \eqref{S3-EQ-24b} and \eqref{S3-EQ-26b}). Specifically, lines~$5$ and $7$ of Algorithm~\ref{Algorithm2} describe the parallel steps (i.e., Steps~$3$-$6$ in Fig.~\ref{Fig3}). Among them, line~$5$ specifies the acceleration steps (i.e., Steps~$3$ and $6$ in Fig.~\ref{Fig3}). Moreover, line~$9$ describes aggregated messages from different tasks (i.e., Step~$7$ in Fig.~\ref{Fig3}). Also, $\mu (t)$ in Algorithm~\ref{Algorithm2} is adaptive to the stepsize $\theta (t)$ to guide convergence more efficiently.
		\begin{algorithm*}[t!]
			\small
			\caption{The accelerated algorithm.}
			\label{Algorithm2}
			\renewcommand{\algorithmicrequire}{\textbf{Input:}}
			\renewcommand{\algorithmicensure}{\textbf{Output:}}
			\begin{algorithmic}[1]
				\REQUIRE Setting $\left( I, \, N, \, K, \, P, \, T, \, B, \, \sigma^{2}, \, \{ \lambda_{i}, \lambda_{p_{i}}, \, \lambda_{\delta_{i}}, \, a_{i}, \, b_{i}, \, V_{i}, \, A_{i} \}_{ i \in \mathcal{I} } \right)$, user set $\mathcal{K}$, channels $\{ \bm{h}_{k} \}_{k \in \mathcal{K}}$, gain matrix $\bm{G}$, gain diagonal matrix $\bm{D}$, learning rate $\eta$, and error tolerance~$\varepsilon$.
				\ENSURE The optimization solution $\hat{ \bm{p} }$.
				\STATE Initialize $t = 0, \, \bm{x}_{p} ( 0 ) = \bm{y}_{p} ( 0 ) = \bm{z}_{p} ( 0 ) = P / K \times \bm{\mathsf{1}}, \, \bm{x}_{\delta} ( 0 ) = \bm{y}_{\delta} ( 0 ) = \bm{z}_{\delta} ( 0 ) = ( \bm{G} - \bm{D} ) \bm{x}_{p} ( 0 ) + \sigma^{2} \bm{\mathsf{1}}, \, \tilde{ \bm{w} } = \bm{\mathsf{1}}, \, \bm{\alpha} (0) = 1 / K \times \bm{\mathsf{1}}, \, \beta (0) = 1, \, \mu (0) = \theta (0) = 1$;
				\REPEAT
				\FOR {$i \in \mathcal{I}$ in parallel}
				\FOR {$\ell \in \{ p, \, \delta \}$ in parallel}
				\STATE Compute $\bm{y}_{\ell_{i}} (t + 1)$, $\bm{z}_{\ell_{i}} (t + 1)$, and $\bm{x}_{\ell_{i}} (t + 1)$ as per \eqref{S3-EQ-32b}, \eqref{S3-EQ-32c}, and \eqref{S3-EQ-22a}, respectively;
				\ENDFOR
				\STATE Compute $\bm{\alpha}_{i}^{\prime} (t + 1)$ and $\tilde{ \bm{w} }_{i}$ as per \eqref{S3-EQ-25a} and \eqref{S3-P1-EQ-6}, respectively;
				\ENDFOR
				\STATE Update $\beta (t + 1)$, $\theta (t + 1)$, and $\mu (t + 1)$ according to \eqref{S3-EQ-25b}, \eqref{S3-EQ-23c}, and \eqref{S3-EQ-27c}, respectively;
				\STATE Compute ${\rm MSE}$ by \eqref{S3-EQ-21M};
				\STATE $ t = t + 1 $;
				\UNTIL{${\rm MSE} \leq \varepsilon$};
				\STATE $\hat{ \bm{p} } = \lfloor \tilde{ \bm{w} } \rceil \circ \bm{x}_{p} ( t )$.
			\end{algorithmic}
		\end{algorithm*}	

\section{Simulation Results and Discussions}
	This section presents simulation results to evaluate the performances of the designed algorithms compared with state-of-the-art benchmark ones. The simulation parameter settings are as follows unless specified otherwise. On the one hand, we use a similar parameter setting for the wireless communication system as \cite{Machine2020Wang}. Specifically, we set the noise power $\sigma^{2} = - 77 \, {\rm dBm}$, the communication bandwidth $B = 180 \, {\rm kHz}$, the path loss of the $k^{\rm th}$ user $\varrho_{k} = - 90 \, {\rm dB}$, and the channel $\bm{h}_{k}$ is generated according to $\mathcal{CN} ( \bm{0}, \, \varrho_{k} \bm{I} )$. Also, we assume that the number of users is identical among different tasks, i.e., $| \mathcal{K}_{1} | = | \mathcal{K}_{2} | = \cdots = | \mathcal{K}_{I} | = 120$.  This is a valid assumption since we consider massive connectivity in large-scale IoT networks. On the other hand, for the task-oriented learning at the edge, we consider a support vector machine (SVM) for the classification of digits dataset in Scikit-learn \cite{Scikit2011Pedregosa}, a $6$-layer convolutional neural network (CNN$6$) for classification of the MNIST dataset \cite{MNIST2012Deng}, a $110$-layer deep residual network (ResNet$110$) using the CIFAR10 dataset \cite{Deep2016Kaiming}, and a PointNet using $3$D point clouds in the ModelNet$40$ dataset \cite{PointNet2017Charles}. In our pertaining simulation experiments, the single-task case $\{ \text{SVM} \}$, two-task case $\{ \text{SVM}, \, \text{CNN}6 \}$, and four-task case $\{ \text{SVM}, \, \text{CNN}6, \, \text{ResNet}110, \, \text{PointNet} \}$ are considered. For ease of tractability, relative learning parameters are summarized in Table~\ref{Table1}. For more details on how to get these learning parameters, the interested reader refers to Section~III of \cite{Machine2020Wang}. Apart from simulation experiments, we also investigate autonomous vehicle perception in the real world to demonstrate the excellent generalization performance of our proposed model.	
	
	In the simulation experiments, we consider seven schemes: a parallel task-oriented power allocation scheme (i.e., Algorithm~\ref{Algorithm1}), an accelerated task-oriented power allocation scheme (i.e., Algorithm~\ref{Algorithm2}), the parallel algorithm without scheduling (Algorithm~\ref{Algorithm1} w/o SH for short), and the accelerated algorithm without scheduling (Algorithm~\ref{Algorithm2} w/o SH for short). In addition to our algorithms, we also simulate two benchmark ones: a sum-rate maximization scheme \cite{Achieving2012Shatri} and an MM-based LCPA scheme \cite{Machine2020Wang}. The sum-rate maximization algorithm is typical in conventional wireless communications but only considers the wireless channel state information without accounting for the learning factors. Finally, for fair comparison of different multi-user scheduling strategies, the user-fair scheduling (UFS) algorithm developed in \cite{9222214} is also accounted for in the simulation experiments.
	
		\begin{table*}[t!]
			\centering		
			\small
			\setstretch{1.2}	
			\caption{Summary of the learning Parameters \cite{Machine2020Wang}.}
			\setlength{ \arraycolsep }{-0.2em}
			\resizebox{\linewidth}{!}{
				\begin{tabular}{!{\vrule width1.2pt}c !{\vrule width1.2pt} c !{\vrule width1.2pt}c !{\vrule width1.2pt}c !{\vrule width1.2pt}c !{\vrule width1.2pt}}
					\Xhline{1.2pt}
					\textbf{Models} & \textbf{Datasets} & \textbf{Symbols} & \textbf{Values} & \textbf{Description} \\
					\Xhline{1.0pt}
					SVM \cite{Scikit2011Pedregosa} & Digits & $( a_{1}, \, b_{1}, \, A_{1}, \, V_{1} )$ & $( 5.2, \, 0.72, \, 200, \, 324 \, {\rm bits})$ & The $1^{\rm st}$ learning task \\
					\hline
					CNN$6$ \cite{MNIST2012Deng} & MNIST & $( a_{2}, \, b_{2}, \, A_{2}, \, V_{2} )$ & $( 7.3, \, 0.69, \, 300, \, 6276 \, {\rm bits})$ & The $2^{\rm nd}$ learning task \\
					\hline
					ResNet$110$ \cite{Deep2016Kaiming} & CIFAR$10$ & $( a_{3}, \, b_{3}, \, A_{3}, \, V_{3} )$ & $( 8.15, \, 0.44, \, 1600, \, 24584 \, {\rm bits})$ & The $3^{\rm rd}$ learning task \\
					\hline
					PointNet \cite{PointNet2017Charles} & ModelNet$40$ & $( a_{4}, \, b_{4}, \, A_{4}, \, V_{4} )$ & $( 0.96, \, 0.24, \, 800, \, 192008 \, {\rm bits})$ & The $4^{\rm th}$ learning task \\
					\Xhline{1.0pt}
				\end{tabular}
			}
			\label{Table1}
		\end{table*}

	\subsection{Convergence Performance and Complexity Analysis}
	In this subsection, the number of antennas $N = 2$, the total transmit power $P = 13 \, {\rm dBm}$ (i.e., $20 \, {\rm mW}$), the transmit time $T = 10 \, {\rm s}$ for the single-task case, $T = 20 \, {\rm s}$ for the two-task case, and $T = 200 \, {\rm s}$ for the four-task case are used in the simulation experiments. The dataset types and task parameters are defined in Table~\ref{Table1}. In particular, as the four-task case is associated with deep networks, $T = 200 \, {\rm s}$ is set to obtain enough data to fine-tune these deep networks. To evaluate the process of convergence, we define a mean squared error (MSE) as
		\begin{align}
			&{\rm MSE} \nonumber \\
			&{}\triangleq \left\| \bm{p} ( t ) - \bm{p} ( t - 1 ) \right\|_{ 2 } + \left\| \bm{\delta} ( t ) - \bm{\delta} ( t - 1 ) \right\|_{ 2 } + \left| \| \bm{p} (t) \|_{1} - P \right| \nonumber \\
			&\quad{} + \, \left\| \left( \bm{G} - \bm{D} \right) \bm{p} (t) - \bm{\delta} (t) + \sigma^{2} \bm{\mathsf{1}} \right\|_{2}. \label{S3-EQ-21M}
		\end{align}	
	
\begin{figure*}[t!]
			\centering
			\subfloat[Single-task case.]{\includegraphics[width=0.32\linewidth]{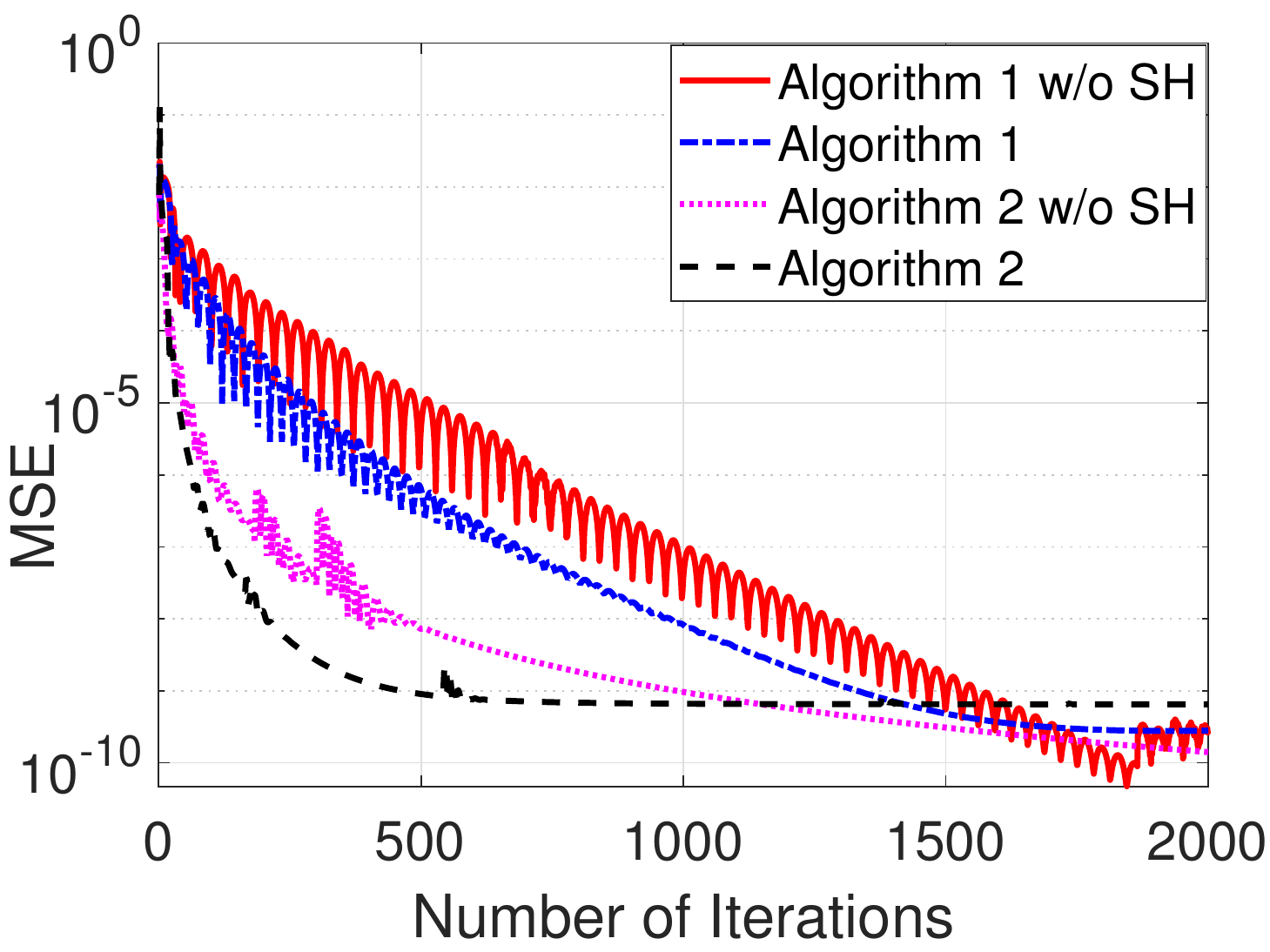} \label{Fig4a}}
			\hfil
			\subfloat[Two-task case.]{\includegraphics[width=0.32\linewidth]{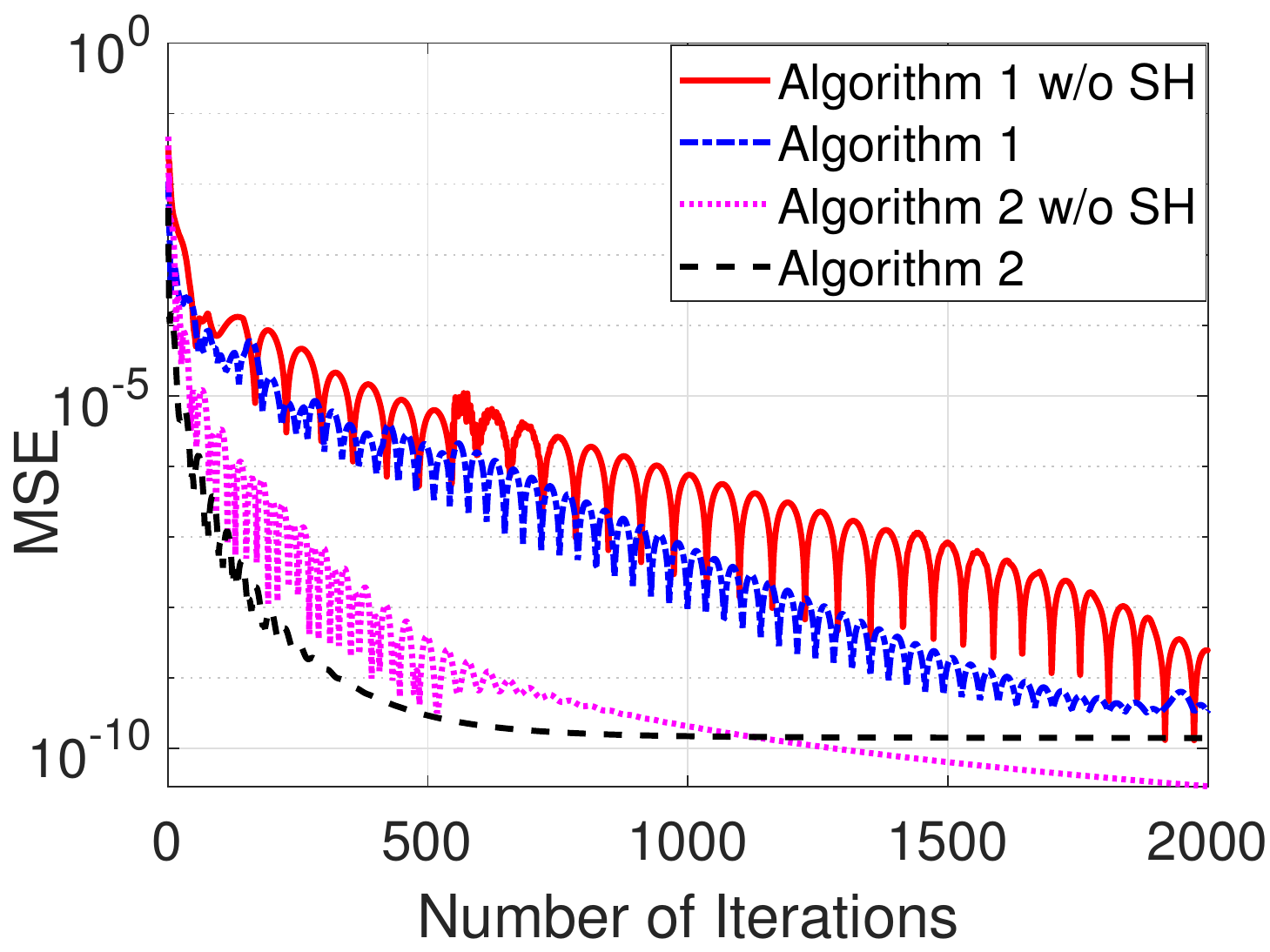} \label{Fig4b}}
			\hfil
			\subfloat[Four-task case.]{\includegraphics[width=0.32\linewidth]{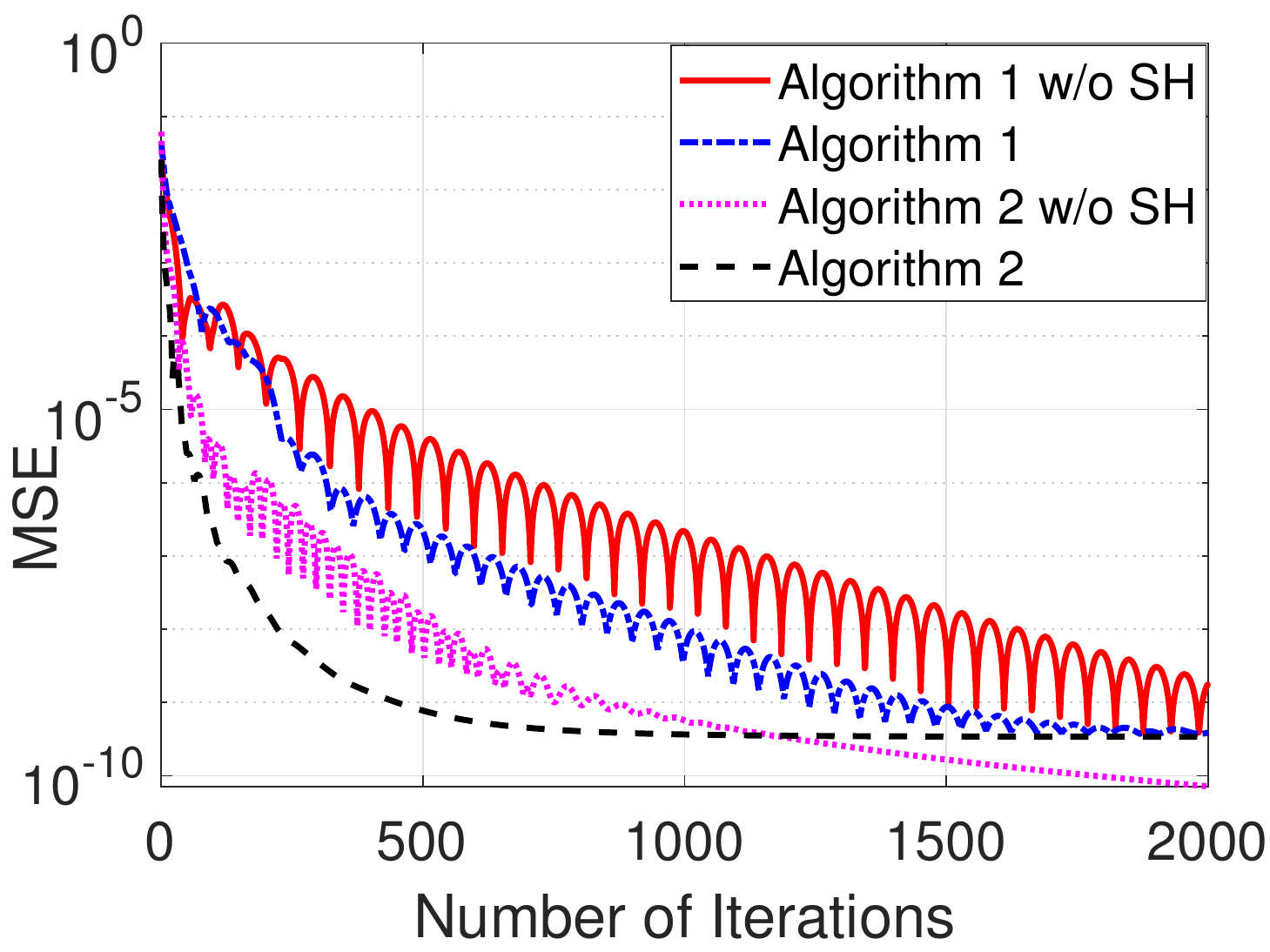} \label{Fig4c}}
			\hfil
			\caption{Mean squared error vs. the number of iterations.}
			\label{Fig4}
		\end{figure*}

	Figure~\ref{Fig4} depicts the MSE computed by \eqref{S3-EQ-21M} versus the number of iterations. On the one hand, we observe from Fig.~\ref{Fig4a} that Algorithms~\ref{Algorithm1} and \ref{Algorithm2} with multi-user scheduling outperform those without it in terms of both convergence speed and MSE performance. The reason behind these observations is that although the redundant variable introduced may slow down convergence and increase instability in the proposed algorithms, the multi-user scheduling strategy activates only a small fraction of users. Thus the dimensionality of the corresponding variable is highly reduced. Therefore, the algorithm with multi-user scheduling is relatively stable and converges faster. On the other hand, Fig.~\ref{Fig4a} also shows that the performance of Algorithm~\ref{Algorithm1} suffers from a slower convergence and more severe stochastic fluctuations than Algorithm~\ref{Algorithm2}. The reason is that Algorithm~\ref{Algorithm2} accelerates the convergence rate from $\mathcal{O} ( 1 / \tau )$ to $\mathcal{O} ( 1 / \tau^{2} )$. Similarly, Figs.~\ref{Fig4b} and \ref{Fig4c} illustrate that multi-scheduling and accelerated algorithms also benefit from faster convergence and lower MSE in the two-task and four-task learning cases, respectively.
	
\begin{figure*}[t!]
			\centering
			\subfloat[Single-task case.]{\includegraphics[width=0.32\linewidth]{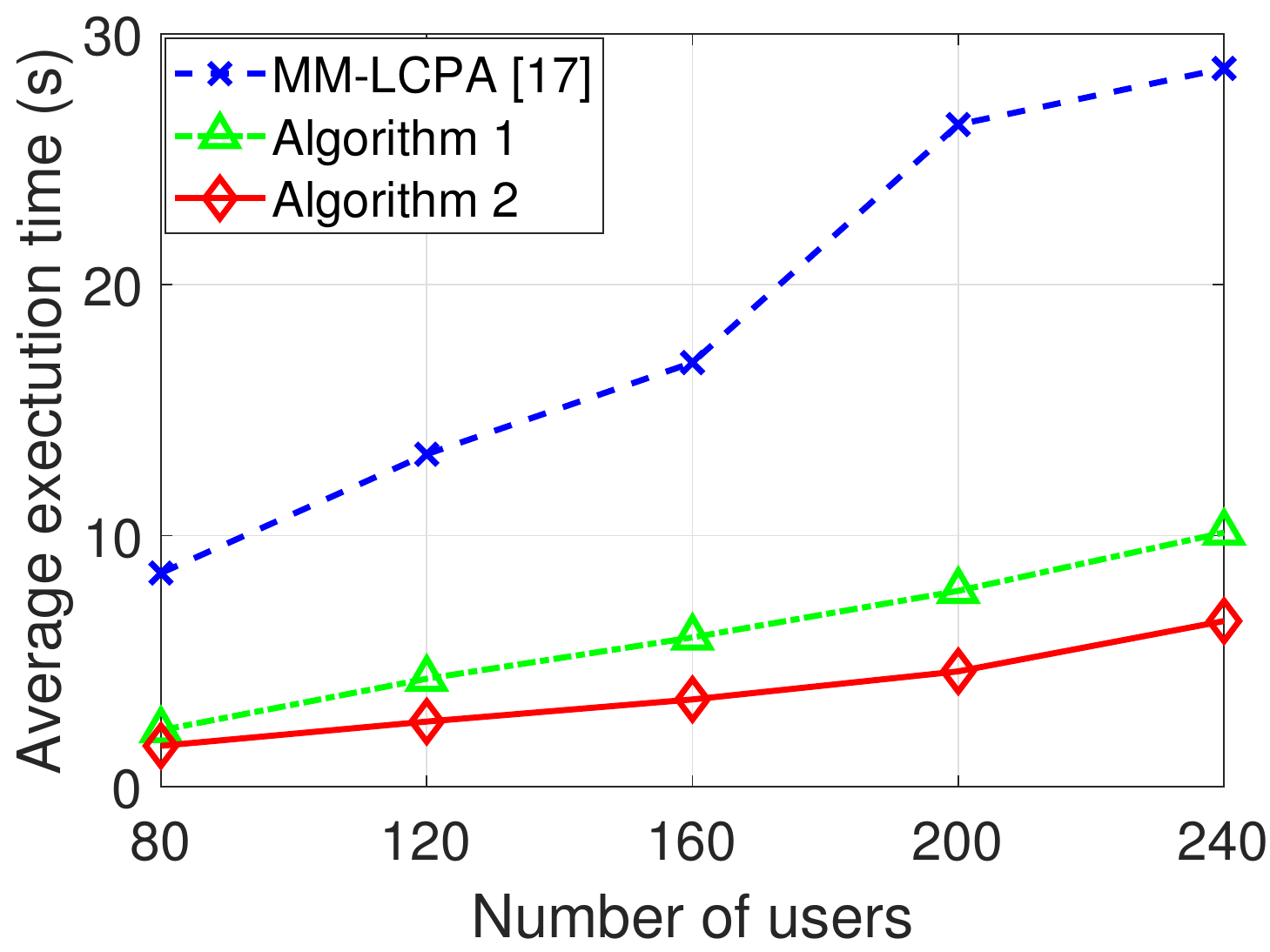} \label{Fig5a}}
			\hfil
			\subfloat[Two-task case.]{\includegraphics[width=0.32\linewidth]{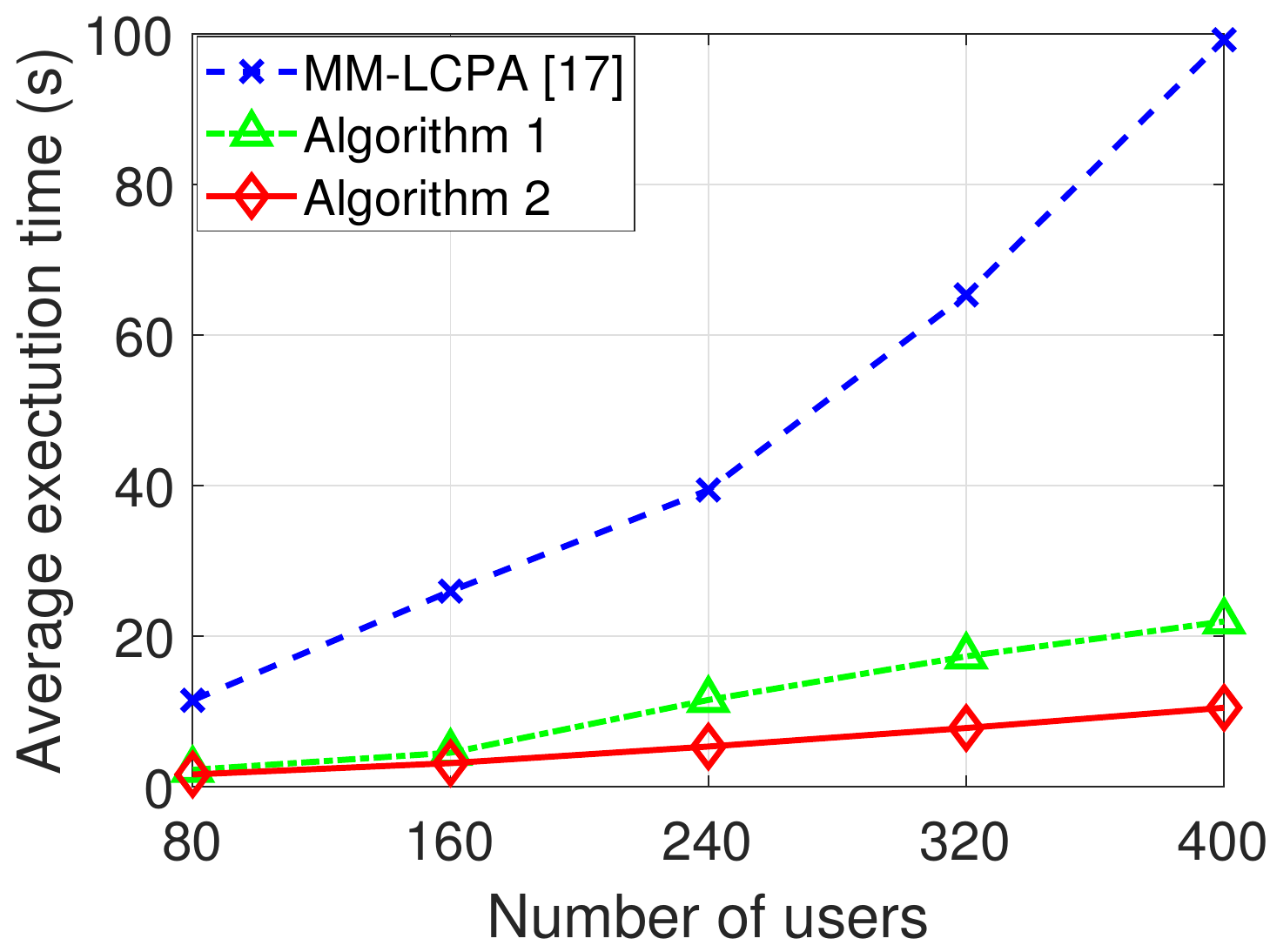} \label{Fig5b}}
			\hfil
			\subfloat[Four-task case.]{\includegraphics[width=0.32\linewidth]{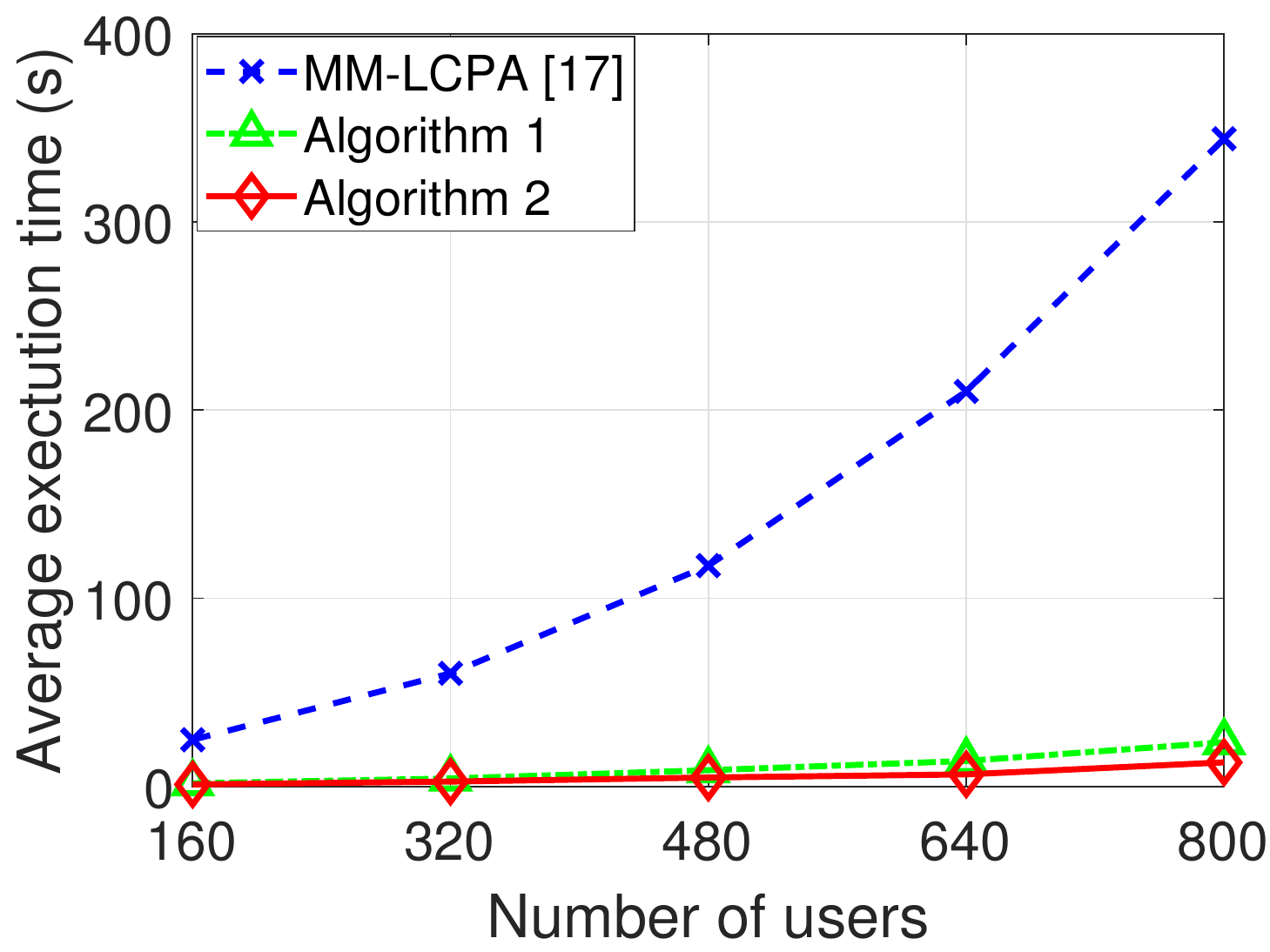} \label{Fig5c}}
			\hfil
			\caption{Average execution time vs. the number of users.}
			\label{Fig5}
		\end{figure*}

	Figure~\ref{Fig5} illustrates the computational complexity in the sense of the average execution time. On the one hand, Fig.~\ref{Fig5a} shows that the MM-LCPA algorithm for the single-task case developed in \cite{Machine2020Wang} has a longer execution time than our algorithms.  Even worse, the MM-LCPA algorithm shows a steeper increment than ours. The reason behind these observations is that, when the number of users $K$ is large, the per-iteration complexity of two proposed algorithms is $\mathcal{O} \left( K^{2} + K \right)$ whereas that of MM-LCPA is as high as $\mathcal{O} \left( ( I + K^{2} + K )^{3.5} \right)$. We also observe that Algorithm~\ref{Algorithm2} has a shorter execution time than Algorithm~\ref{Algorithm1}. It is because the accelerated algorithm speeds up the convergence rate from $\mathcal{O} \left( 1 / \tau \right)$ to $\mathcal{O} \left( 1 / \tau^{2} \right)$. Hence it decreases the number of iterations, specifically for large-scale IoT networks. On the other hand, Figs.~\ref{Fig5b}~and~\ref{Fig5c} show that the execution time of our algorithms remains almost the same as the number of tasks changes from two to four, compared with Fig.~\ref{Fig5}a. For example, the computational time is approximately computed by $10 \, {\rm s}$ for $K=200$ in the single-task case, $K=400$ in the two-task case, and $K=800$ in the four-task case  (i.e., each task has the same number of users). The reason behind these observations is that the per-iteration complexity of our parallel algorithm is reduced from $\mathcal{O} \left( K^{2} + K \right)$ to $\mathcal{O} \left( ( K^{2} + K ) / I \right)$. In other words, if the value of $K$ is fixed, the computational complexity of our algorithms decreases with the number of tasks $I$. Thus, we infer that the proposed parallel algorithms efficiently solve the task-oriented power allocation problem.
	
	\subsection{Learning Error Performance}
\begin{figure*}[t!]
			\centering
			\subfloat[Single-task case.]{\includegraphics[width=0.32\linewidth]{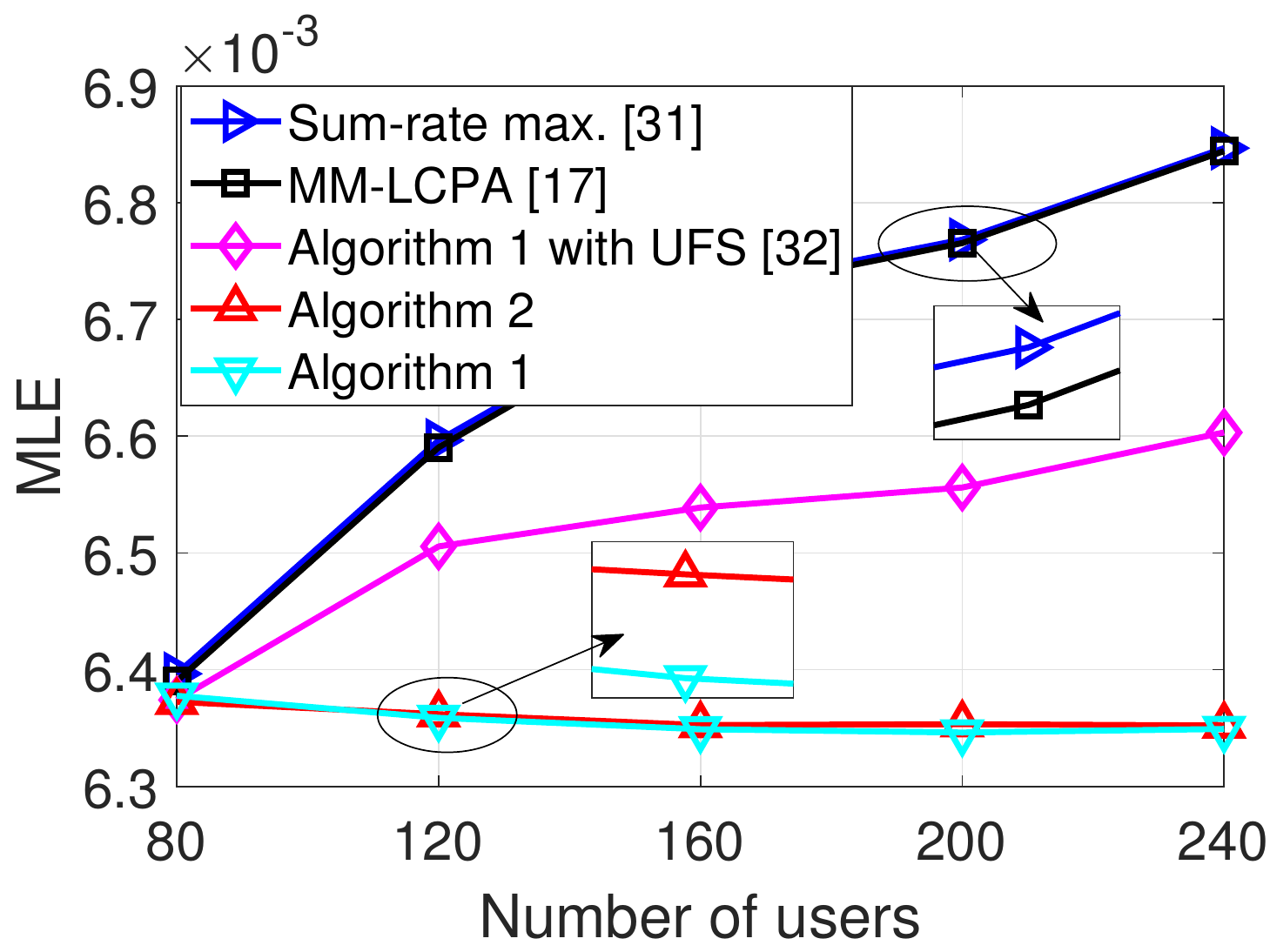} \label{Fig6a}}
			\hfil
			\subfloat[Two-task case.]{\includegraphics[width=0.32\linewidth]{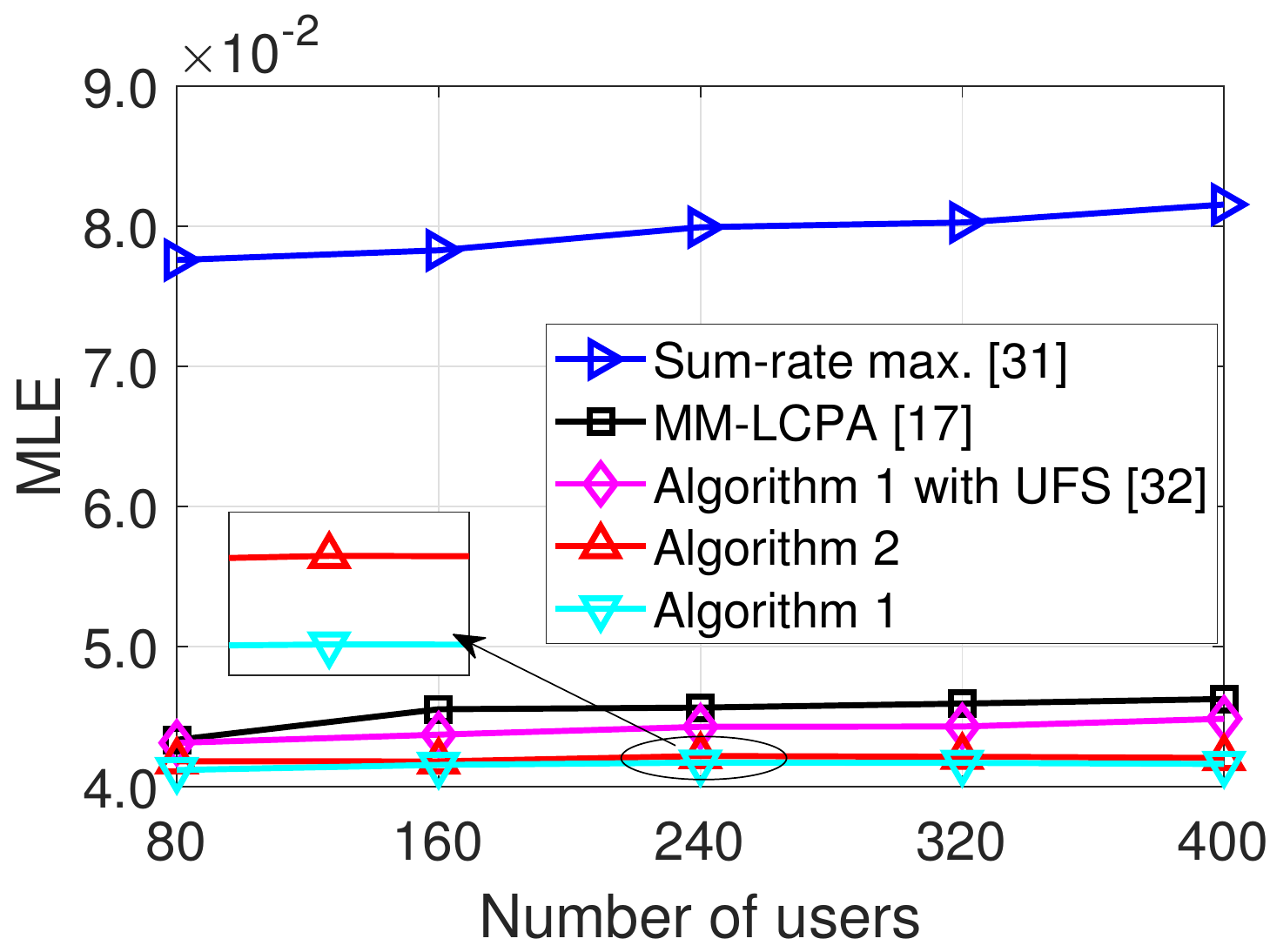} \label{Fig6b}}
			\hfil
			\subfloat[Four-task case.]{\includegraphics[width=0.32\linewidth]{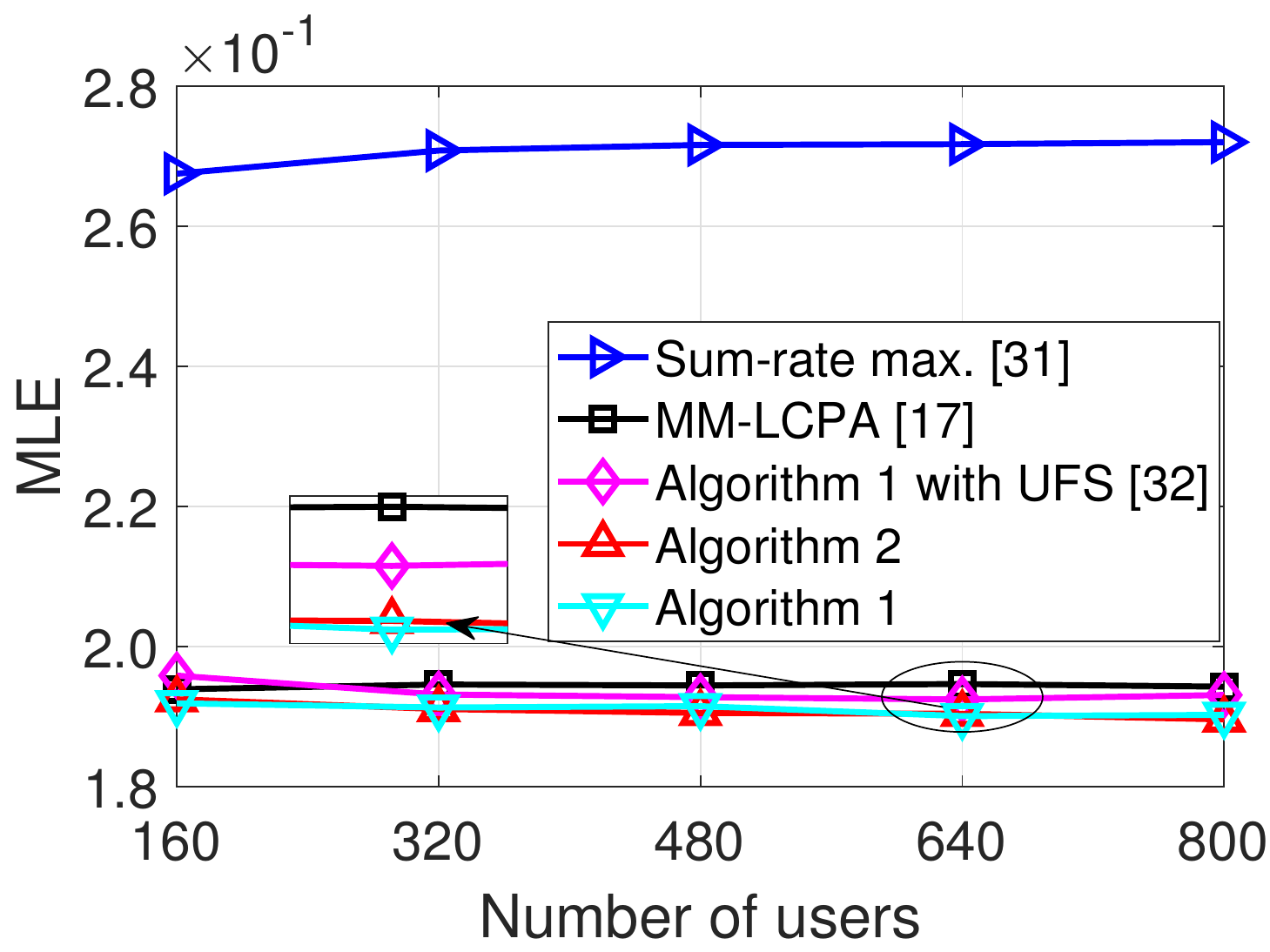} \label{Fig6c}}
			\caption{Mean learning error vs. the number of users.}	
			\label{Fig6}
		\end{figure*}

	Figure~\ref{Fig6} depicts the mean learning error (MLE) computed by \eqref{S1-EQ-3}. On the one hand, Fig.~\ref{Fig6a} shows that the MM-LCPA algorithm developed in \cite{Machine2020Wang} performs similarly to the sum-rate maximization algorithm developed in \cite{Achieving2012Shatri}, and they both underperform our algorithms. The reason behind these observations is that in the case of single-task, the objective function of the MM-LCPA algorithm degenerates into that of the sum-rate maximization algorithm due to the monotonicity of the learning error function, such that they have similar performance. Instead, as multi-user scheduling eliminates CCI in dense networks, the proposed algorithm outperforms the others. On the other hand, in the different task-oriented learning cases, Figs.~\ref{Fig6b} and \ref{Fig6c} show that the MLE of the MM-LCPA algorithm is superior to that of the sum-rate maximization algorithm due to the joint design of efficient task-oriented communications for different learning models. Also, it is seen that our algorithms have a smaller MLE than the MM-LCPA and the UFS algorithm developed in \cite{9222214}: the former is due to the multi-user scheduling and task fairness of our algorithms, whereas the latter is caused by the fact that the UFS algorithm concentrates on user fairness but degrades learning performance.
	
	In summary, Table~\ref{Table2} compares the four algorithms discussed above in terms of computational complexity, convergence rate, parallelization capability, and MLE. Our designed Algorithms~\ref{Algorithm1} and \ref{Algorithm2} are effective for task-oriented power allocation, thanks to their low computational complexity, fast convergence rate, high parallel capability, and low learning error. In particular, the former applies to small- or medium-scale IoT networks in terms of lower MLE, whereas the latter adapts to large-scale ones thanks to its faster convergence rate.

	\subsection{Experimental Validation for Autonomous Vehicle Perception}
\begin{figure}[t!]
			\centering
			\includegraphics[width=0.95\linewidth]{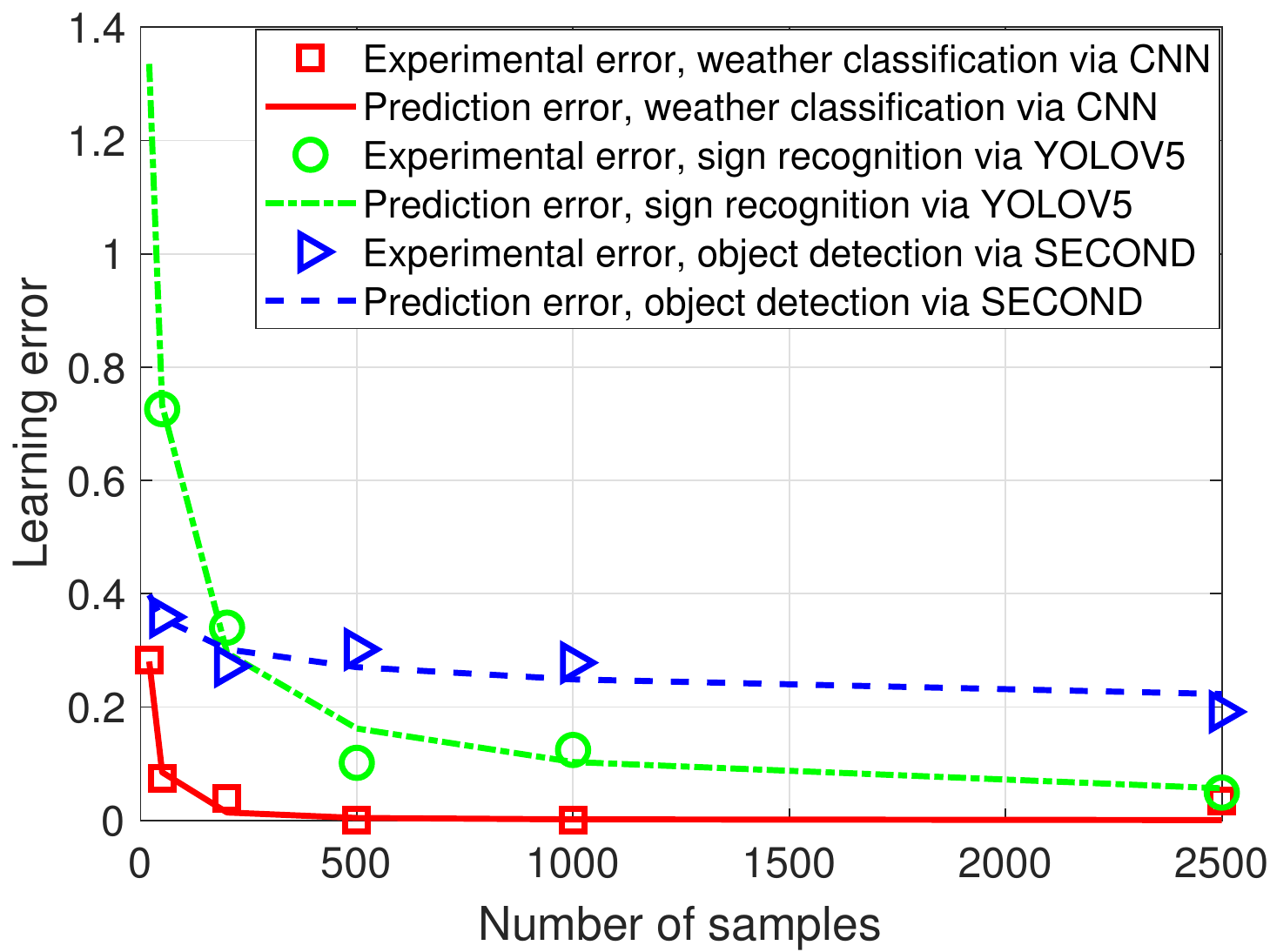}
			\caption{Learning error vs. the number of samples.}	
			\label{Fig7}
		\end{figure}

	To verify the robustness of the proposed algorithms in real-world applications, we consider three perception tasks in autonomous driving \cite{abs-2206-01748}, and they are  Task~$1$: weather classification using the RGB images and CNN; Task~$2$: traffic sign detection using the RGB images and YOLOV$5$, and Task~$3$: object detection using the point cloud data and sparsely embedded convolutional detection object detection (SECOND). In the pertaining experiments, all the datasets are generated by the CarlaFLCAV framework, which is an open-source autonomous driving simulation platform and online available at \url{https://github.com/SIAT-INVS/CarlaFLCAV}. In particular, the transmit time $T = 500 \, {\rm s}$ is set for this autonomous vehicle perception. The size of each RGB image sample is $V_{1} = V_{2} = 0.7 \, {\rm MB}$ and that of each point cloud sample is $V_{3} = 1.6 \, {\rm MB}$. The number of historical data samples is $A_{1} = A_{2} = A_{3} = 300$. By fitting the error function to the historical data, we obtain the model parameters $(a_{1}, \, b_{1}) = (10.34, \, 1.2)$, $(a_{2}, \, b_{2}) = (8.89, \, 0.64)$, and $(a_{3}, \, b_{3}) = (0.5, \, 0.1)$ for Tasks $1$, $2$ and $3$, respectively. It can be seen from Fig.~\ref{Fig7} that the three fitting curves match the experimental data very well. Note that with a smaller $A_{i}$, the estimated parameters $(a_{i}, \, b_{i})$ may be less accurate. However, such parameters can still perform the power allocation efficiently since our goal is to distinguish different tasks rather than accurately predict learning errors.
	
	
		\begin{table*}[t!]
			\centering
			\setstretch{1.2}		
			\caption{Performance comparison.}
			\setlength{ \arraycolsep }{-0.2em}
			\begin{threeparttable}[!t]
				\begin{tabular}{!{\vrule width1.2pt}c !{\vrule width1.2pt} c !{\vrule width1.2pt}c !{\vrule width1.2pt}c !{\vrule width1.2pt}c !{\vrule width1.2pt}}
					\Xhline{1.2pt}
					\textbf{Algorithm} & \textbf{Complexity} & \textbf{Convergence rate} & \textbf{Parallelism}\tnote{a} & \textbf{MLE} \\
					\Xhline{1.0pt}
					Sum-rate max. \cite{Achieving2012Shatri} & $\mathcal{O} \left( ( K - 1 )^{7} \right)$ & $\mathcal{O} \left( 1 / \log \tau \right)$ & \XSolidBrush & high \\
					\hline
					MM-based LCPA \cite{Machine2020Wang} & $\mathcal{O} \left( ( I + K^{2} + K )^{3.5} \right)$ & $\mathcal{O} \left( 1 / \log \tau \right)$ & \XSolidBrush & low \\
					\hline
					Algorithm~\ref{Algorithm1} & $\mathcal{O} \left( ( K^{2} + K ) / I \right)$ & $\mathcal{O} \left( 1 / \tau \right)$ & \Checkmark & low \\
					\hline
					Algorithm~\ref{Algorithm2} & $\mathcal{O} \left( ( K^{2} + K ) / I \right)$ & $\mathcal{O} \left( 1 / \tau^{2} \right)$ & \Checkmark & low \\
					\Xhline{1.0pt}
				\end{tabular}
				\label{Table2}
				{\scriptsize
					\begin{tablenotes}
						\item[a] The tick ``\Checkmark'' indicates a functionality supported, whereas the cross ``\XSolidBrush" indicates not supported.
				\end{tablenotes}}
			\end{threeparttable}
		\end{table*}
	
		\begin{figure*}[t!]
			\centering
			\includegraphics[width=0.75\linewidth]{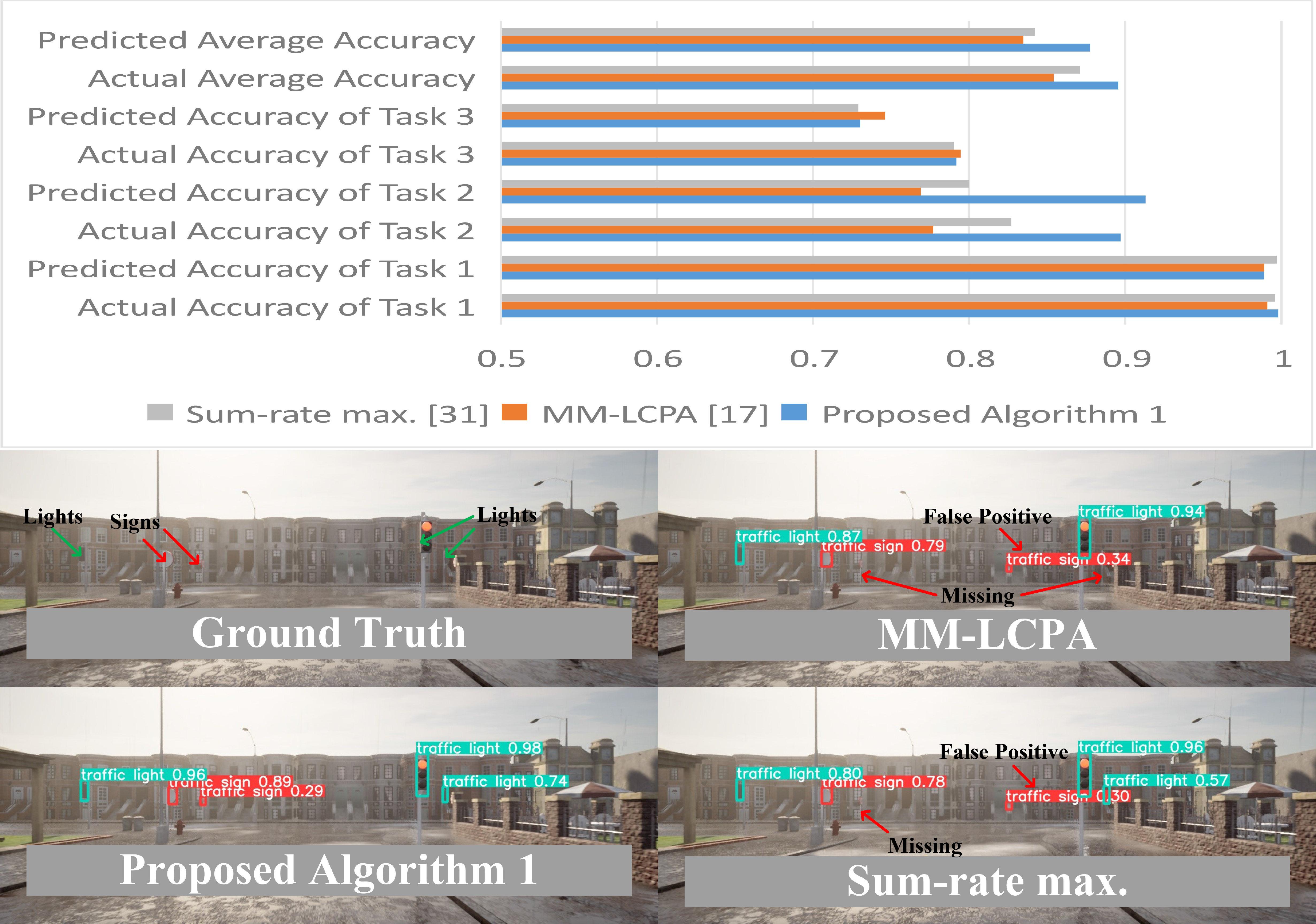}
			\caption{Qualitative and quantitative results of multi-task perception for autonomous driving.}	
			\label{Fig8}
		\end{figure*}
						
	The top panel of Fig.~\ref{Fig8} compares the perception accuracies of the proposed and benchmark algorithms. Firstly, it is seen that the actual perception accuracies obtained from the machine learning experiments coincide with the predicted perception accuracies obtained from the error functions for all the tasks and simulated schemes. Secondly, the proposed algorithm achieves significantly higher average perception accuracy than the MM-LCPA and sum-rate maximization schemes. This is because the proposed algorithm is a task-oriented scheme, which computes the ``learning curve'', i.e., the derivative of the learning error w.r.t. the number of samples, for each task by leveraging the associated fitted error functions. As such, it automatically allocates more power resources to the task with a more significant learning curve since it needs more samples to train the learning model. In our experiment, Task~$2$ has the steepest ``learning curve'' as seen from Fig.~\ref{Fig7}. Accordingly, the proposed algorithm allocates more power to Task $2$ and achieves the highest perception accuracy. In contrast, the MM-LCPA and sum-rate maximization schemes give more power resources to Tasks~$1$ and $3$, whose learning errors are saturated when the number of samples exceeds $400$. Therefore, these benchmark schemes are less learning-efficient than the proposed scheme.
	
	Lastly, the qualitative results of different schemes are shown in the bottom panel of Fig.~\ref{Fig8}. It can be seen that there are three traffic lights and two traffic signs at the T-junction. The proposed Algorithm~$1$ successfully detects all the objects in the image. The MM-LCPA scheme fails to detect a far-away traffic sign and a traffic light (impeded by the wall) while misclassifying a door as a traffic sign. The sum-rate maximization scheme fails to detect a far-away traffic sign and misclassifies a door as a traffic sign. The reason behind these observations is that the proposed Algorithm~$1$ can obtain more samples for multiple tasks in task-oriented principle than other schemes. However, the MM-LCPA algorithm only focuses on one of these tasks, even if this task is unimportant. The sum-rate maximization scheme may not obtain data for multiple tasks as it ignores task-irrelevant information. 

\section{Conclusions}
	This paper has developed a task-oriented power allocation model to process distinct learning datasets for large-scale IoT networks, especially for multi-task multi-modal scenarios. To deal with massive connectivity, a multi-user scheduling algorithm has been designed to mitigate co-channel interference and decouple multi-user scheduling and power allocation. Moreover, highly parallel and accelerated algorithms have been designed to solve multi-objective and large-scale optimization problems. Extensive experimental results have shown that multi-user scheduling could effectively mitigate the influence of interference in dense networks. The parallel algorithm and its accelerated version enable different learning tasks efficiently, including the real-world multi-task multi-modal scenario for autonomous vehicle perception. In real-world applications, the proposed algorithms can be deployed at the edge, e.g., the gateway of a large-scale IoT network, which can then inform the users of their transmit powers and other parameters through the downlink control channel, e.g., the narrowband physical downlink control channel in NB-IoT networks. However, as the offline-learning mode is not adaptive to a real-time wireless environment, developing an online-learning mode is promising for future work.

{\appendices
\numberwithin{equation}{section}

\section{Proof of Proposition~\ref{S3-P1}} \label{SA-A}
	Substituting $\delta_{k} \triangleq \sum_{ \ell \in \mathcal{K} \setminus k } \tilde{G}_{k, \ell} p_{\ell} + \sigma^{2}$ and $\tilde{G}_{k, k} \triangleq w_{k} {G}_{k, k} $ into the cost function of $\mathcal{P}2$, and performing some algebraic manipulations, we obtain
			\begin{subequations}
				\begin{align}
					\mathcal{P}2a : \min_{ \{ w_{k} \}_{ k \in \mathcal{K}_{i} } } \, & a_{i} \left( \dfrac{B T}{ V_{i} } \sum_{ k \in \mathcal{K}_{i} } w_{k} \log_{2} \left( 1 + \dfrac{ \tilde{G}_{k, k} p_{k} }{ w_{k} \delta_{k} } \right) + A_{i} \right)^{ - b_{ i } } \\
					{\rm s.t.} \ & 0 < w_{k} \leq 1 , \label{SA-EQ-A1a} \\
					& \sum_{ k \in \mathcal{K}_{i} } w_{k} \leq N_{i}. \label{SA-EQ-A1b}
				\end{align}
			\end{subequations}
	In light of the non-increasing characteristices of $a_{i} x^{ - b_{i} }$ where $x > 0$, and the sparsity constraint \eqref{SA-EQ-A1b}, $\mathcal{P}2a$ can be transformed into its equivalent penalized form:
		\begin{subequations}
			\begin{align}\label{P2B}
				\mathcal{P}2b : \min_{ \bm{w}_{i} } \, & - \sum_{ k \in \mathcal{K}_{i} } w_{k} \ln \left( 1 + \dfrac{ \tilde{G}_{k, k} p_{k} }{ \delta_{k} w_{k} } \right) + \nu_{i} \sum_{ k \in \mathcal{K}_{i} } w_{k} \\
				{\rm s.t.} \ & \eqref{SA-EQ-A1a}, \, \eqref{SA-EQ-A1b}, \nonumber
			\end{align}
		\end{subequations}
	where $\bm{w}_{i} \triangleq \left[ w_{i_{1}}, w_{i_{2}}, \cdots, w_{i_{\left|\mathcal{K}_{i}\right|}} \right]^{T}$, and $\nu_{i} > 0$ is a tuning parameter for the sparsity regulation. 
	
	Next, by setting the objective function of \eqref{P2B} as $J ( w_{k} ) \triangleq - w_{k} \ln \left( 1 + { \tilde{G}_{k, k} p_{k} }/{ (\delta_{k} w_{k}) } \right) + \nu_{i} w_{k}$, it follows that ${ \partial J ( w_{k} ) }/{ \partial w_{k} } = - \ln \left( 1 + { G_{k, k} \tilde{p}_{k} }/{ (\delta_{k} w_{k}) } \right) +  { G_{k, k} p_{k} }/{(G_{k, k} p_{k} + \delta_{k} )} + \nu_{i}$. Let $ \partial J ( w_{k} ) / \partial w_{k} = 0$, and we obtain
		\begin{equation}
			\hat{w}_{k} = \dfrac{ \tilde{G}_{k, k} p_{k} }{ \delta_{k} \left( \exp \left( \frac{ G_{k, k} p_{k} }{ \delta_{k} + G_{k, k} p_{k} } + \nu_{i} \right) - 1 \right) }.
		\end{equation}	
	
	Considering $0 < w_{k} \leq 1$, there are three cases of $\hat{\omega}_{k}$ to account for:
		\begin{itemize}
			\item [1)] If $\hat{w}_{k} \leq \epsilon$, the minimization of $J ( w_{k} )$ is obtained at $w_{k} = \epsilon$;
			\item [2)] If $\epsilon < \hat{w}_{k} < 1$, the minimization of $J ( w_{k} )$ is obtained at $w_{k} = \hat{w}_{k}$;
			\item [3)] If $\hat{w}_{k} \geq 1$, the minimization of $J ( w_{k} )$ is obtained at $w_{k} = 1$.
		\end{itemize}
	As a result, the optimization point is given by \eqref{S3-P1-EQ-6}. This completes the proof.

\section{Proof of Proposition~\ref{S3-T1}} \label{SA-C}	
	The Lagrange multiplier $\beta_{i}$ for $\bm{\mathsf{1}}^{T} \bm{p}_{i} = P_{i}$ is given by
		\begin{equation} \label{SA-EQ-C1}
			\beta_{i} ( t + 1 ) = \beta_{i} ( t ) + \mu ( t ) \left( \bm{\mathsf{1}}^{T} \bm{p}_{i} ( t + 1 ) - P_{i} ( t + 1 ) \right),
		\end{equation}	
	where $\bm{p}_{i} ( t + 1 )$ and $P_{i} ( t + 1 )$ are obtained by the minimization of the ALF given by \eqref{S1-EQ-10}. These minimizations concerning $\bm{p}_{i}$ and $P_{i}$ are computed iteratively:
			\begin{subequations}
				\begin{align}
					\bm{p}_{i} & = \underset{ \bm{p}_{i} \succeq \bm{0} }{ \rm argmin } \, \lambda_{i} \Phi_{i} ( \bm{p}_{i} ) + \beta_{i} ( t ) \left( \bm{\mathsf{1}}^{T} \bm{p}_{i} - P_{i} \right)  \nonumber \\
							& \quad {}+ \dfrac{ \mu ( t ) }{2} \left( \bm{\mathsf{1}}^{T} \bm{p}_{i} - P_{i} \right)^{2}, \label{SA-EQ-C2a} \\
					P_{i} & = \underset{ \left\{ P_{i} \left| \underset{ i \in \mathcal{I} }{ \sum } P_{i} = P \right. \right\} } { \rm argmin } \hspace{-5pt} \left\{ - \sum_{ i \in \mathcal{I} } \beta_{i} ( t ) P_{i} + \dfrac{ \mu ( t ) }{2} \sum_{i \in \mathcal{I}} \left( \bm{\mathsf{1}}^{T} \bm{p}_{i} - P_{i} \right)^{2} \right\} , \label{SA-EQ-C2b}
				\end{align}
			\end{subequations}
	where
		\begin{equation} \nonumber
			\Phi_{i} ( \bm{p}_{i} ) \triangleq a_{i} \left(  \dfrac{B T}{ V_{i} } \sum_{ k \in \mathcal{K}_{i} } \tilde{w}_{k} \tilde{R}_{k} + A_{i} \right)^{ - b_{ i } }.
		\end{equation}
	Note that the minimization with respect to $\{ P_{i} \left| i \in \mathcal{I} \right. \}$ in \eqref{SA-EQ-C2b} involves a separable quadratic cost and a single equality constraint and can be carried out analytically. Given the optimization values $\bm{p}_{i} (t + 1)$, the optimization value $P_{i} (t + 1)$ in \eqref{SA-EQ-C2b} is analytically given by
		\begin{equation} \label{SA-EQ-C3}
			P_{i} ( t + 1 ) = \bm{\mathsf{1}}^{T} \bm{p}_{i} ( t + 1 ) + \dfrac{ \beta_{i} ( t ) - \beta ( t + 1 ) }{ \mu ( t ) },
		\end{equation}
	where  $\beta (t + 1)$ is a scalar Lagrange multiplier subject to $\sum_{i \in \mathcal{I}} P_{i} = P$, and it is determined by
		\begin{equation} \label{SA-EQ-C4}
			\beta ( t + 1 ) = \dfrac{1}{I} \sum_{i \in \mathcal{I}} \beta_{i} ( t ) + \dfrac{ \mu ( t ) }{I} \left( \bm{\mathsf{1}}^{T} \bm{p} ( t + 1 ) - P \right).
		\end{equation}		
	By comparing \eqref{SA-EQ-C3} with \eqref{SA-EQ-C1}, we see that
		\begin{equation} \label{SA-EQ-C5}
			\beta_{i} ( t + 1 ) = \beta ( t + 1 ).
		\end{equation}
	Then, summing \eqref{SA-EQ-C1} up for all $i \in \mathcal{I}$ yields
		\begin{subequations} \label{SA-EQ-C6}
			\begin{align}
				\beta ( t + 1 ) & = \beta ( t ) + \dfrac{ \mu ( t ) }{I} \sum_{i \in \mathcal{I}} \left( \bm{\mathsf{1}}^{T} \bm{p}_{i} ( t + 1 ) - P_{i} ( t + 1 ) \right) \label{SA-EQ-C6a} \\
				& = \beta ( t ) + \left( \beta ( t + 1 ) - \dfrac{1}{I} \sum_{i \in \mathcal{I}} \beta_{i} (t) \right) \label{SA-EQ-C6b} \\
				& = \beta ( t ) + \dfrac{ \mu ( t ) }{I} \left( \bm{\mathsf{1}}^{T} \bm{p} ( t + 1 ) - P \right), \label{SA-EQ-C6c}
			\end{align}
		\end{subequations}
	where \eqref{SA-EQ-C6b}-\eqref{SA-EQ-C6c} are derived by \eqref{SA-EQ-C3}-\eqref{SA-EQ-C4}, respectively, and $P_{i}$ is updated by
		\begin{equation} \label{SA-EQ-C7}
			P_{i} ( t + 1 ) = \bm{\mathsf{1}}^{T} \bm{p}_{i} ( t + 1 ) - \dfrac{1}{I} \left( \bm{\mathsf{1}}^{T} \bm{p} ( t + 1 ) - P \right),
		\end{equation}	
	where \eqref{SA-EQ-C7} is derived from \eqref{SA-EQ-C3} and \eqref{SA-EQ-C4}. Hence, \eqref{S3-T1-EQ-11a} and \eqref{S3-T1-EQ-12a} are immediately proved.
	
	Next, we derive \eqref{S3-T1-EQ-11b} and \eqref{S3-T1-EQ-12b}. Similar to \eqref{SA-EQ-C1}, we consider Lagrange multipliers $\bm{\alpha}^{\prime}_{i}$. The method of multipliers consists of
		\begin{equation} \label{SA-EQ-C8}
			\bm{\alpha}^{\prime}_{i} ( t + 1 ) = \bm{\alpha}^{\prime}_{i} ( t ) + \mu ( t ) \left(\bm{\Delta} \left( :, \, \mathcal{K}_{i} \right) {\bm{p}}_{i} ( t + 1 ) - \bm{z}_{i} ( t + 1 ) \right),
		\end{equation}
	where $\bm{p}_{i} ( t + 1 )$, $\bm{\delta}_{i} ( t + 1 )$, and $\bm{z}_{i} ( t + 1 )$ are obtained by the minimization of the ALF \eqref{S1-EQ-10}. 
	Similar to \eqref{SA-EQ-C4}, a Lagrange multiplier  vector $\bm{\alpha}$ is shown below:
		\begin{subequations} \label{SA-EQ-C10}
			\begin{align}
				& \bm{\alpha} (t + 1) \nonumber \\
				& = \dfrac{1}{I} \sum_{i \in \mathcal{I}} \bm{\alpha}^{\prime}_{i} ( t ) + \dfrac{ \mu ( t ) }{I} \left( \bm{\Delta} {\bm{p}} (t + 1) - \bm{\delta}  (t + 1) + \sigma^{2} \bm{\mathsf{1}} \right) \\
				& = \bm{\alpha} ( t ) + \dfrac{ \mu ( t ) }{I} \left( \bm{\Delta} {\bm{p}}  (t + 1) - \bm{\delta}  (t + 1) + \sigma^{2} \bm{\mathsf{1}} \right), \label{SA-EQ-C10a}
			\end{align}
		\end{subequations}			
	where \eqref{SA-EQ-C10a} is obtained by $\bm{\alpha}^{\prime}_{i}  (t + 1) = \bm{\alpha}  (t + 1)$. Moreover, we obtain the following optimization solution involving $\sum_{i = 1}^{I} \bm{z}_{i} = \bm{\delta} - \sigma^{2} \bm{\mathsf{1}}$:
		\begin{subequations} \label{SA-EQ-C11}
			\begin{align}
				& \bm{z}_{i} ( \bm{\delta}  (t + 1) ) \nonumber \\
				& = \bm{\Delta} \left( :, \, \mathcal{K}_{i} \right) {\bm{p}}_{i} ( t + 1 ) + \dfrac{ 1 }{ \mu ( t ) } ( \bm{\alpha}^{\prime}_{i} ( t ) - \bm{\alpha}  (t + 1) ) \\
				& = \bm{\Delta} \left( :, \, \mathcal{K}_{i} \right) {\bm{p}}_{i} ( t + 1 ) + \dfrac{1}{ \mu ( t ) } \left( \bm{\alpha} (t) - \bm{\alpha}  (t + 1) \right) \label{SA-EQ-C11a} \\
				& = \bm{\Delta} \left( :, \, \mathcal{K}_{i} \right) {\bm{p}}_{i} ( t + 1 ) - \dfrac{ 1 }{I} \left( \bm{\Delta} {\bm{p}}  (t + 1) - \bm{\delta}  (t + 1) + \sigma^{2} \bm{\mathsf{1}} \right) , \label{SA-EQ-C11b}
			\end{align}
		\end{subequations}	
	where \eqref{SA-EQ-C11a} is obtained by $\bm{\alpha}^{\prime}_{i}  (t + 1) = \bm{\alpha} (t + 1)$, and \eqref{SA-EQ-C11b} by \eqref{SA-EQ-C10a}.

\section{ Proof of Lemma~\ref{S3-L1} } \label{SA-E}
	First, we prove part i) of Lemma~\ref{S3-L1}: 
		\begin{equation} \label{SA-EQ-Lemma1-a}
			\begin{aligned}
				&\| \nabla_{ \bm{x} } \Phi_{i} ( \bm{x} | \bm{\delta}_{i}^{*} ) - \nabla_{ \bm{y} } \Phi_{i} ( \bm{y} | \bm{\delta}_{i}^{*} ) \|_{2} \\
				&\leq N_{1} \| \nabla_{ \bm{x} } \Phi_{i} ( \bm{x} | \bm{\delta}_{i}^{*} ) - \nabla_{ \bm{y} } \Phi_{i} ( \bm{y} | \bm{\delta}_{i}^{*} ) \|_{\infty} \leq L_{p} \| \bm{x} - \bm{y} \|_{2},
			\end{aligned}
		\end{equation}
	where $N_{1} \triangleq \| \nabla_{ \bm{x} } \Phi_{i} ( \bm{x} | \bm{\delta}_{i}^{*} ) - \nabla_{ \bm{y} } \Phi_{i} ( \bm{y} | \bm{\delta}_{i}^{*} ) \|_{0}^{ 1 / 2 }$ is a bounded constant and $L_{p}$ is a positive constant. By recalling the definition of \cite[Lemma~1]{Machine2020Wang}, \eqref{SA-EQ-Lemma1-a} can be obtained in a straightforward manner.
	
	To prove part ii) of Lemma 1, we notice that $\nabla_{ x_{k} } \Phi_{i} ( \bm{p}_{i}^{*} | \bm{x} )$ can be rewritten as $\nabla_{ x_{k} } \Phi_{i} ( \bm{p}_{i}^{*} | \bm{x} ) = h ( \bm{x} ) g ( x_{k} )$, with the auxiliary functions 
		\begin{subequations}
			\begin{align}
				& h ( \bm{x} ) = b_{i}  a_{i} \left( \sum_{ \ell \in \mathcal{K}_{i} } \dfrac{B T}{ V_{i} } \tilde{w}_{\ell} \log_{2} \left( 1 + \dfrac{ G_{\ell, \ell} p_{\ell}^{*} }{ x_{\ell} } \right) + A_{i} \right)^{ - b_{i} - 1 }, \\
				& g ( x_{k} ) = \dfrac{ B T \tilde{w}_{k} G_{k, k} p_{k}^{*} }{ \ln (2) V_{i} x_{k} ( x_{k} + G_{k, k} p_{k}^{*} ) }.
			\end{align}
		\end{subequations}
	where $x_{k}$ denotes the $k^{\rm th}$ entry of $\bm{x}$. The assumption $\Phi_{i} ( \bm{p}_{i}^{*} | \bm{\delta}_{i} ) \leq u_{0}$ gives
		\begin{equation} \label{SA-EQ-E4}
			\sum_{ \ell \in \mathcal{K}_{i} } \dfrac{B T}{ V_{i} } \tilde{w}_{\ell} \log_{2} \left( 1 + \dfrac{ G_{\ell, \ell} p_{\ell}^{*} }{ x_{\ell} } \right) + A_{i} \geq \left( \dfrac{ a_{i}}{u_{0}} \right)^{1 / b_{i}} ,
		\end{equation}
	then, we have
		\begin{subequations}
			\begin{align}
				& | h ( \bm{x} ) | \leq a_{i} b_{i} \left( \dfrac{ u_{0} }{ a_{i} } \right)^{1 + 1 / b_{i}} , \label{SA-EQ-E5a} \\
				& | g ( x_{k} ) | \leq \dfrac{ B T U_{0} }{ \ln (2) V_{i} \sigma^{2} ( \sigma^{2} + U_{0} ) }, \label{SA-EQ-E5b}
			\end{align}
		\end{subequations}
	where \eqref{SA-EQ-E5b} is derived by $U_{0} \geq G_{k, k} p_{k}$ and $x_{k} \geq \sigma^{2}$. Here, $G_{k, \ell}$ satisfies Gaussian distribution and $p_{k} \leq P$, hence we obtain an upper bound $U_{0}$ of $G_{k, k} p_{k}$ with a high probability \cite{8523680}. Furthermore, according to Lipschitz conditions \cite{Convex2015S} of $h$ and $g$, they satisfy
		\begin{subequations}
			\begin{align}
				& \left| h ( \bm{x} ) - h ( \bm{y} ) \right| \nonumber \\
				& \leq \sup_{ \bm{x} \succeq \sigma^{2} \bm{\mathsf{1}} } \| \nabla_{\bm{x}} h ( \bm{x} ) \|_{2} \times \| \bm{x} - \bm{y} \|_{2} \nonumber \\
				& \leq \dfrac{ K a_{i} b_{i} ( b_{i} + 1 ) B T U_{0} }{ \ln (2) I V_{i} \sigma^{2} ( \sigma^{2} + U_{0} ) } \left( \dfrac{ u_{0} }{ a_{i} } \right)^{1 + 2 / b_{i}} \| \bm{x} - \bm{y} \|_{2} , \label{SA-EQ-E6a} \\
				& | g ( x_{k} ) - g ( y_{k} ) | \nonumber \\
				& \leq \sup_{ x_{k} \geq \sigma^{2} } | \nabla_{x_{k}} g ( x_{k} ) | \times | x_{k} - y_{k} | \nonumber \\
				& \leq \dfrac{ B T U_{0} ( 2 \sigma^{2} + U_{0} ) }{ \ln (2) V_{i} \sigma^{4} ( \sigma^{2} + U_{0} )^{2} } | x_{k} - y_{k} | \nonumber \\
				& \leq \dfrac{ B T U_{0} ( 2 \sigma^{2} + U_{0} ) }{ \ln (2) V_{i} \sigma^{4} ( \sigma^{2} + U_{0} )^{2} } \| \bm{x} - \bm{y} \|_{2} . \label{SA-EQ-E6b}
			\end{align}
		\end{subequations}
	As a result, the following inequality is obtained:
		\begin{equation} \nonumber
			\begin{aligned}
				&\| \nabla_{ \bm{x} } \Phi_{i} ( \bm{p}_{i}^{*} | \bm{x} ) - \nabla_{ \bm{y} } \Phi_{i} ( \bm{p}_{i}^{*} | \bm{y} ) \|_{\infty} \\
				&\leq \sup_{k \in \mathcal{K}_{i}} | h ( \bm{x} ) | | g ( x_{k} ) - g ( y_{k} ) | + \left| h ( \bm{x} ) - h ( \bm{y} ) \right| | g ( x_{k} ) | \\
				&\leq L_{2} \| \bm{x} - \bm{y} \|_{2},
			\end{aligned}
		\end{equation}		
	where the first inequality is due to $| a b + c d | \leq |a| |b| + |c| |d|$, and the second inequality is obtained from \eqref{SA-EQ-E5a}, \eqref{SA-EQ-E5b}, \eqref{SA-EQ-E6a}, and \eqref{SA-EQ-E6b}; also, $L_{2}$ is defined as
		\begin{equation}
			\begin{aligned}
				L_{2} &\triangleq \dfrac{ a_{i} b_{i} B T U_{0} ( 2 \sigma^{2} + U_{0} ) }{ \ln (2) V_{i} \sigma^{4} ( \sigma^{2} + U_{0} )^{2} } \left( \dfrac{ u_{0} }{ a_{i} } \right)^{1 + 1 / b_{i}} \\
				&\quad{} + \, \dfrac{ K a_{i} b_{i} ( b_{i} + 1 ) B^{2} T^{2} U_{0}^{2} }{ \ln^{2} (2) I V_{i}^{2} \sigma^{4} ( \sigma^{2} + U_{0} )^{2} } \left( \dfrac{ u_{0} }{ a_{i} } \right)^{1 + 2 / b_{i}} .
			\end{aligned}
		\end{equation}
	Thus, the gradient function $\nabla_{ \bm{\delta}_{i} } \Phi_{i} ( \bm{p}_{i}^{*} | \bm{\delta} )$ satisfies the following inequality: 
		\begin{equation} \label{SA-EQ-E8}
			\| \nabla_{ \bm{x} } \Phi_{i} ( \bm{p}_{i}^{*} | \bm{x} ) - \nabla_{ \bm{y} } \Phi_{i} ( \bm{p}_{i}^{*} | \bm{y} ) \|_{\infty} \leq L_{2} \| \bm{x} - \bm{y} \|_{2} .
		\end{equation}
	Based on \eqref{SA-EQ-E8}, we have 
		\begin{equation}
			\begin{aligned}
				&\| \nabla_{ \bm{x} } \Phi_{i} ( \bm{p}_{i}^{*} | \bm{x} ) - \nabla_{ \bm{y} } \Phi_{i} ( \bm{p}_{i}^{*} | \bm{y} ) \|_{2} \\
				&\leq N_{2} \| \nabla_{ \bm{x} } \Phi_{i} ( \bm{p}_{i}^{*} | \bm{x} ) - \nabla_{ \bm{y} } \Phi_{i} ( \bm{p}_{i}^{*} | \bm{y} ) \|_{\infty} \leq L_{f} \| \bm{x} - \bm{y} \|_{2},
			\end{aligned}
		\end{equation}
	where $N_{2} \triangleq \| \nabla_{ \bm{x} } \Phi_{i} ( \bm{p}_{i}^{*} | \bm{x} ) - \nabla_{ \bm{y} } \Phi_{i} ( \bm{p}_{i}^{*} | \bm{y} ) \|_{0}^{ 1 / 2 }$ and $L_{f} \triangleq N_{2} L_{2}$. This completes the proof.	
}

	\bibliographystyle{IEEEtran}
	\bibliography{ref}

\begin{thebibliography}{10}
\providecommand{\url}[1]{#1}
\csname url@samestyle\endcsname
\providecommand{\newblock}{\relax}
\providecommand{\bibinfo}[2]{#2}
\providecommand{\BIBentrySTDinterwordspacing}{\spaceskip=0pt\relax}
\providecommand{\BIBentryALTinterwordstretchfactor}{4}
\providecommand{\BIBentryALTinterwordspacing}{\spaceskip=\fontdimen2\font plus
\BIBentryALTinterwordstretchfactor\fontdimen3\font minus
  \fontdimen4\font\relax}
\providecommand{\BIBforeignlanguage}[2]{{%
\expandafter\ifx\csname l@#1\endcsname\relax
\typeout{** WARNING: IEEEtran.bst: No hyphenation pattern has been}%
\typeout{** loaded for the language `#1'. Using the pattern for}%
\typeout{** the default language instead.}%
\else
\language=\csname l@#1\endcsname
\fi
#2}}
\providecommand{\BIBdecl}{\relax}
\BIBdecl

\bibitem{8664630}
S.~Wang, T.~Tuor, T.~Salonidis, K.~K. Leung, C.~Makaya, T.~He, and K.~Chan,
  ``Adaptive federated learning in resource-constrained edge computing
  systems,'' \emph{{IEEE} J. Sel. Areas Commun.}, vol.~37, no.~6, pp.
  1205--1221, Jun. 2019.

\bibitem{Toward2017Dawy}
Z.~{Dawy}, W.~{Saad}, A.~{Ghosh}, J.~G. {Andrews}, and E.~{Yaacoub}, ``Toward
  massive machine type cellular communications,'' \emph{IEEE Wireless Commun.},
  vol.~24, no.~1, pp. 120--128, Feb. 2017.

\bibitem{8970161}
G.~Zhu, D.~Liu, Y.~Du, C.~You, J.~Zhang, and K.~Huang, ``Toward an intelligent
  edge: Wireless communication meets machine learning,'' \emph{{IEEE} Commun.
  Mag.}, vol.~58, no.~1, pp. 19--25, Jan. 2020.

\bibitem{Edge2020Deng}
S.~Deng, H.~Zhao, W.~Fang, J.~Yin, S.~Dustdar, and A.~Y. Zomaya, ``Edge
  intelligence: The confluence of edge computing and artificial intelligence,''
  \emph{IEEE Internet Things J.}, vol.~7, no.~8, pp. 7457--7469, Aug. 2020.

\bibitem{2024789118}
M.~Chen, N.~Shlezinger, H.~V. Poor, Y.~C. Eldar, and S.~Cui,
  ``Communication-efficient federated learning,'' \emph{Proc. Nat. Acad. Sci.
  USA}, vol. 118, no.~17, Apr. 2021, {A}rt. no. e2024789118.

\bibitem{9606720}
K.~B. Letaief, Y.~Shi, J.~Lu, and J.~Lu, ``Edge artificial intelligence for
  {6G}: {Vision}, enabling technologies, and applications,'' \emph{{IEEE} J.
  Sel. Areas Commun.}, vol.~40, no.~1, pp. 5--36, Jan. 2022.

\bibitem{9380667}
A.~Badi and I.~Mahgoub, ``\text{ReapIoT}: Reliable, energy-aware network
  protocol for large-scale {Internet-of-Things} ({IoT}) applications,''
  \emph{IEEE Internet Things J.}, vol.~8, no.~17, pp. 13\,582--13\,592, Sept.
  2021.

\bibitem{Spatial2019Cui}
W.~{Cui}, K.~{Shen}, and W.~{Yu}, ``Spatial deep learning for wireless
  scheduling,'' \emph{IEEE J. Sel. Areas Commun.}, vol.~37, no.~6, pp.
  1248--1261, Jun. 2019.

\bibitem{Edge2020Li}
E.~{Li}, L.~{Zeng}, Z.~{Zhou}, and X.~{Chen}, ``Edge {AI}: On-demand
  accelerating deep neural network inference via edge computing,'' \emph{IEEE
  Trans. Wireless Commun.}, vol.~19, no.~1, pp. 447--457, Jan. 2020.

\bibitem{4027772}
Q.~Cheng, B.~Chen, and P.~K. Varshney, ``Detection performance limits for
  distributed sensor networks in the presence of nonideal channels,''
  \emph{{IEEE} Trans. Wireless Commun.}, vol.~5, no.~11, pp. 3034--3038, Nov.
  2006.

\bibitem{9344705}
D.~Ciuonzo, P.~S. Rossi, and P.~K. Varshney, ``Distributed detection in
  wireless sensor networks under multiplicative fading via generalized score
  tests,'' \emph{{IEEE} Internet Things J.}, vol.~8, no.~11, pp. 9059--9071,
  Jun. 2021.

\bibitem{8809196}
X.~Cheng, D.~Ciuonzo, and P.~S. Rossi, ``Multibit decentralized detection
  through fusing smart and dumb sensors based on {Rao} test,'' \emph{{IEEE}
  Trans. Aerosp. Electron. Syst.}, vol.~56, no.~2, pp. 1391--1405, Apr. 2020.

\bibitem{9606667}
J.~Shao, Y.~Mao, and J.~Zhang, ``Learning task-oriented communication for edge
  inference: An information bottleneck approach,'' \emph{{IEEE} J. Sel. Areas
  Commun.}, vol.~40, no.~1, pp. 197--211, Nov. 2022.

\bibitem{2207-00969}
\BIBentryALTinterwordspacing
D.~Wen, P.~Liu, G.~Zhu, Y.~Shi, J.~Xu, Y.~C. Eldar, and S.~Cui, ``Task-oriented
  sensing, computation, and communication integration for multi-device edge
  \text{AI},'' 2022. [Online]. Available:
  \url{https://doi.org/10.48550/arXiv.2207.00969}
\BIBentrySTDinterwordspacing

\bibitem{2206-05949}
\BIBentryALTinterwordspacing
P.~Liu, G.~Zhu, S.~Wang, W.~Jiang, W.~Luo, H.~V. Poor, and S.~Cui, ``Toward
  ambient intelligence: Federated edge learning with task-oriented sensing,
  computation, and communication integration,'' 2022. [Online]. Available:
  \url{https://doi.org/10.48550/arXiv.2206.05949}
\BIBentrySTDinterwordspacing

\bibitem{9653664}
H.~Xie, Z.~Qin, and G.~Y. Li, ``Task-oriented multi-user semantic
  communications for \text{VQA},'' \emph{IEEE Wireless Comm. Lett.}, vol.~11,
  no.~3, pp. 553--557, Mar. 2022.

\bibitem{Machine2020Wang}
S.~{Wang}, Y.-C. {Wu}, M.~{Xia}, R.~{Wang}, and H.~V. {Poor}, ``Machine
  intelligence at the edge with learning-centric power allocation,'' \emph{IEEE
  Trans. Wireless Commun.}, vol.~19, no.~11, pp. 7293--7308, Nov. 2020.

\bibitem{9252948}
H.~Xie and Z.~Qin, ``A lite distributed semantic communication system for
  {Internet} of {Things},'' \emph{{IEEE} J. Sel. Areas Commun.}, vol.~39,
  no.~1, pp. 142--153, Jan. 2021.

\bibitem{456056}
H.~S. Seung, H.~Sompolinsky, and N.~Tishby, ``Statistical mechanics of learning
  from examples,'' \emph{Phys. Rev.}, vol.~45, no.~8, p. 6056, Apr. 1991.

\bibitem{Joint2020Shi}
W.~{Shi}, S.~{Zhou}, Z.~{Niu}, M.~{Jiang}, and L.~{Geng}, ``Joint device
  scheduling and resource allocation for latency constrained wireless federated
  learning,'' \emph{IEEE Trans. Wireless Commun.}, vol.~20, no.~1, pp.
  453--467, Jan. 2021.

\bibitem{Energy2020Zeng}
Q.~{Zeng}, Y.~{Du}, K.~{Huang}, and K.~K. {Leung}, ``Energy-efficient radio
  resource allocation for federated edge learning,'' in \emph{Proc. Int. Conf.
  Commun. (ICC)}, Jun. 2020, pp. 1--6.

\bibitem{ADMM2018Lu}
C.~{Lu}, J.~{Feng}, S.~{Yan}, and Z.~{Lin}, ``A unified alternating direction
  method of multipliers by majorization minimization,'' \emph{IEEE Trans.
  Pattern Anal. Mach. Intell.}, vol.~40, no.~3, pp. 527--541, Mar. 2018.

\bibitem{ADMM2012He}
B.~He, M.~Tao, and X.~Yuan, ``Alternating direction method with \text{G}aussian
  back substitution for separable convex programming,'' \emph{SIAM J.
  Optimiz.}, vol.~22, no.~2, pp. 313--340, May 2012.

\bibitem{Parallel1997Dimitri}
D.~P. Bertsekas and J.~N. Tsitsiklis, \emph{Parallel and Distributed
  Computation: Numerical Methods}.\hskip 1em plus 0.5em minus 0.4em\relax
  Athena Scientific, Belmont, Massachusetts, 1997.

\bibitem{Activity2019Li}
Y.~{Li}, M.~{Xia}, and Y.-C. {Wu}, ``Activity detection for massive
  connectivity under frequency offsets via first-order algorithms,'' \emph{IEEE
  Trans. Wireless Commun.}, vol.~18, no.~3, pp. 1988--2002, Mar. 2019.

\bibitem{lu2015fast}
C.~Lu, H.~Li, Z.~Lin, and S.~Yan, ``Fast proximal linearized alternating
  direction method of multiplier with parallel splitting,'' in \emph{Proc.
  Conf. Artif. Intell. (AAAI)}, Feb. 2016, pp. 739--745.

\bibitem{Scikit2011Pedregosa}
F.~Pedregosa, G.~Varoquaux, A.~Gramfort, and et~al., ``Scikit-learn: Machine
  learning in {Python},'' \emph{J. Mach. Learn. Res.}, vol.~12, pp. 2825--2830,
  Nov. 2011.

\bibitem{MNIST2012Deng}
L.~{Deng}, ``The \text{MNIST} database of handwritten digit images for machine
  learning research,'' \emph{IEEE Signal Process. Mag.}, vol.~29, no.~6, pp.
  141--142, Nov. 2012.

\bibitem{Deep2016Kaiming}
K.~He, X.~Zhang, S.~Ren, and J.~Sun, ``Deep residual learning for image
  recognition,'' in \emph{Proc. Comp. Vis. Pat. Rec. (CVPR)}, Jun. 2016, pp.
  770--778.

\bibitem{PointNet2017Charles}
C.~R. Qi, H.~Su, K.~Mo, and L.~J. Guibas, ``Pointnet: Deep learning on point
  sets for 3\text{D} classification and segmentation,'' in \emph{Proc. Comp.
  Vis. Pat. Rec. (CVPR)}, Jun. 2017, pp. 77--85.

\bibitem{Achieving2012Shatri}
H.~{Al-Shatri} and T.~{Weber}, ``Achieving the maximum sum rate using
  \text{D.C.} programming in cellular networks,'' \emph{IEEE Trans. Signal
  Process.}, vol.~60, no.~3, pp. 1331--1341, Mar. 2012.

\bibitem{9222214}
J.~A. de~Carvalho, D.~B. da~Costa, L.~Yang, G.~C. Alexandropoulos, R.~Oliveira,
  and U.~S. Dias, ``User fairness in wireless powered communication networks
  with non-orthogonal multiple access,'' \emph{IEEE Wireless Commun. Lett.},
  vol.~10, no.~1, pp. 189--193, Jan. 2021.

\bibitem{abs-2206-01748}
\BIBentryALTinterwordspacing
S.~Wang, C.~Li, Q.~Hao, C.~Xu, D.~W.~K. Ng, Y.~C. Eldar, and H.~V. Poor,
  ``Federated deep learning meets autonomous vehicle perception: Design and
  verification.'' [Online]. Available:
  \url{https://doi.org/10.48550/arXiv.2206.01748}
\BIBentrySTDinterwordspacing

\bibitem{8523680}
L.~Meng, G.~Li, J.~Yan, and Y.~Gu, ``A general framework for understanding
  compressed subspace clustering algorithms,'' \emph{{IEEE} J. Sel. Top. Signal
  Process.}, vol.~12, no.~6, pp. 1504--1519, Dec. 2018.

\bibitem{Convex2015S}
S.~{Bubeck}, ``Convex optimization: Algorithms and complexity,'' \emph{Found.
  Trends Mach. Learn.}, vol.~8, no. 3-4, pp. 231--357, Nov. 2015.

\end{thebibliography}

		\begin{IEEEbiography}
			[{\includegraphics[width=1in, height=1.25in, clip, keepaspectratio]{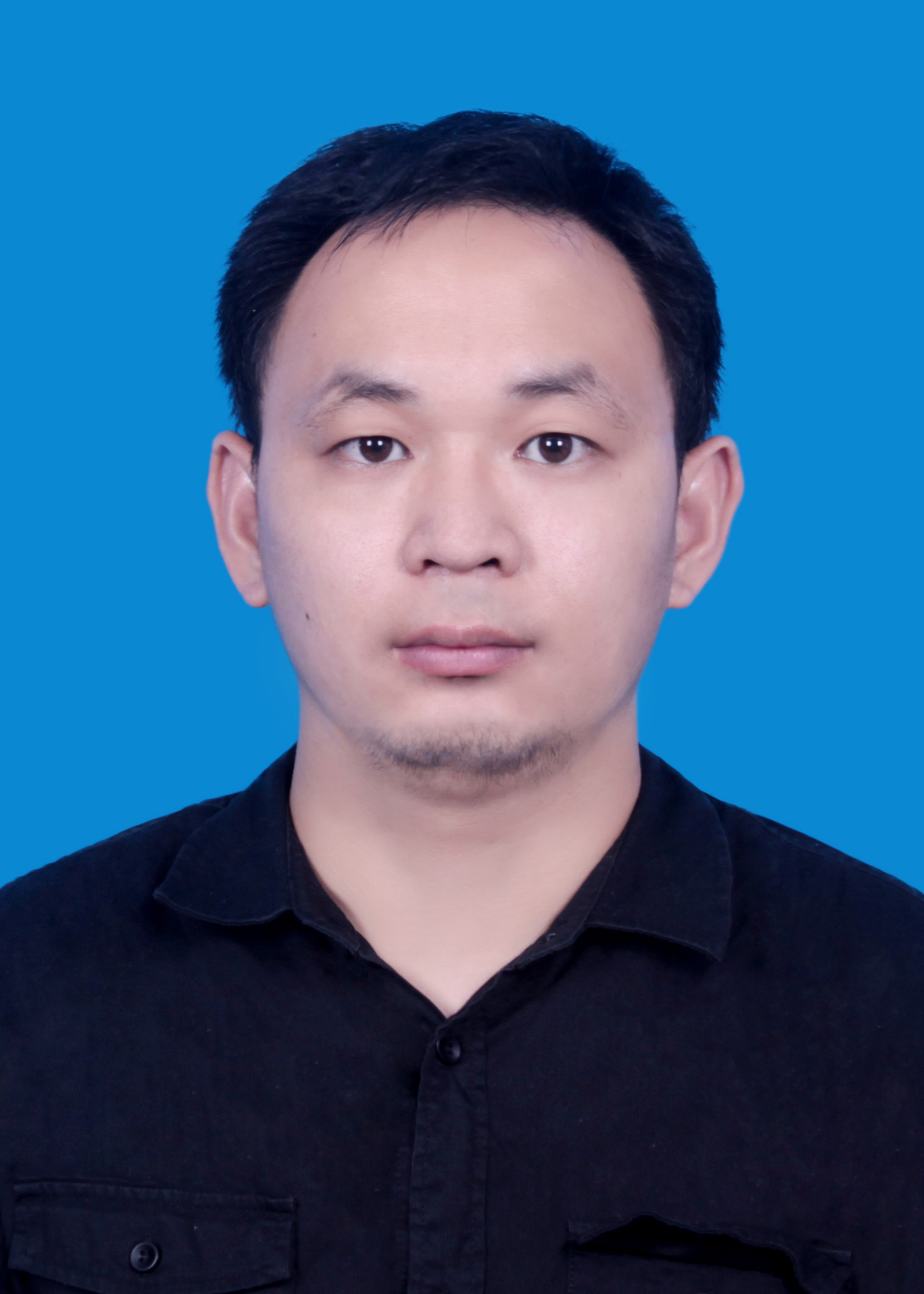}}]{Haihui Xie} received the B.S. degree and the M.S. degree in photonic and electronic engineering from Fujian Normal University, Fuzhou, China, in 2014 and 2016, respectively. Currently, he is pursuing the Ph.D. degree in information and communication engineering at Sun Yat-sen University, Guangzhou, China. His research interests include edge learning, optimization, and their applications in wireless communications.
		\end{IEEEbiography}

		\begin{IEEEbiography}
			[{\includegraphics[width=1in, height=1.25in, clip, keepaspectratio]{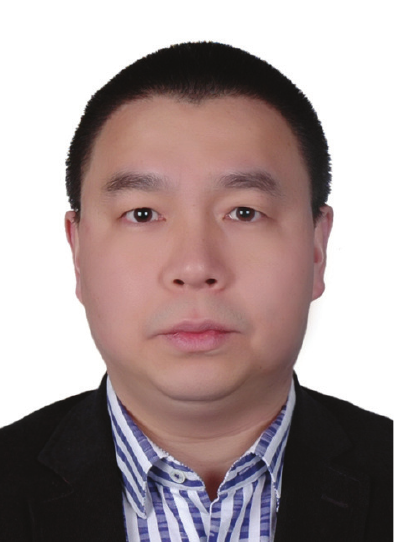}}]{Minghua Xia} (Senior Member, IEEE) received the Ph.D. degree in Telecommunications and Information Systems from Sun Yat-sen University, Guangzhou, China, in 2007.
			
			From 2007 to 2009, he was with the Electronics and Telecommunications Research Institute (ETRI) of South Korea, Beijing R\&D Center, Beijing, China, where he worked as a member and then as a senior member of the engineering staff. From 2010 to 2014, he was in sequence with The University of Hong Kong, Hong Kong, China; King Abdullah University of Science and Technology, Jeddah, Saudi Arabia; and the Institut National de la Recherche Scientifique (INRS), University of Quebec, Montreal, Canada, as a Postdoctoral Fellow. Since 2015, he has been a Professor at Sun Yat-sen University. Since 2019, he has also been an Adjunct Professor with the Southern Marine Science and Engineering Guangdong Laboratory (Zhuhai). His research interests are in the general areas of wireless communications and signal processing.
		\end{IEEEbiography}

		\begin{IEEEbiography}
			[{\includegraphics[width=1in, height=1.25in, clip, keepaspectratio]{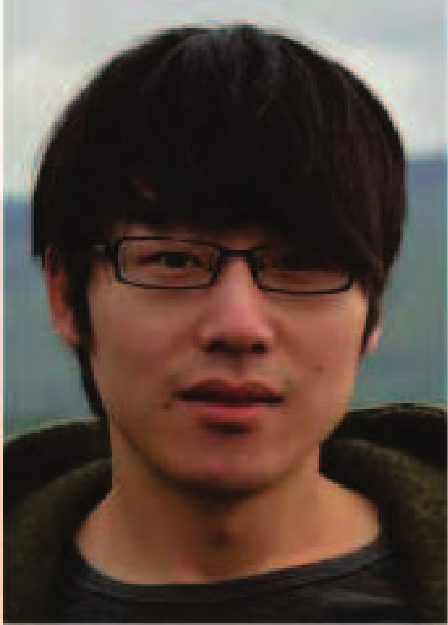}}]{Peiran Wu } (Member, IEEE) received the Ph.D. degree in electrical and computer engineering from The University of British Columbia (UBC), Vancouver, Canada, in 2015.
			
			From October 2015 to December 2016, he was a Post-Doctoral Fellow at UBC. In the Summer of 2014, he was a Visiting Scholar with the Institute for Digital Communications, Friedrich-Alexander-University Erlangen-Nuremberg (FAU), Erlangen, Germany. Since February 2017, he has been with Sun Yat-sen University, Guangzhou, China, where he is currently an Associate Professor. Since 2019, he has been an Adjunct Associate Professor with the Southern Marine Science and Engineering Guangdong Laboratory, Zhuhai, China. His research interests include mobile edge computing, wireless power transfer, and energy-efficient wireless communications. 
			
			Dr. Wu was a recipient of the Fourth-Year Fellowship in 2010, the C. L. Wang Memorial Fellowship in 2011, the Graduate Support Initiative (GSI) Award from UBC in 2014, the German Academic Exchange Service (DAAD) Scholarship in 2014, and the Chinese Government Award for Outstanding Self-Financed Students Abroad in 2014.
		\end{IEEEbiography}
	
		\begin{IEEEbiography}
			[{\includegraphics[width=1in, height=1.25in, clip, keepaspectratio]{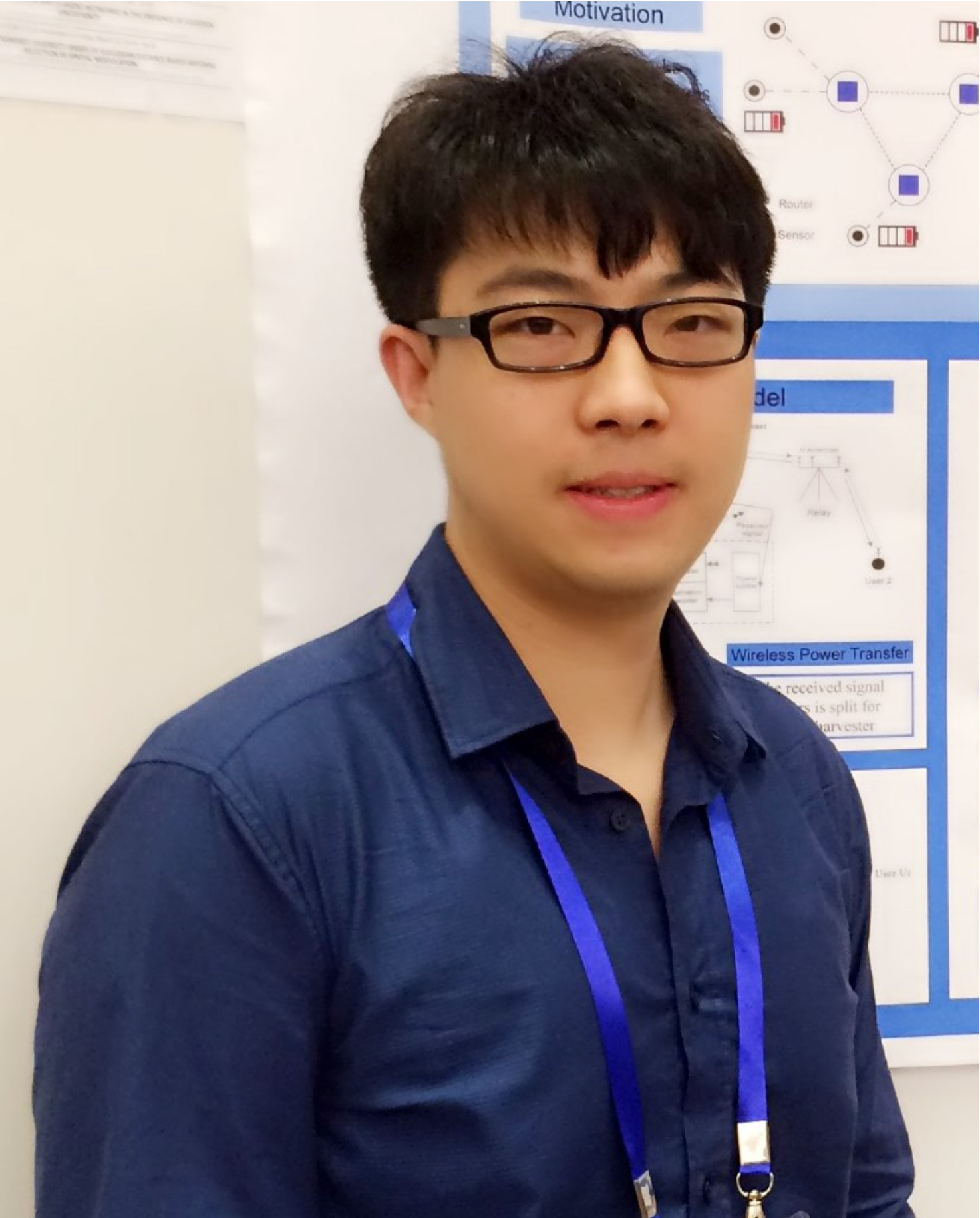}}]{Shuai Wang} (M'19)
			received the Ph.D. degree in Electrical and Electronic Engineering from The University of Hong Kong (HKU) in 2018. From 2018 to 2021, he was a Postdoc Fellow at HKU and then a Research Assistant Professor at the Southern University of Science and Technology (SUSTech). Currently, he is an Associate Professor with the Shenzhen Institute of Advanced Technology (SIAT), Chinese Academy of Sciences. His research interests include autonomous driving, machine learning, and communication networks. Dr. Wang has published 40+ journal papers and 20+ conference papers. He has received various awards from IEEE ICC, IEEE SPCC, IEEE ICCCS, IEEE TWC, IEEE WCL, and National 5G Competition.
		\end{IEEEbiography}
	
		\begin{IEEEbiography}
			[{\includegraphics[width=1in, height=1.25in, clip, keepaspectratio]{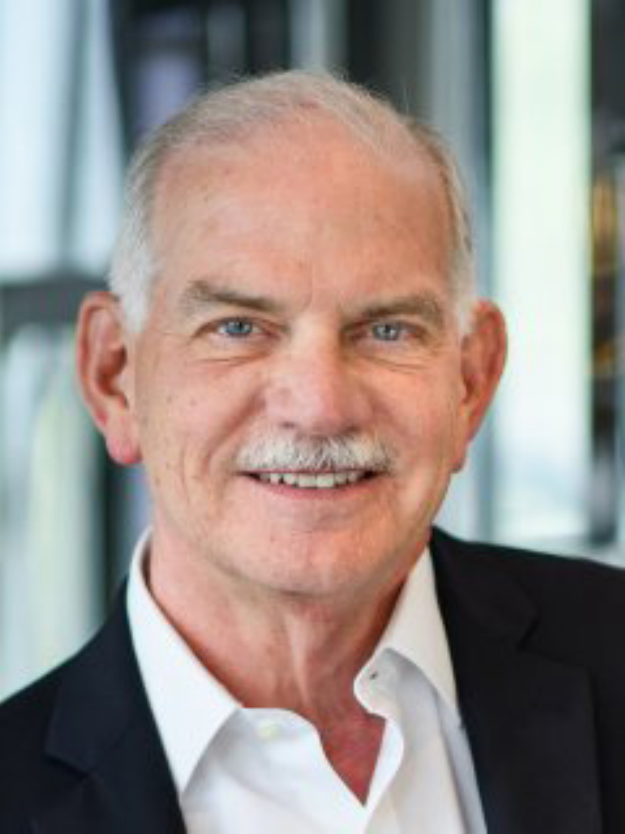}}]{H. Vincent Poor} (S’72, M’77, SM’82, F’87) received the Ph.D. degree in EECS from Princeton University in 1977.  From 1977 until 1990, he was on the faculty of the University of Illinois at Urbana-Champaign. Since 1990 he has been on the faculty at Princeton, where he is currently the Michael Henry Strater University Professor. During 2006 to 2016, he served as the dean of Princeton’s School of Engineering and Applied Science. He has also held visiting appointments at several other universities, including most recently at Berkeley and Cambridge. His research interests are in the areas of information theory, machine learning and network science, and their applications in wireless networks, energy systems and related fields. Among his publications in these areas is the recent book Machine Learning and Wireless Communications.  (Cambridge University Press, 2022). Dr. Poor is a member of the National Academy of Engineering and the National Academy of Sciences and is a foreign member of the Chinese Academy of Sciences, the Royal Society, and other national and international academies. He received the IEEE Alexander Graham Bell Medal in 2017.
		\end{IEEEbiography}

\end{document}